\documentclass[1p]{elsarticle}
\usepackage{pstricks,graphicx,color,tikz}
\usepackage{enumerate}
\usepackage{amsmath,amssymb,latexsym,url,amsthm}
\usepackage{subfigure}
\usepackage{longtable}
\usepackage{array}
\usepackage{enumitem}

\newcommand{\be}{\begin{equation}}
\newcommand{\ee}{\end{equation}}
\newcommand{\ba}{\begin{eqnarray}}
\newcommand{\ea}{\end{eqnarray}}
\newcommand{\ban}{\begin{eqnarray*}}
\newcommand{\ean}{\end{eqnarray*}}
\newcommand{\bitem}{\begin{itemize}}
\newcommand{\eitem}{\end{itemize}}
\newcommand{\benum}{\begin{enumerate}}
\newcommand{\eenum}{\end{enumerate}}

\def\mc{\mathcal{C}} 
\def\me{\mathcal{E}} 
\def\mm{\mathcal{M}} 
\def\mn{\mathcal{N}} 

\def\ro{{\rm out}}
\def\ri{{\rm in}}

\def\hl{\medskip\noindent\textbf}
\def\rms{\medskip\noindent\textit}

\newcolumntype{x}[1]{>{\centering\arraybackslash\hspace{0pt}}p{#1}}

\def\onesize{0.35}

\def\system		{\includegraphics[scale=0.4]{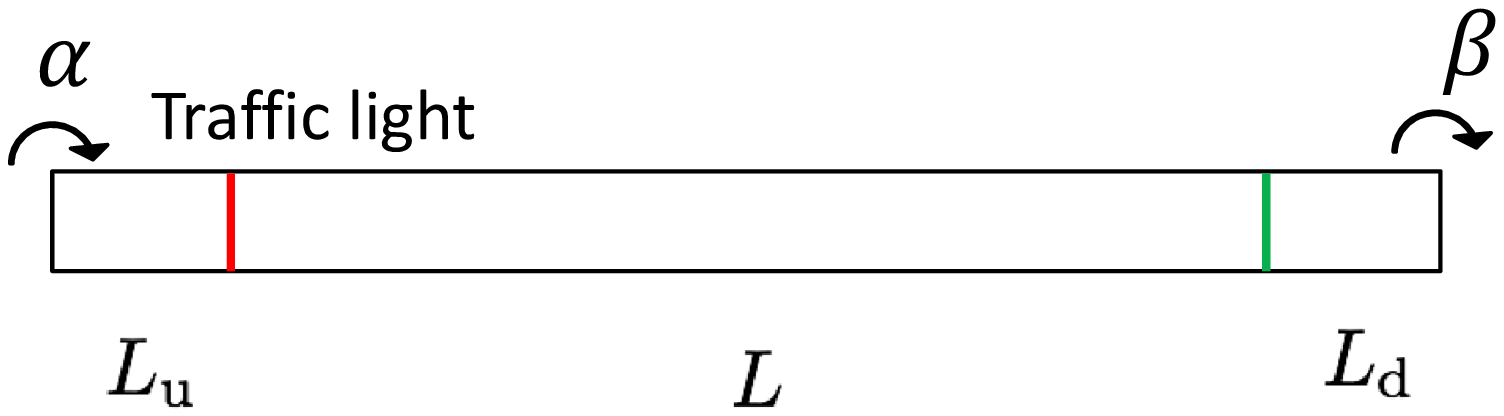}}%
\def\naschFD		{\includegraphics[scale=0.38]{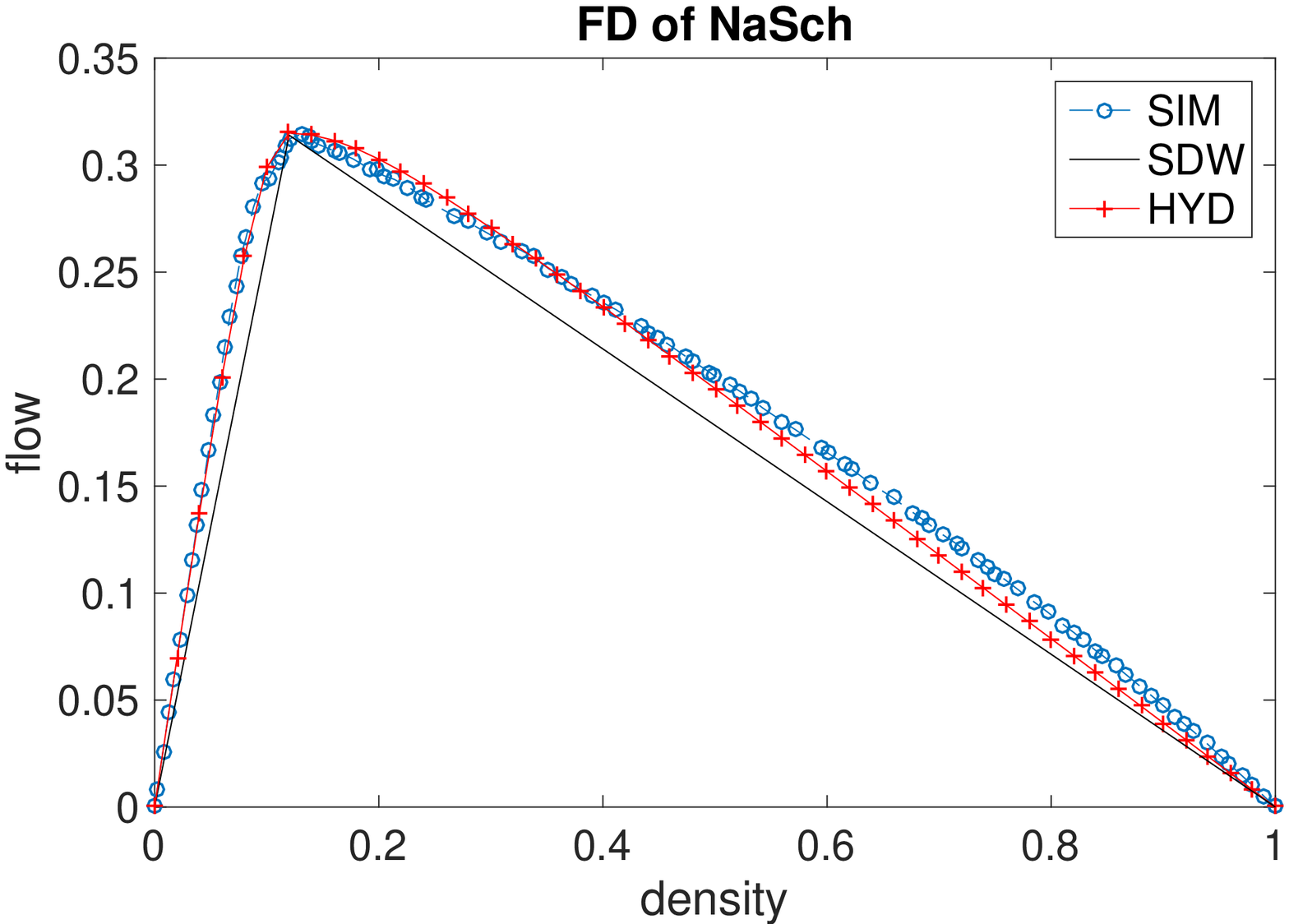}}%
\def\asepTrajectoryLarged		{\includegraphics[scale=0.26]{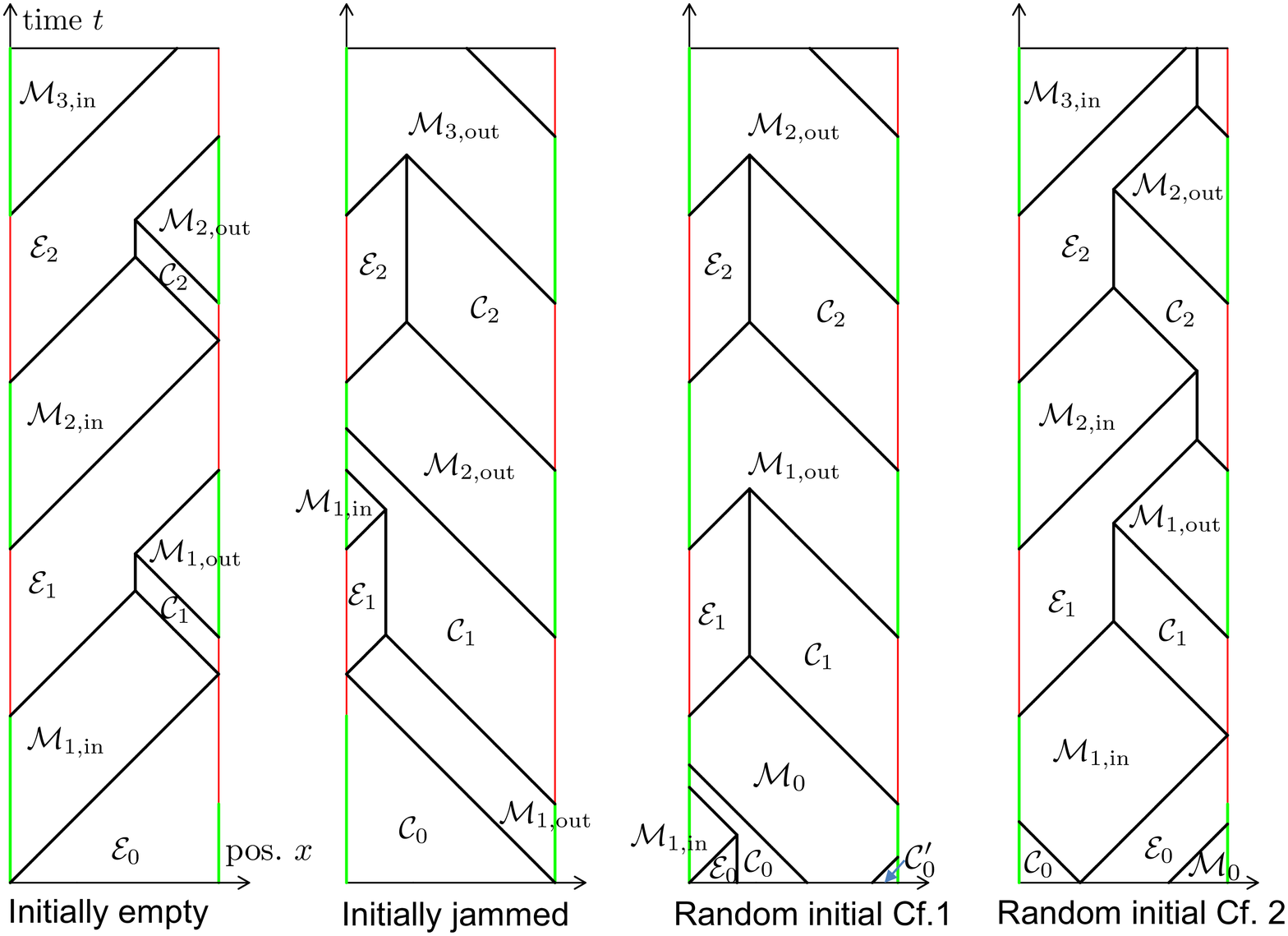}}%
\def\asepTrajectorySmallAlld		{\includegraphics[scale=0.42]{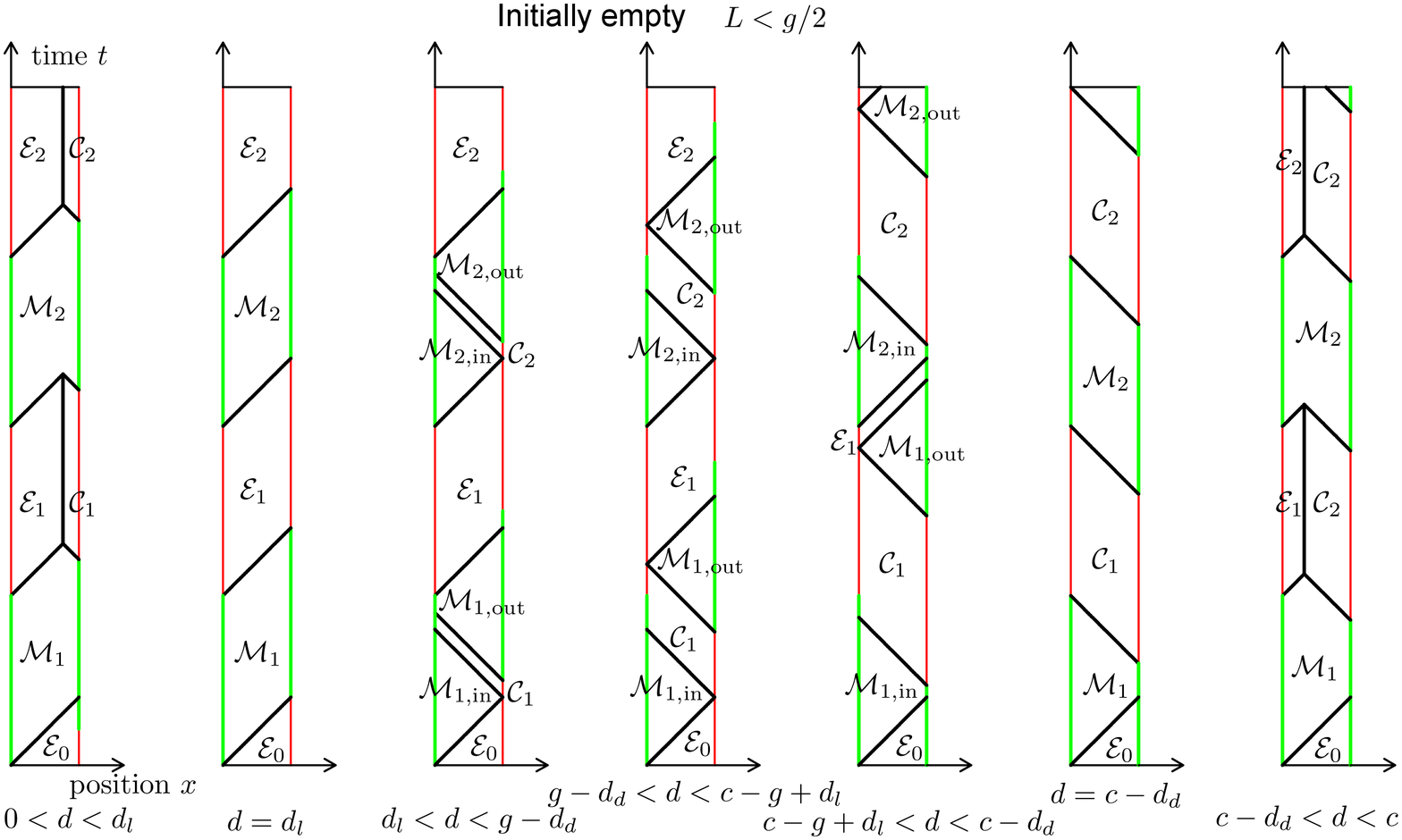}}%
\def\naschTrajectorycmp		{\includegraphics[scale=0.16]{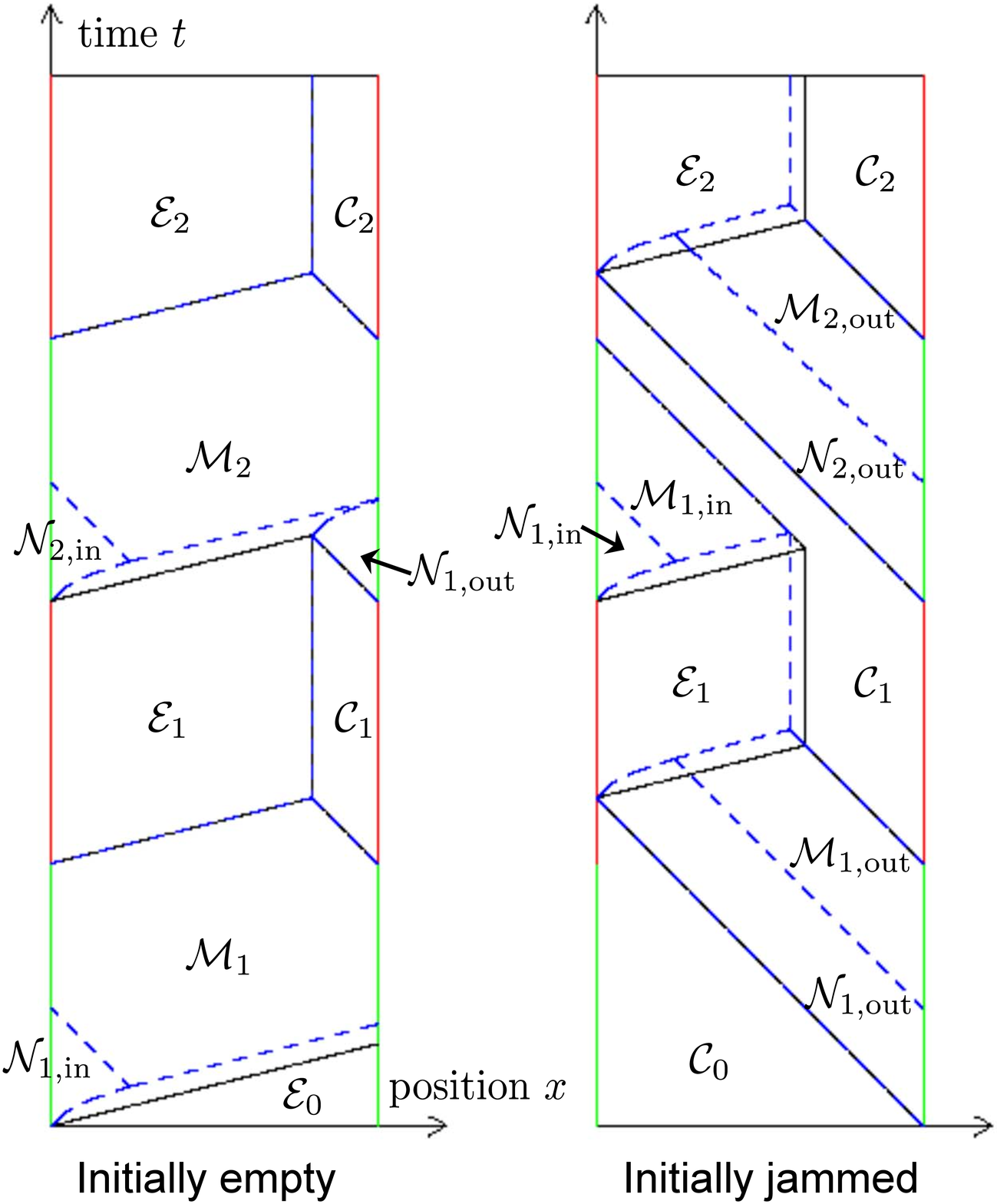}}%

\def\ASEPflowcycleC			{\includegraphics[scale=\onesize]{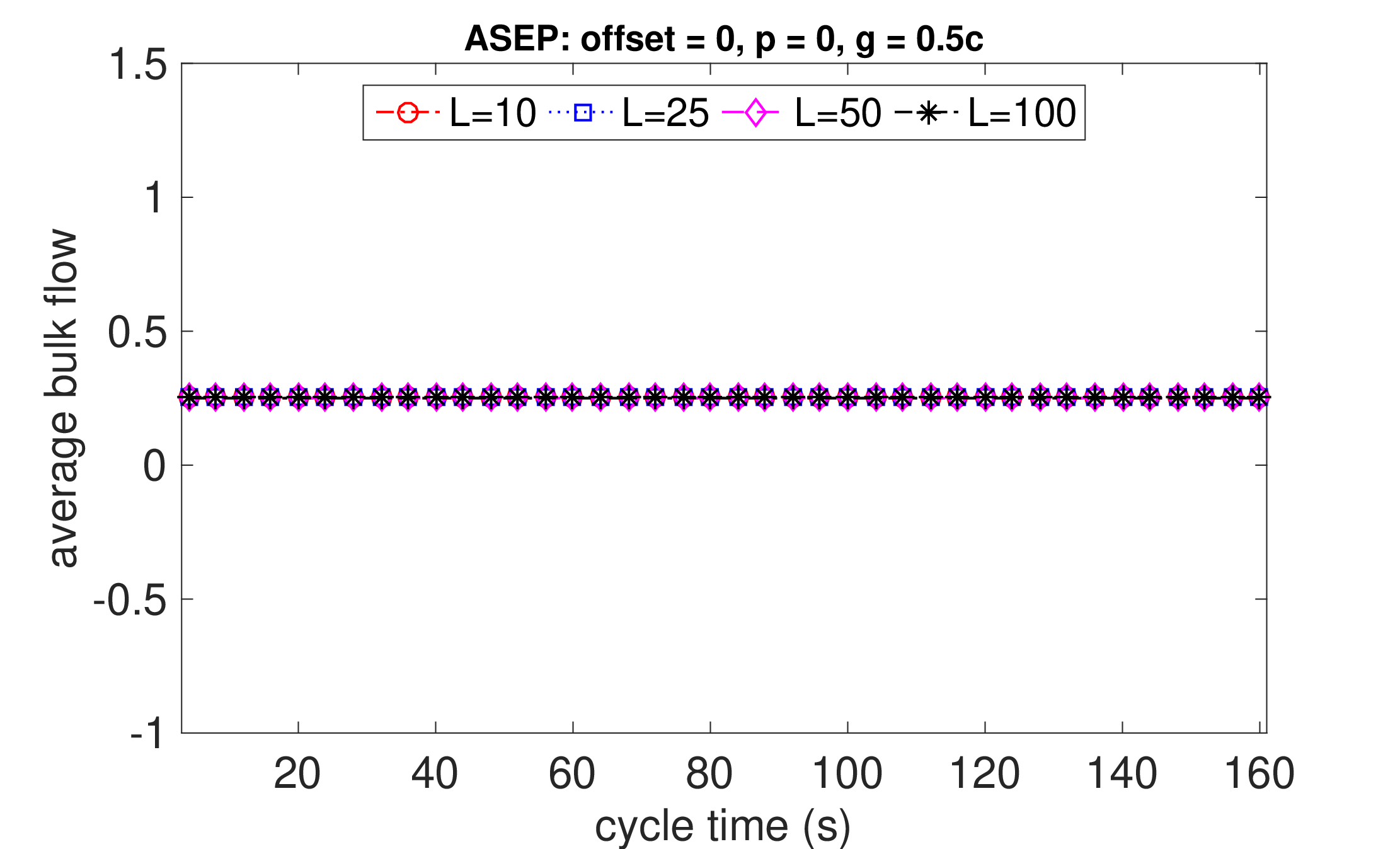}}%
\def\ASEPflowoffset				{\includegraphics[scale=0.3]{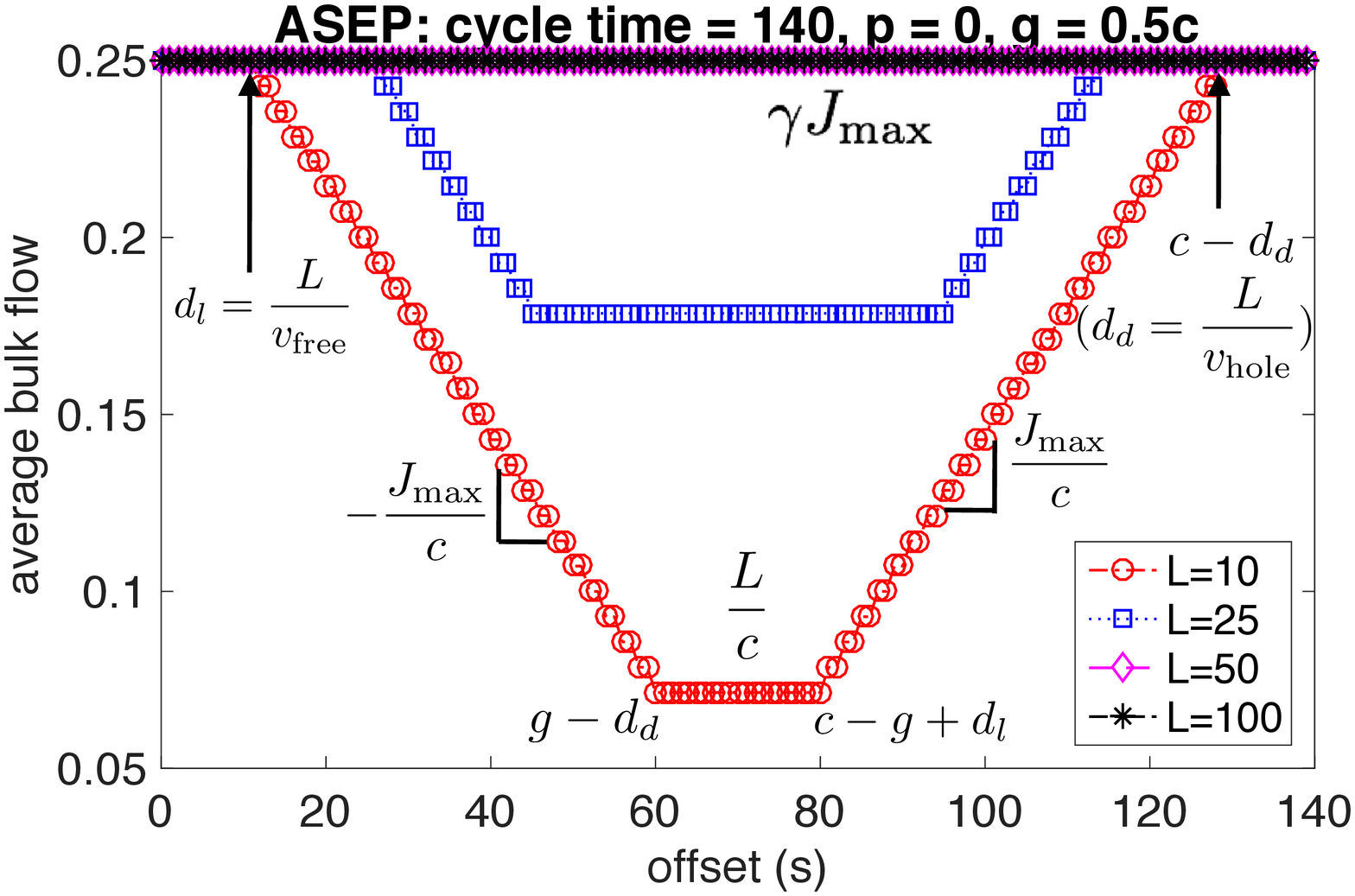}}%
\def\ASEPflowratio				{\includegraphics[scale=0.29]{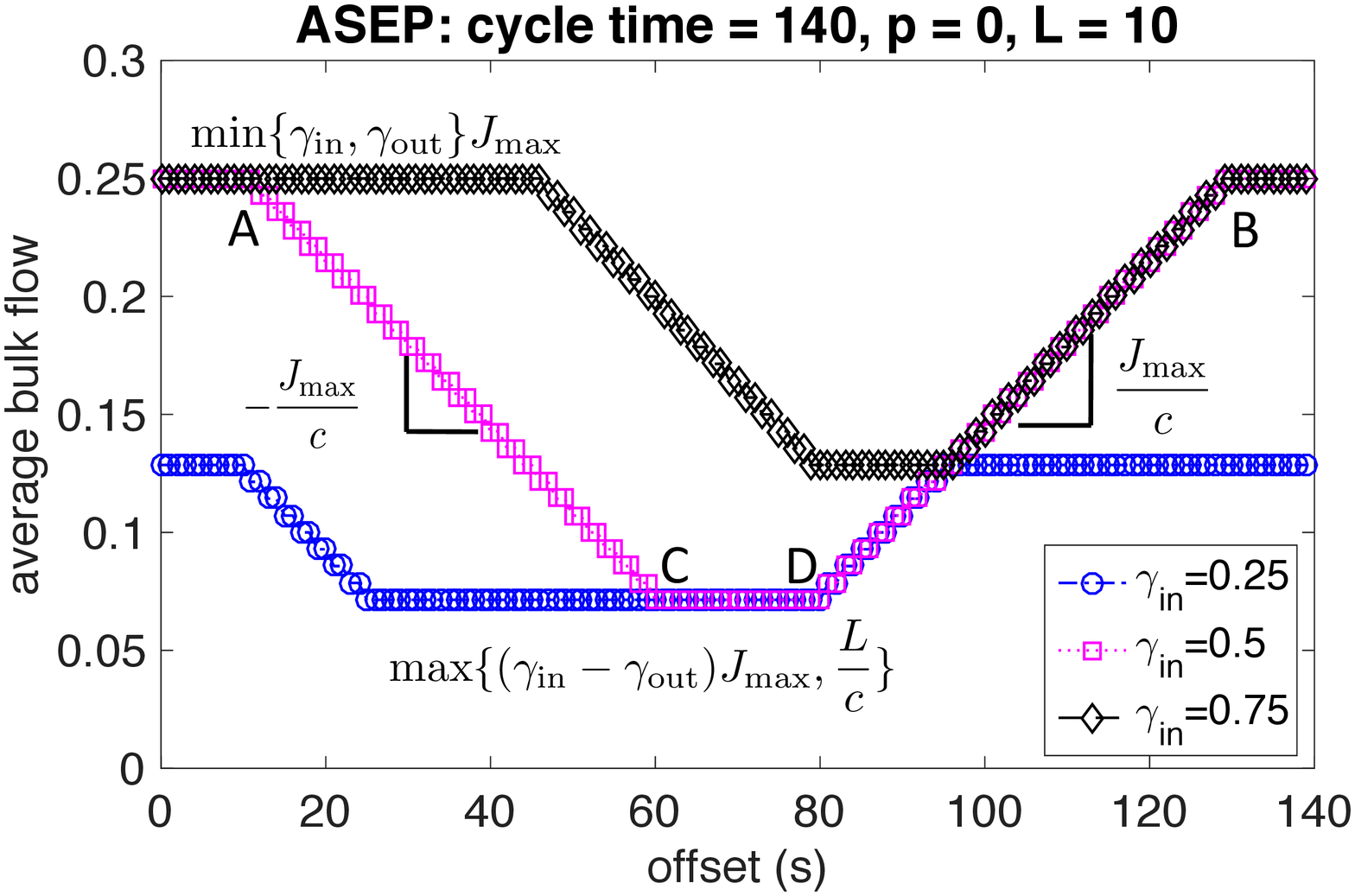}}%
\def\NaSchflowcycle				{\includegraphics[scale=\onesize]{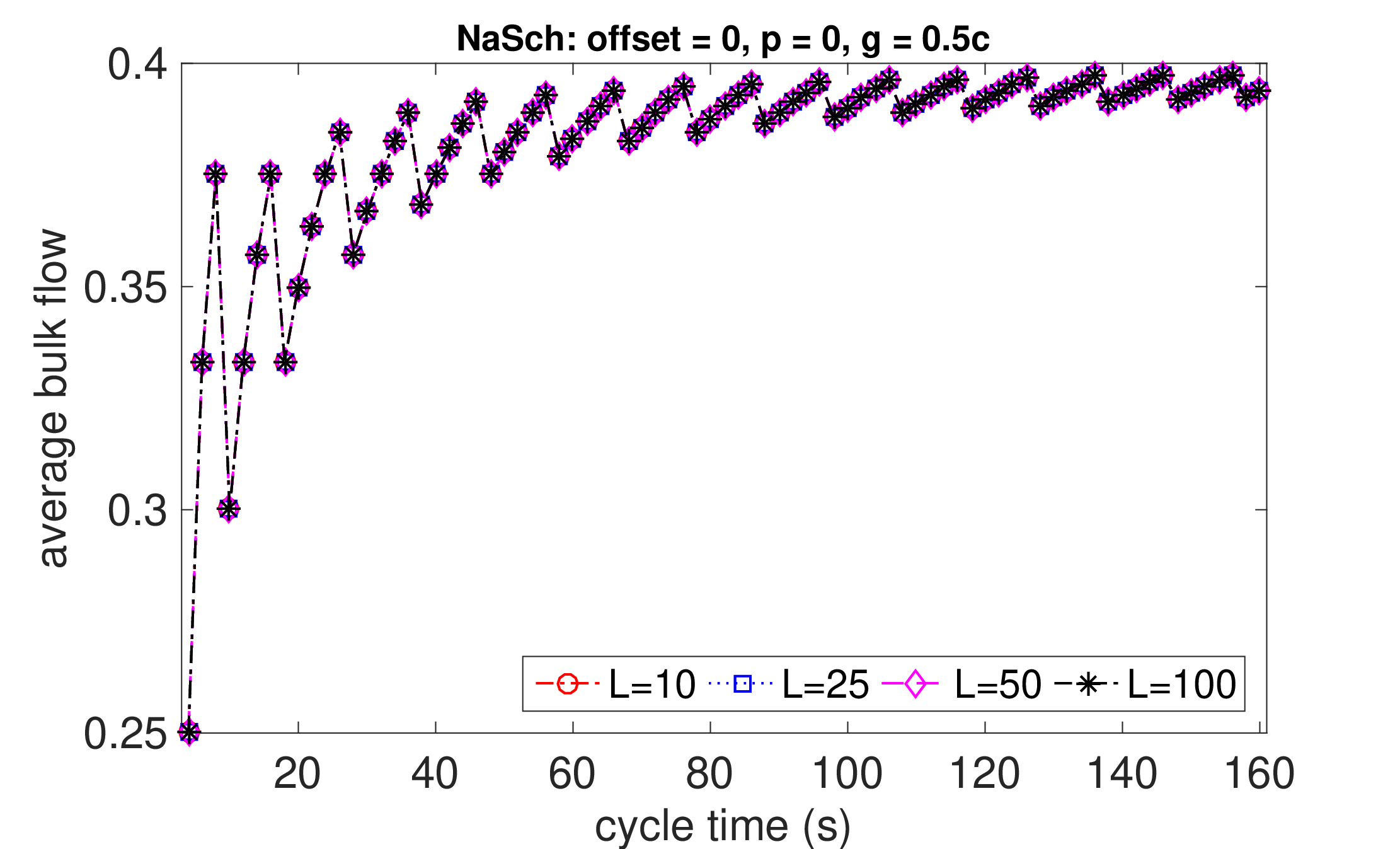}}%
\def\NaSchflowoffset				{\includegraphics[scale=\onesize]{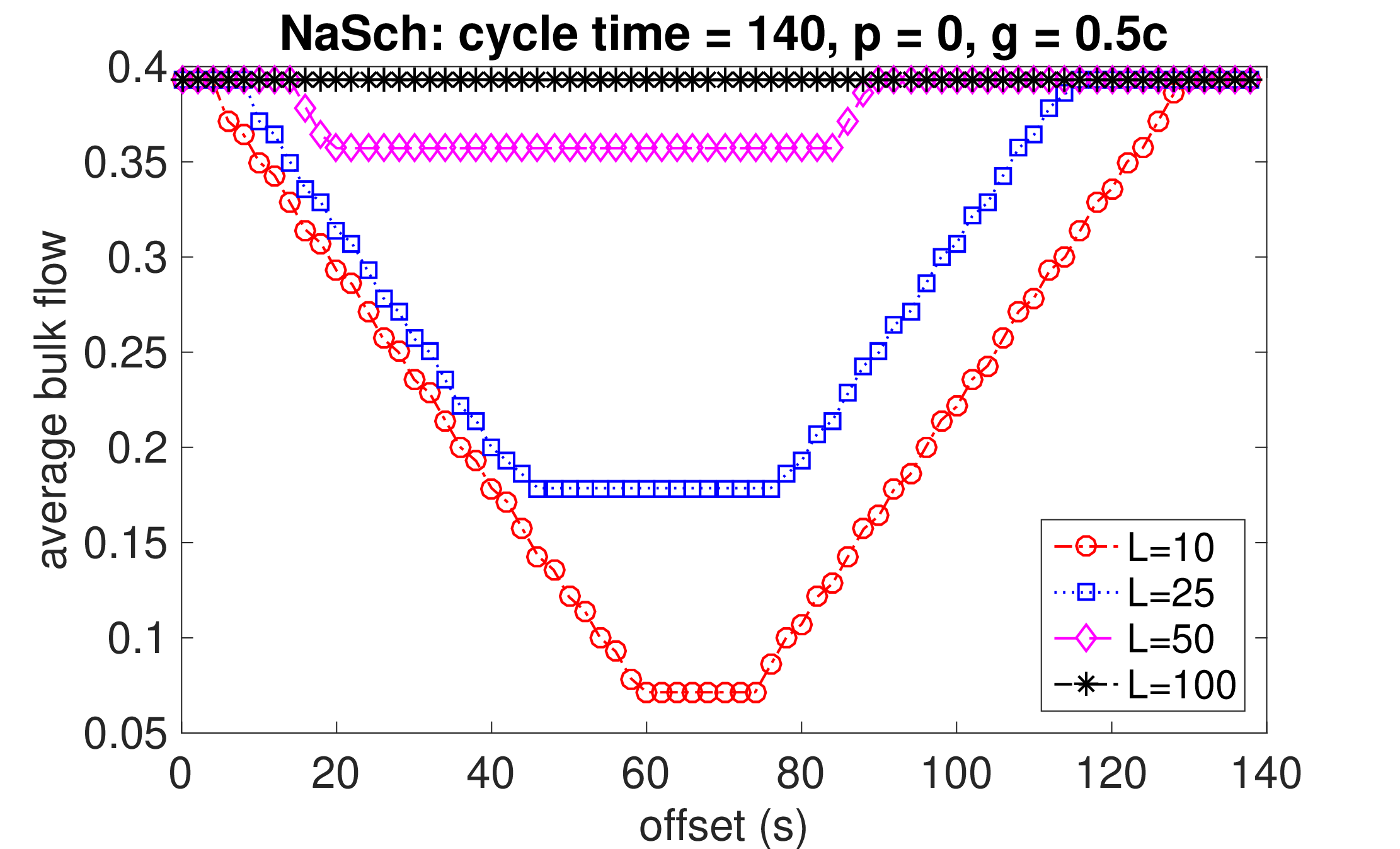}}%
\def\NaSchflowratio				{\includegraphics[scale=\onesize]{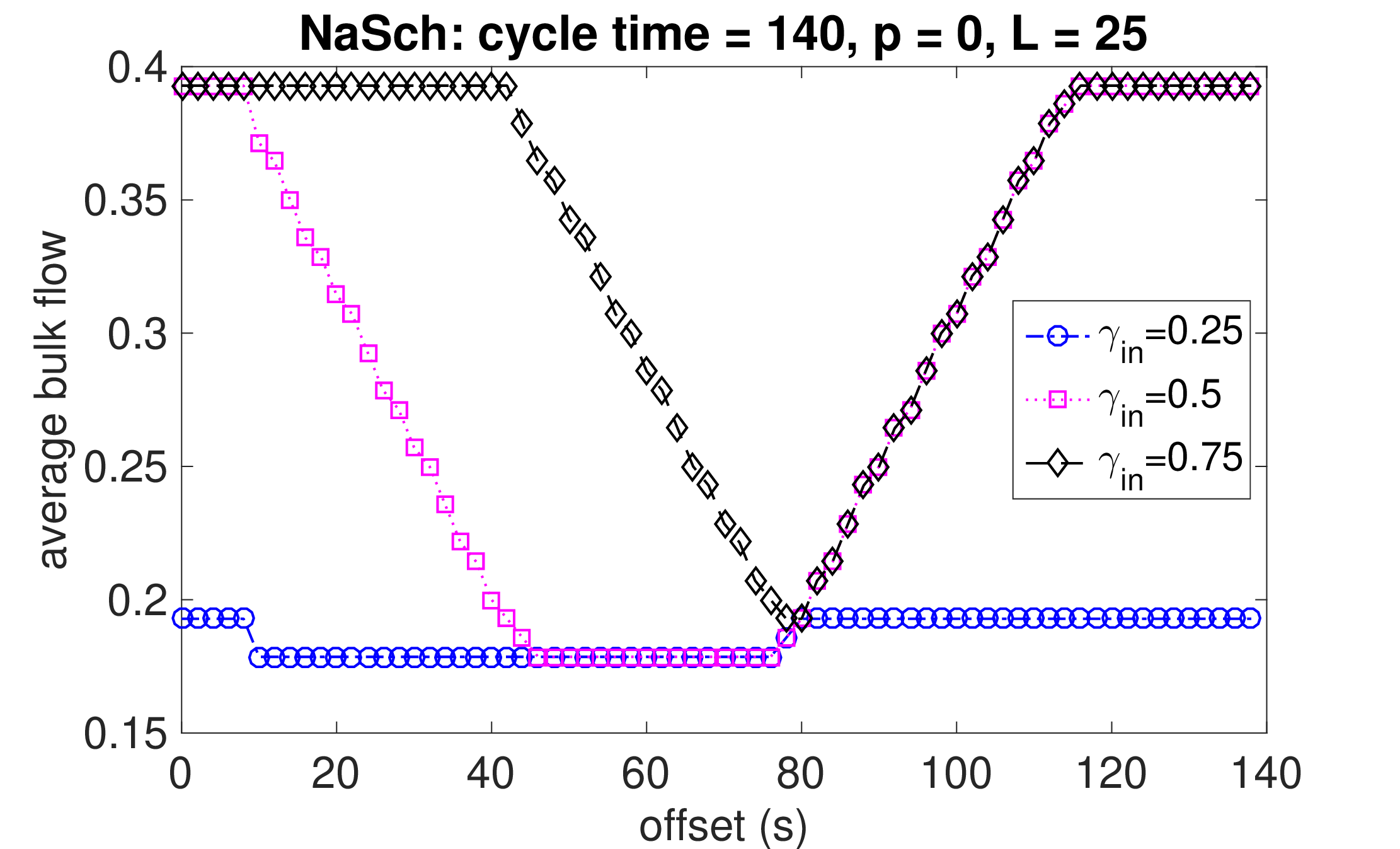}}%
\def\NaSchflowcycleLB			{\includegraphics[scale=\onesize]{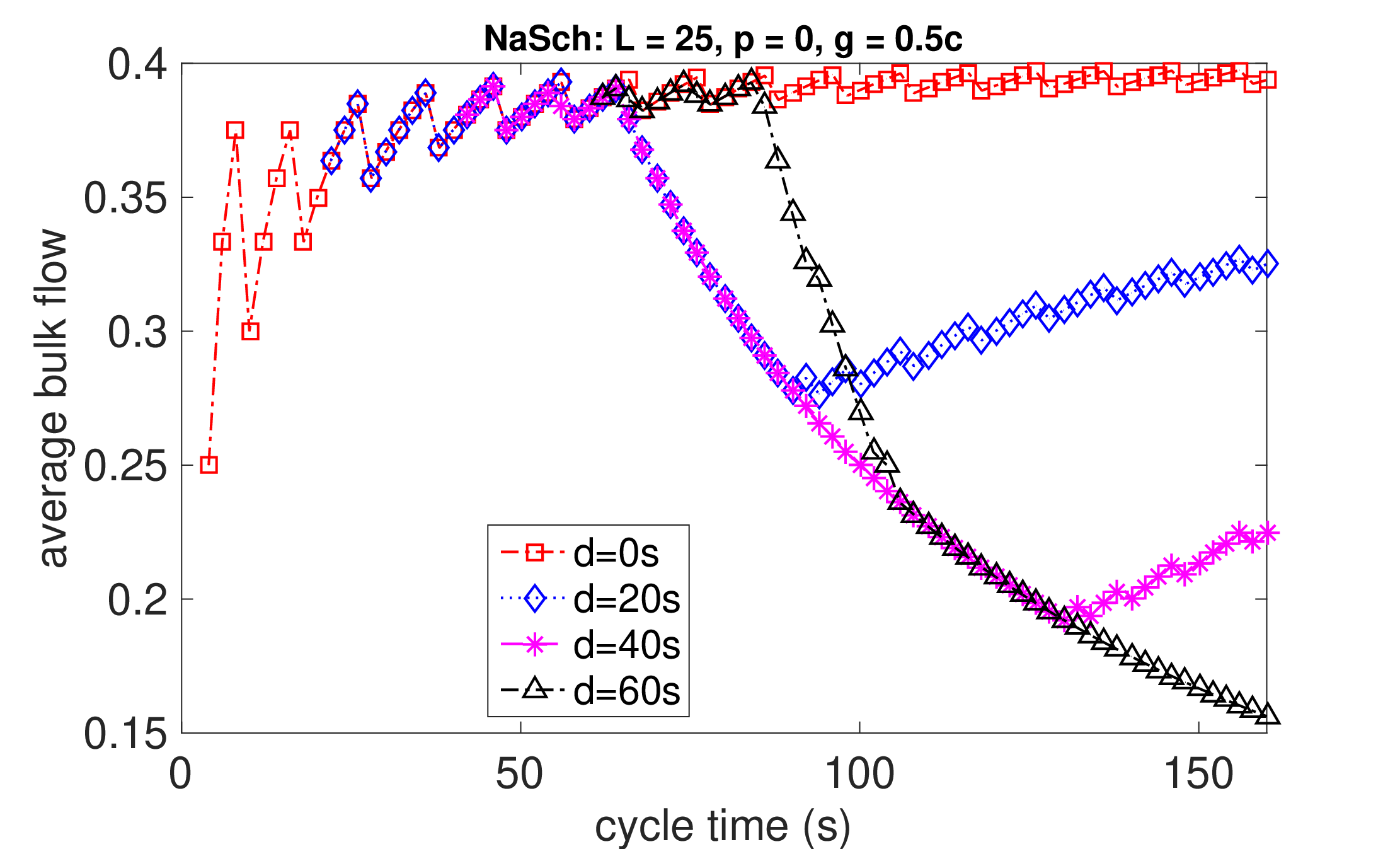}}%
\def\ASEPStflowcycle			{\includegraphics[scale=\onesize]{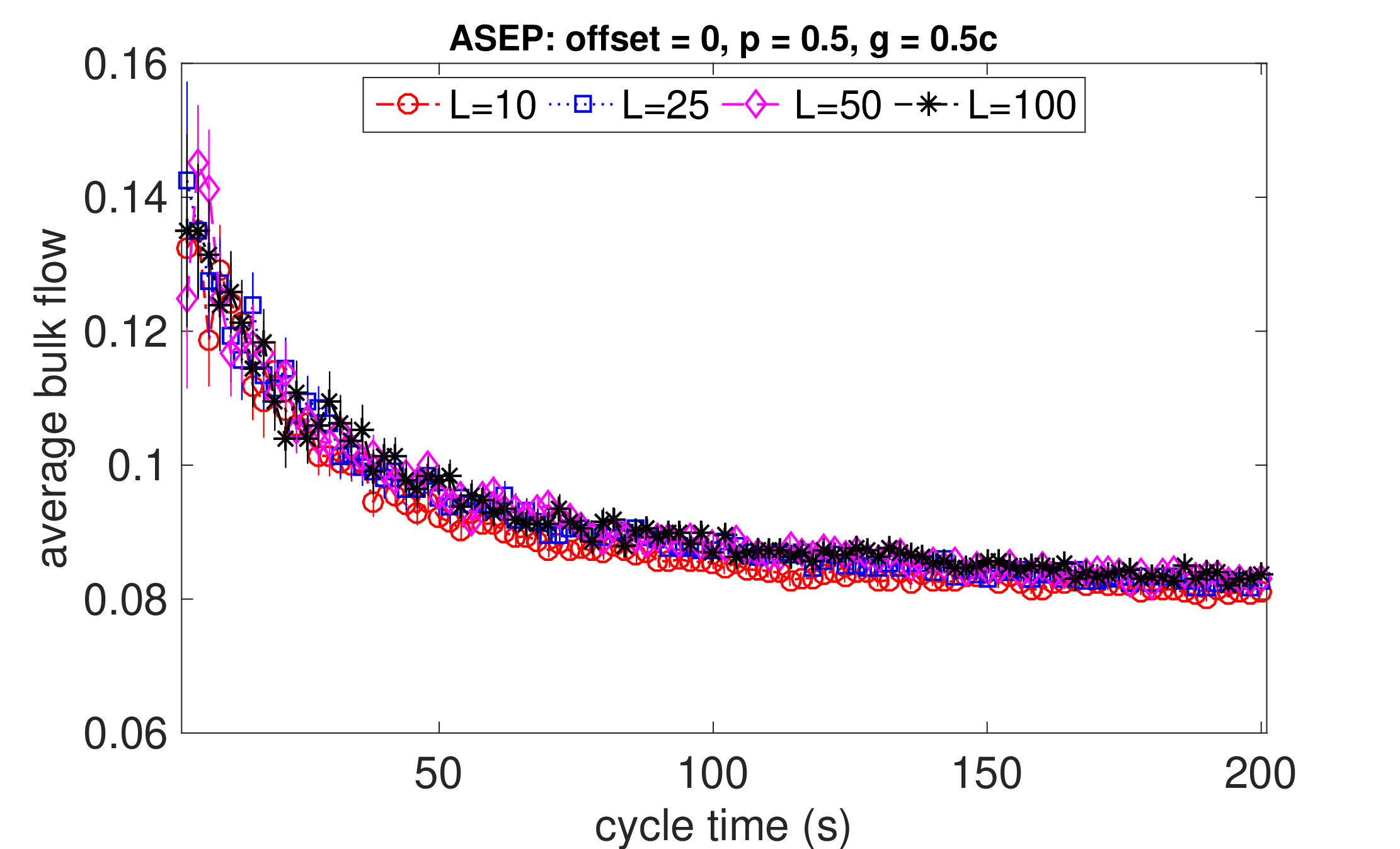}}%
\def\NaSchStflowcycle			{\includegraphics[scale=\onesize]{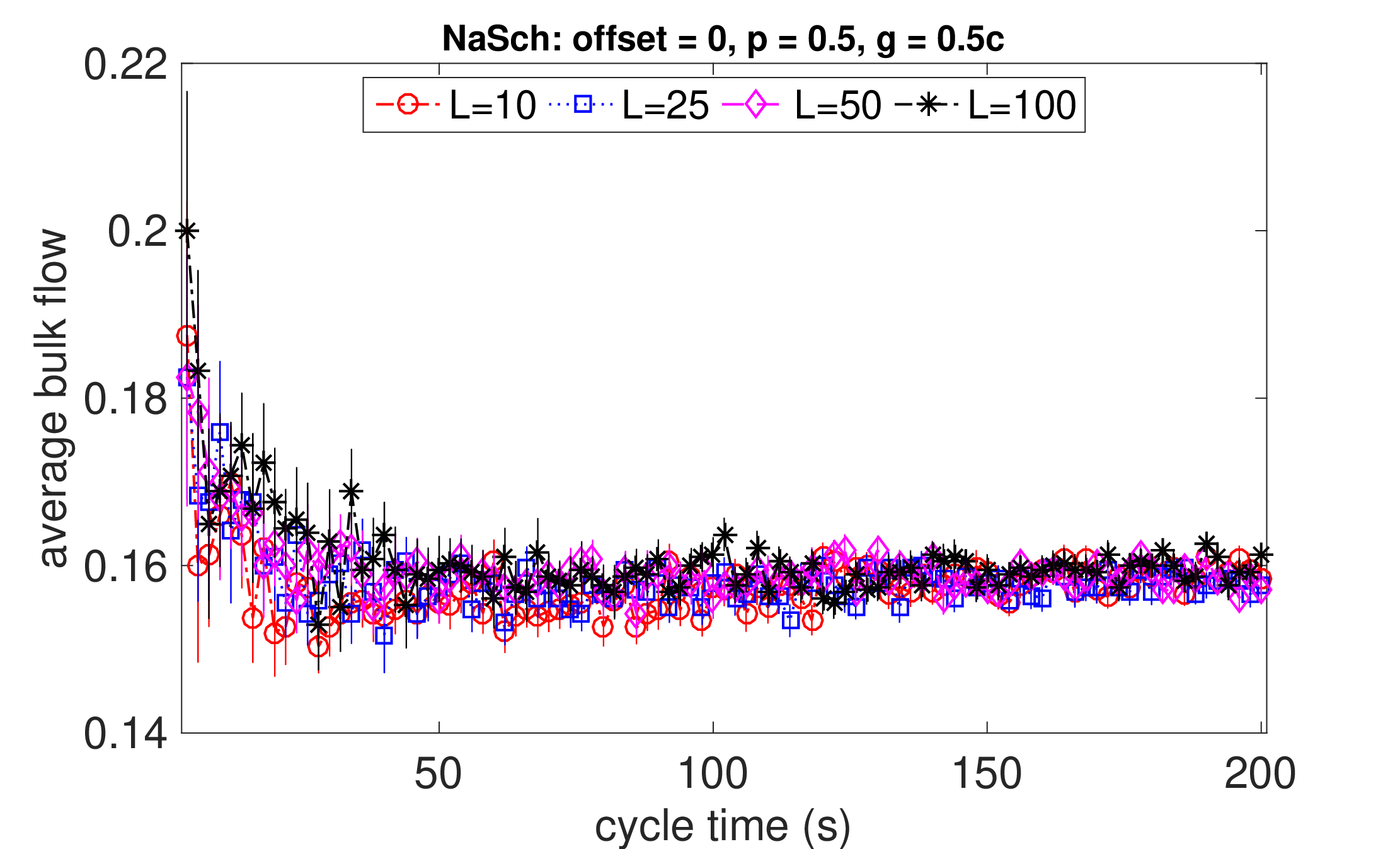}}%
\def\NaSchStflowoffset			{\includegraphics[scale=\onesize]{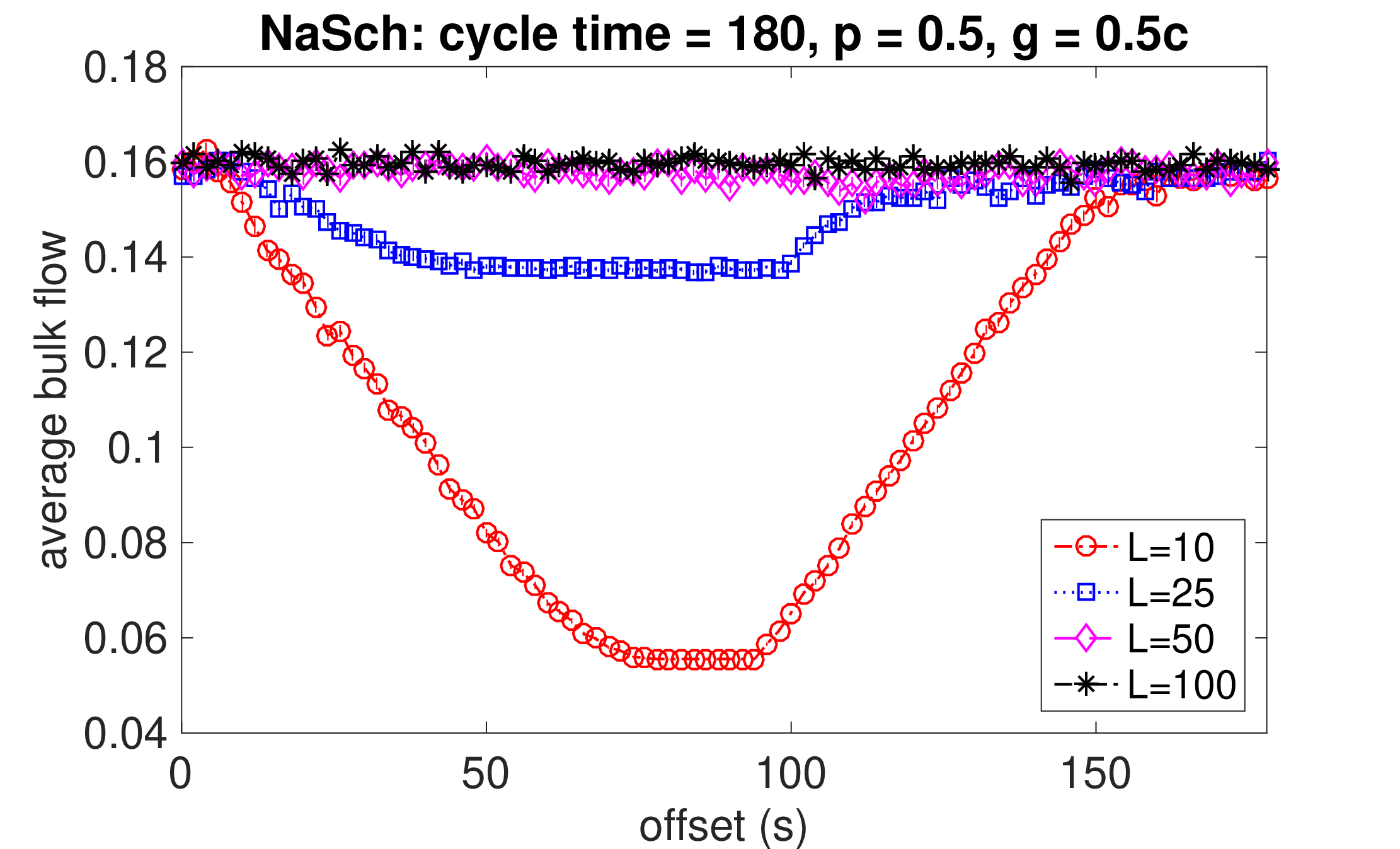}}%
\def\NaSchStflowratioLA			{\includegraphics[scale=\onesize]{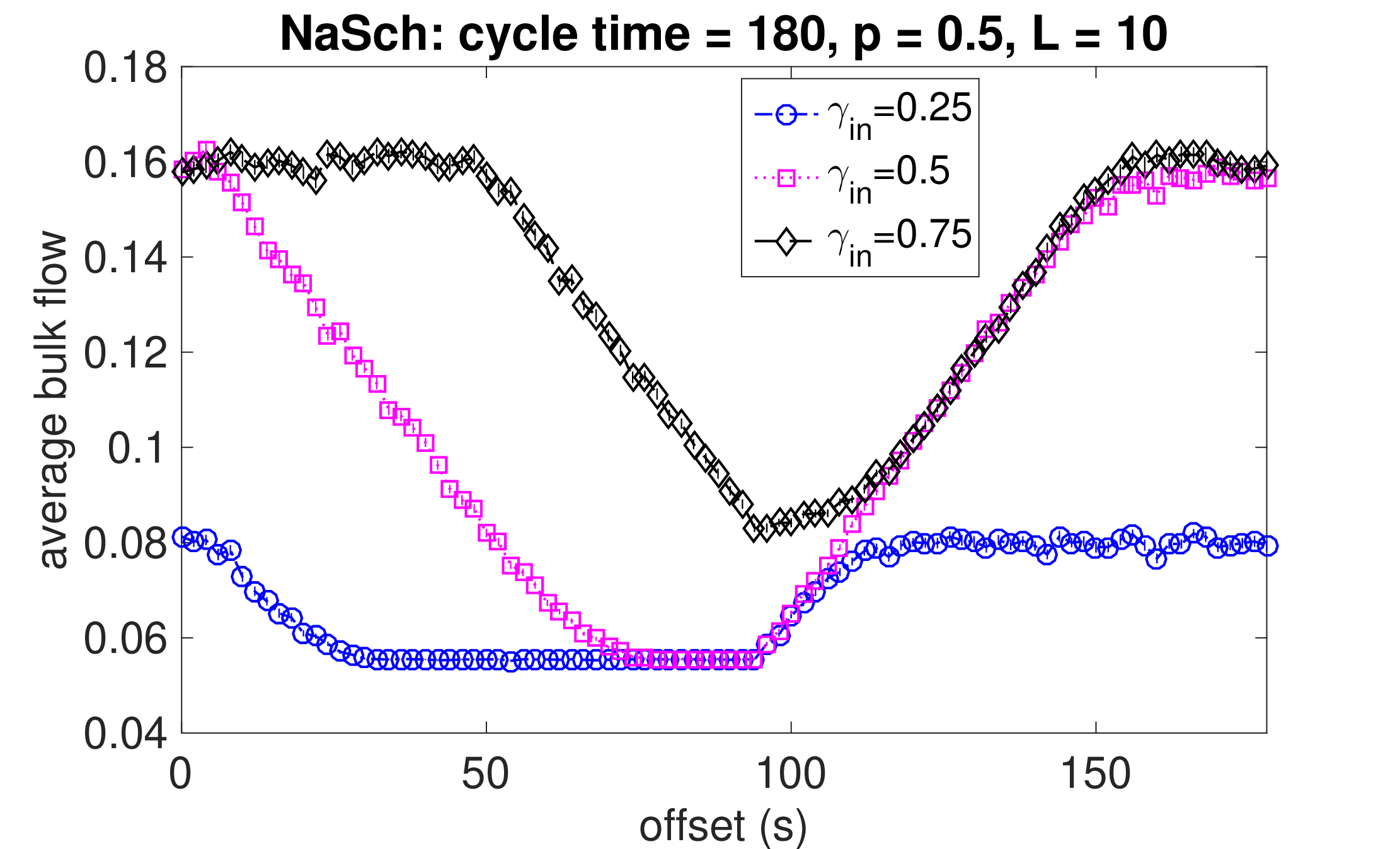}}%
\def\NaSchStflowcycleLA			{\includegraphics[scale=\onesize]{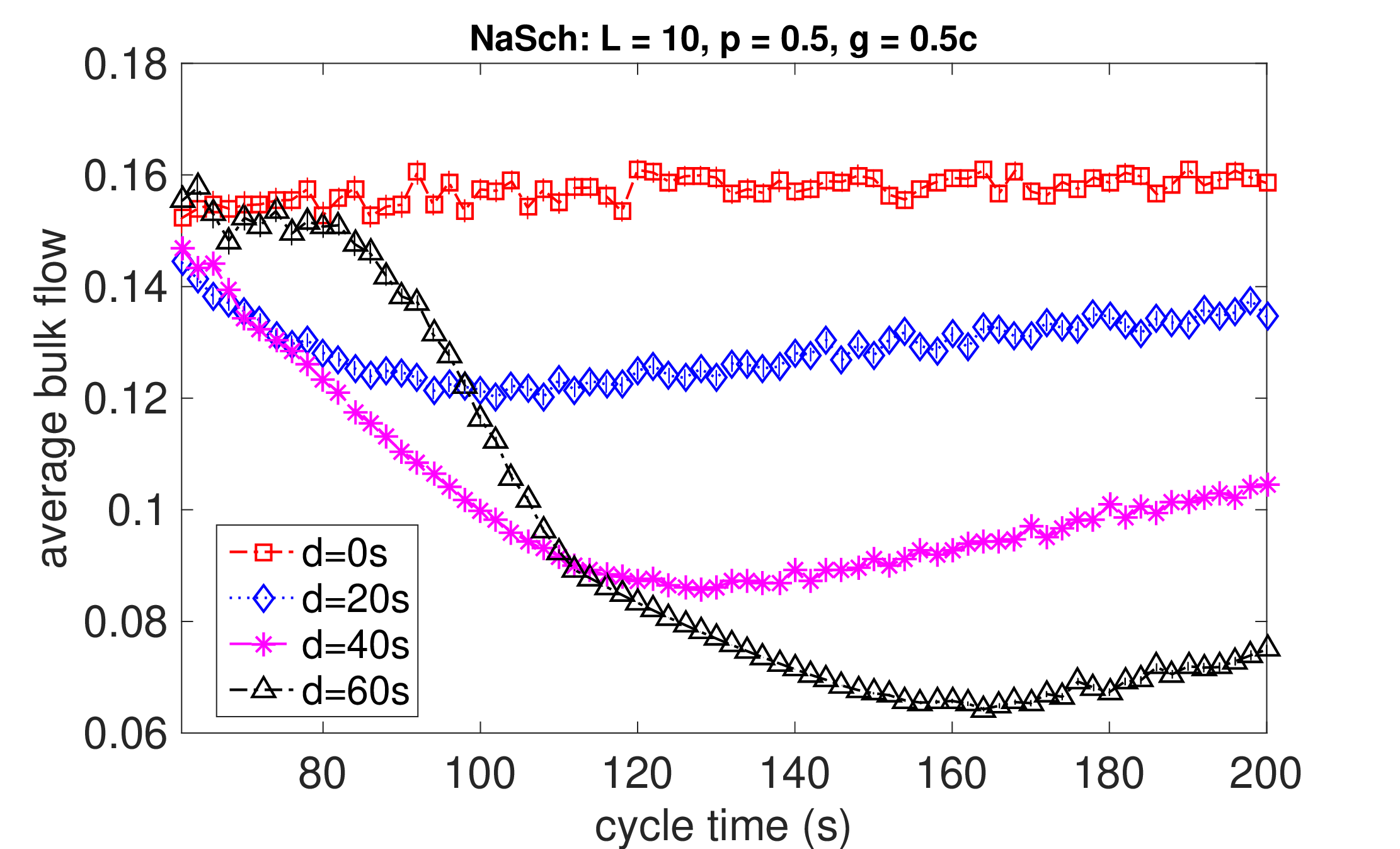}}%
\def\ASEPDensityPfiveCmpGreen			{\includegraphics[scale=\onesize]{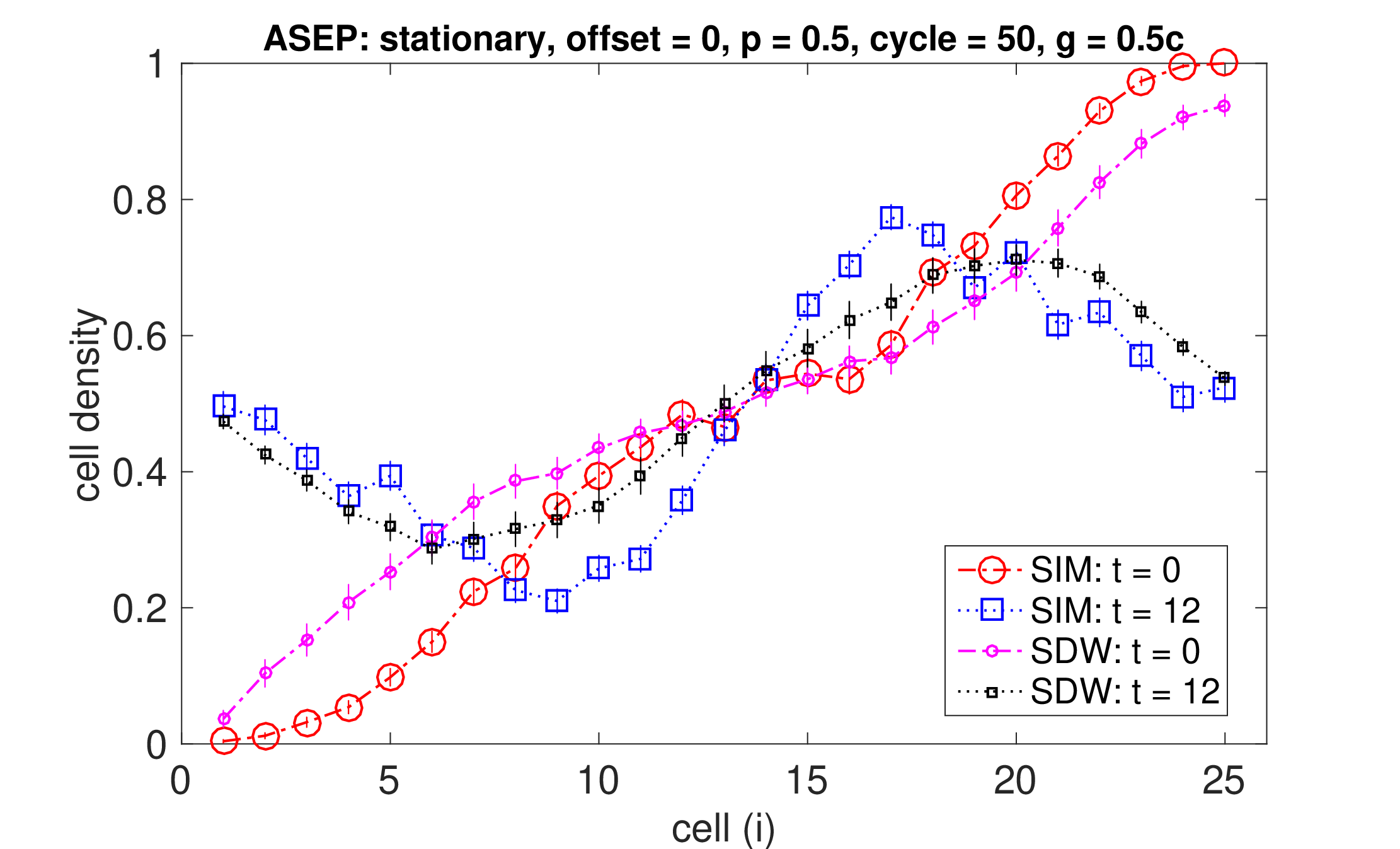}}%
\def\ASEPDensityPfiveCmpLongGreen			{\includegraphics[scale=\onesize]{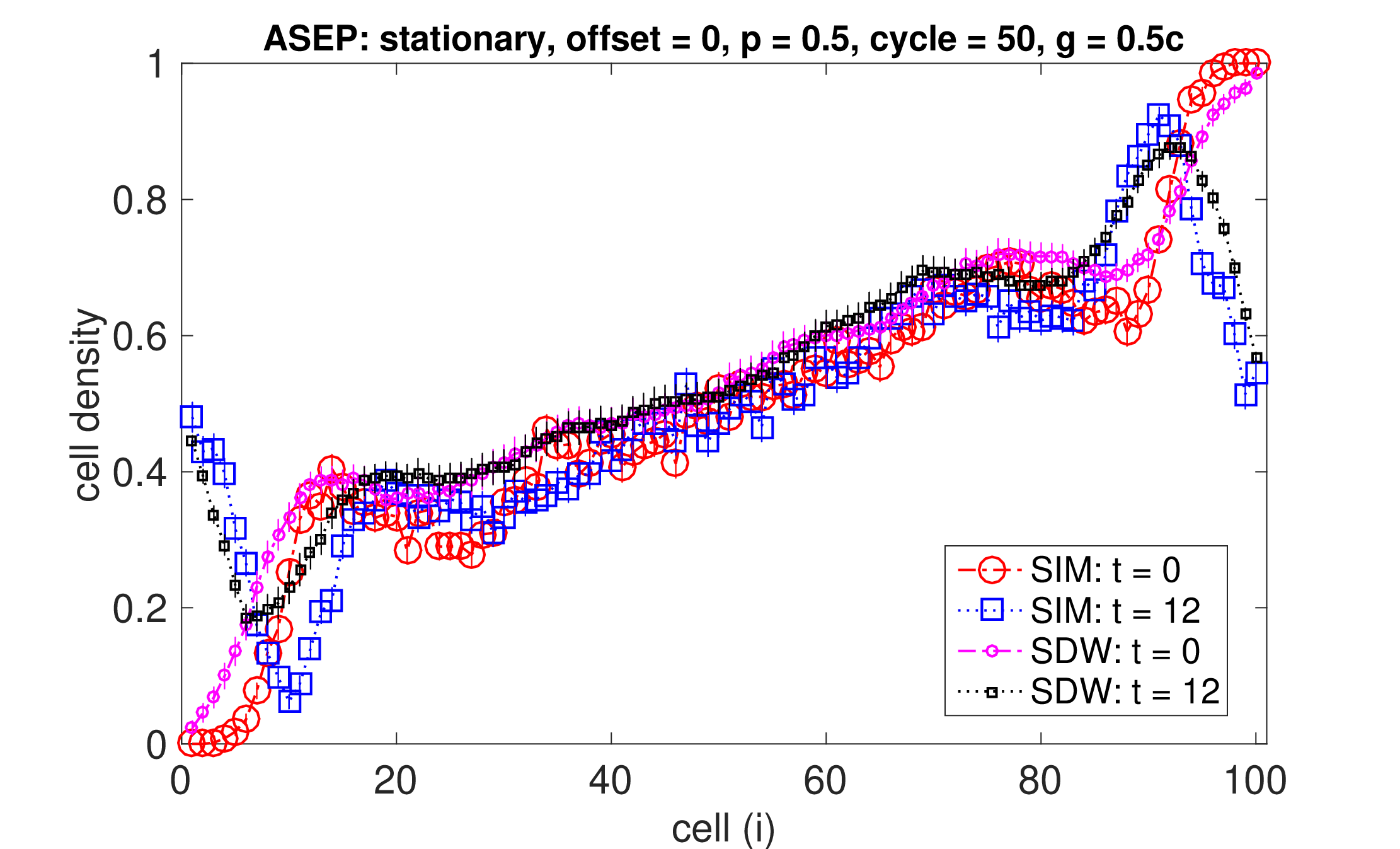}}%
\def\ASEPDensityPfiveCmpLargeRed			{\includegraphics[scale=\onesize]{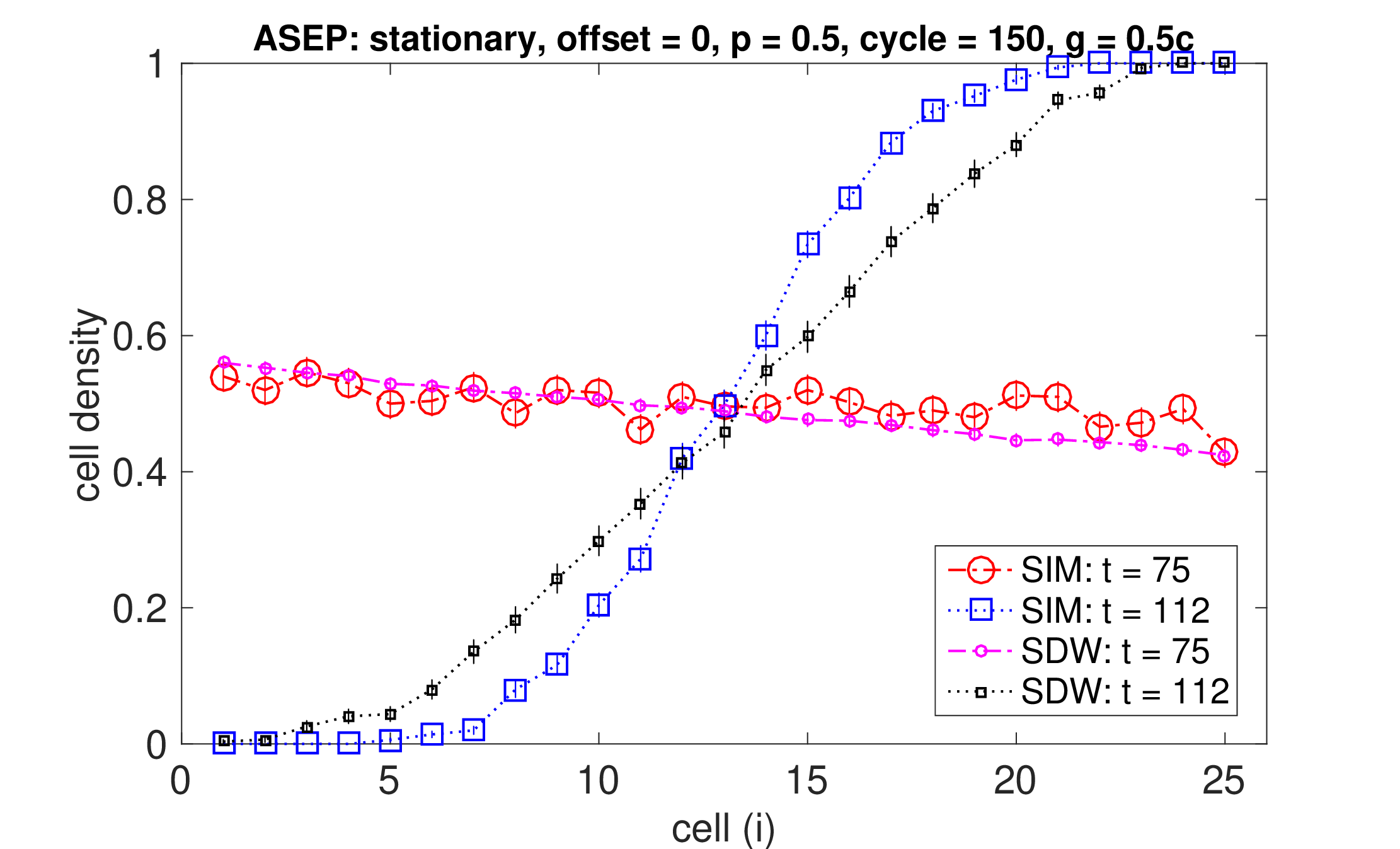}}%
\def\ASEPDensityPfiveCmpLongLargeRed		{\includegraphics[scale=\onesize]{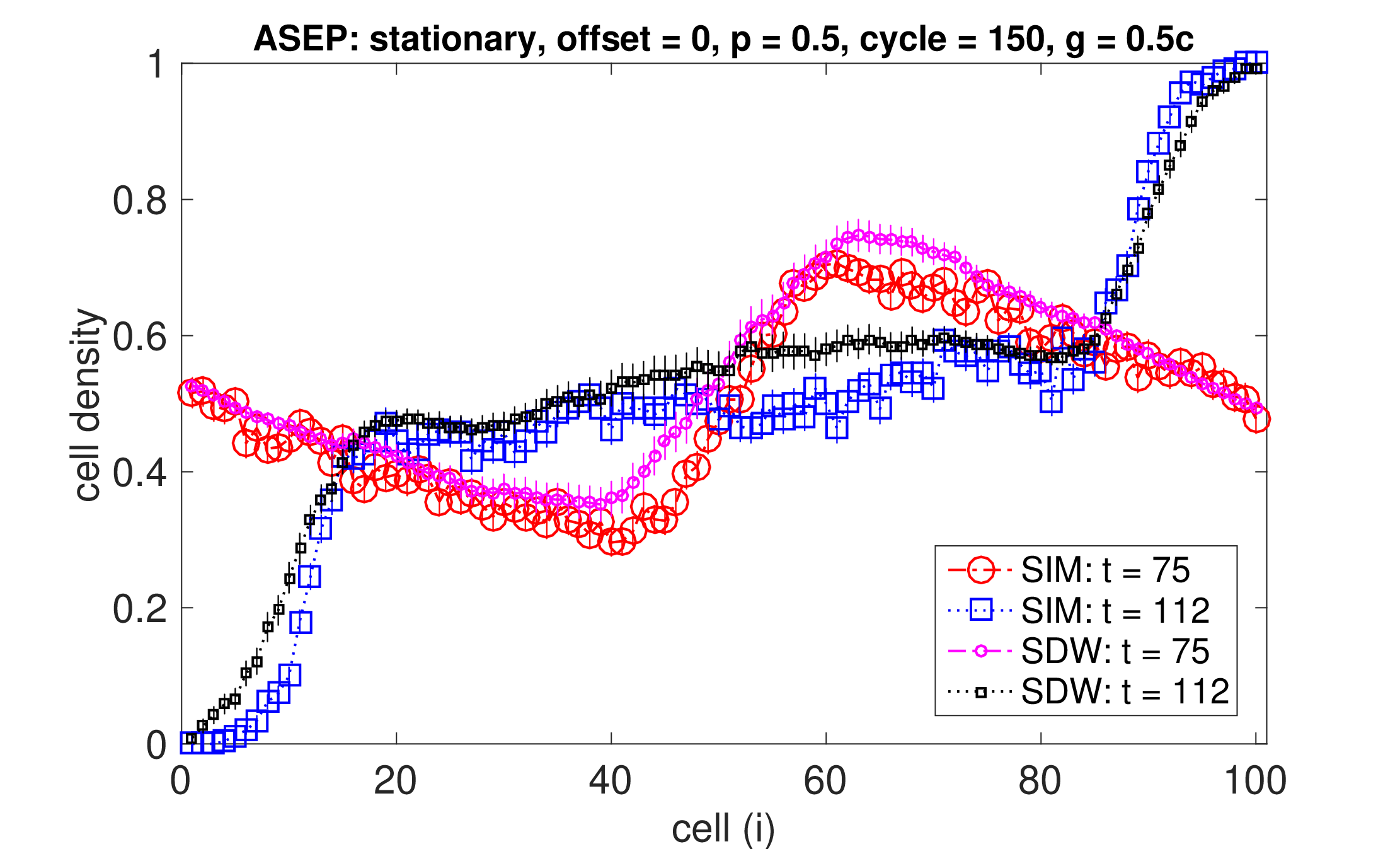}}%
\def\NaSchDensityPfiveCmpLargeRed			{\includegraphics[scale=\onesize]{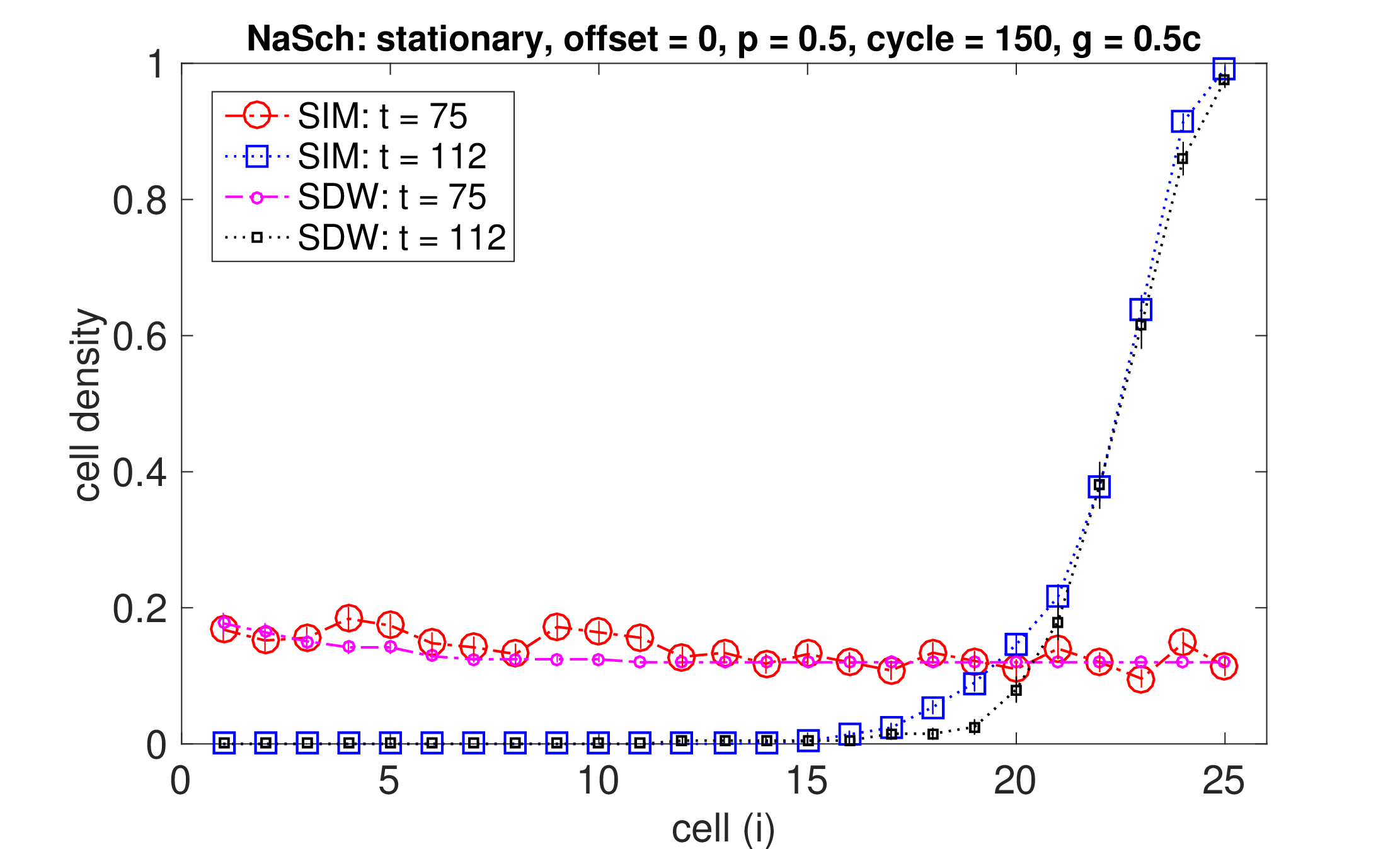}}%
\def\NaSchDensityPfiveCmpLongLargeRed		{\includegraphics[scale=\onesize]{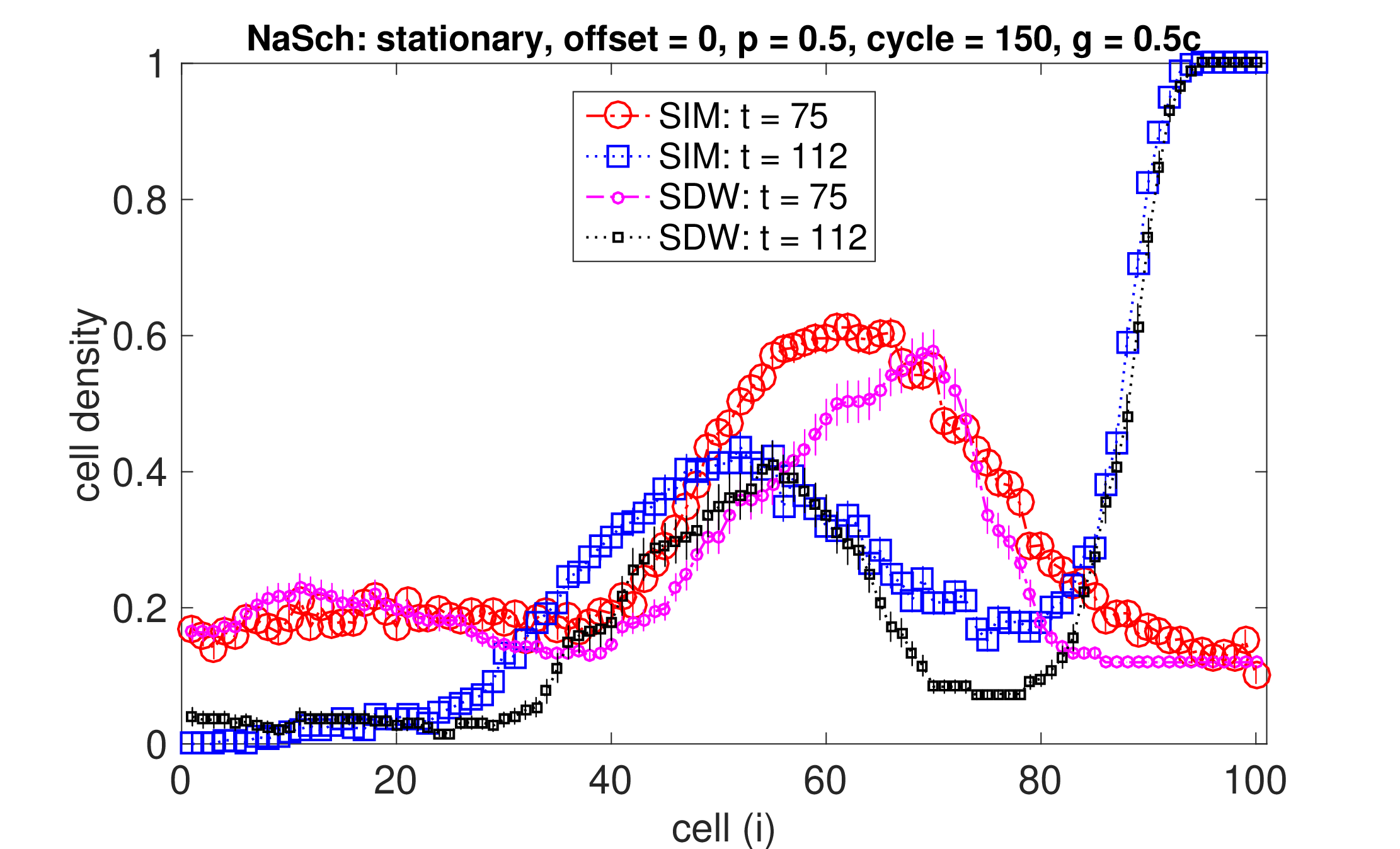}}%
\def\ASEPDensityPfiveEmptyTransient			{\includegraphics[scale=\onesize]{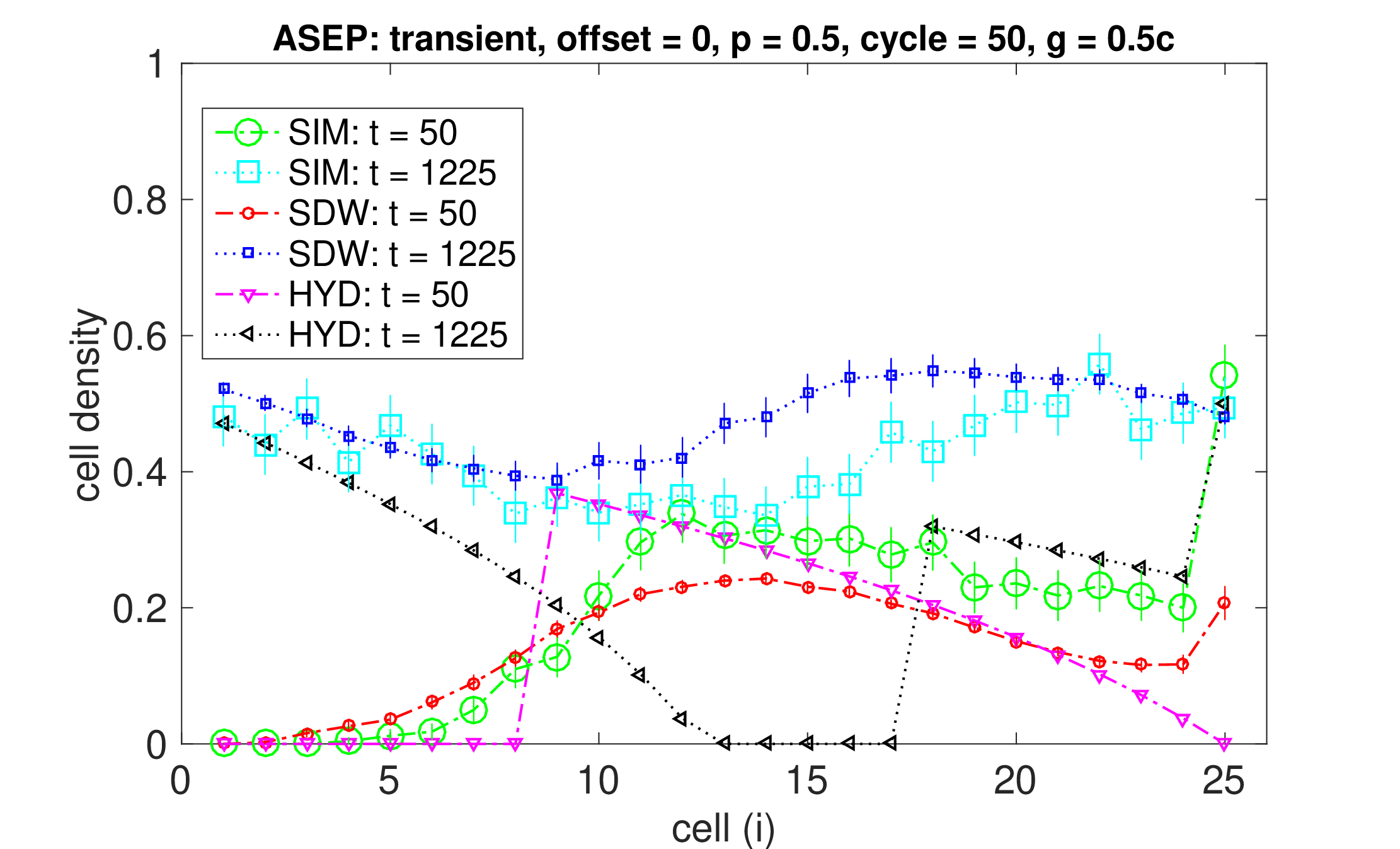}}%
\def\ASEPDensityPfiveEmptyTransientLongLarge	{\includegraphics[scale=\onesize]{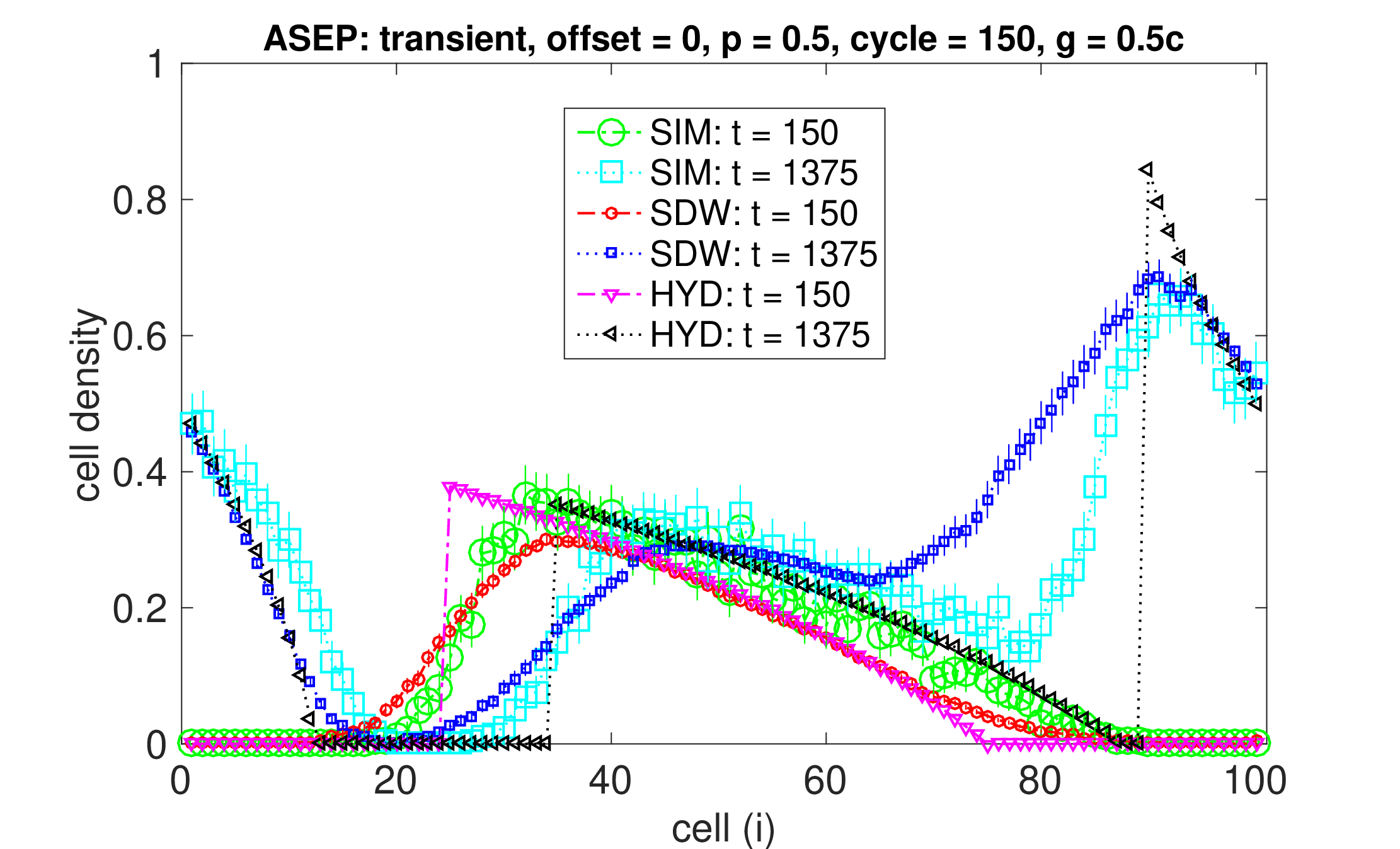}}%
\def\ASEPDensityPfiveFullTransient			{\includegraphics[scale=\onesize]{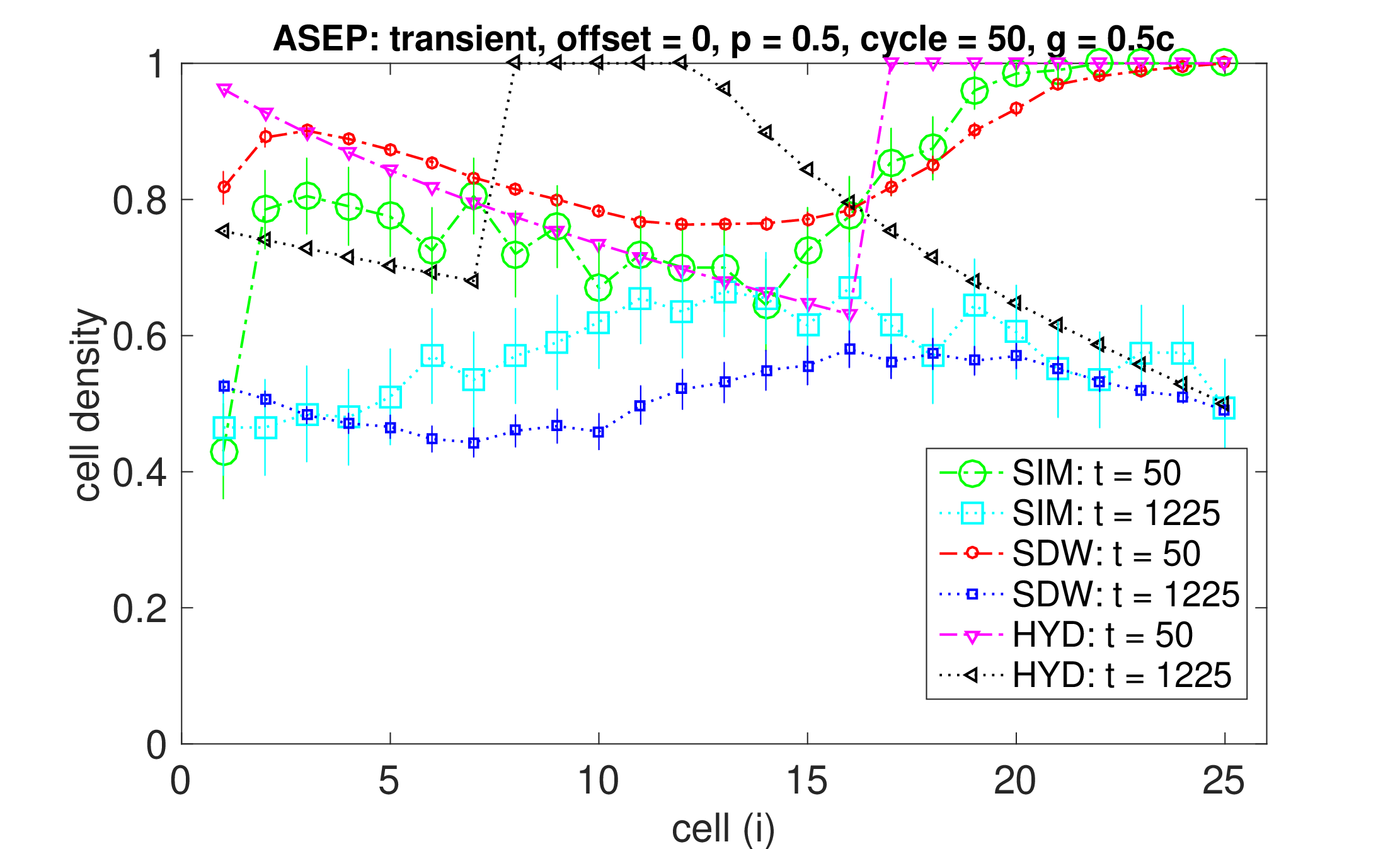}}%
\def\ASEPDensityPfiveFullTransientLongLarge		{\includegraphics[scale=\onesize]{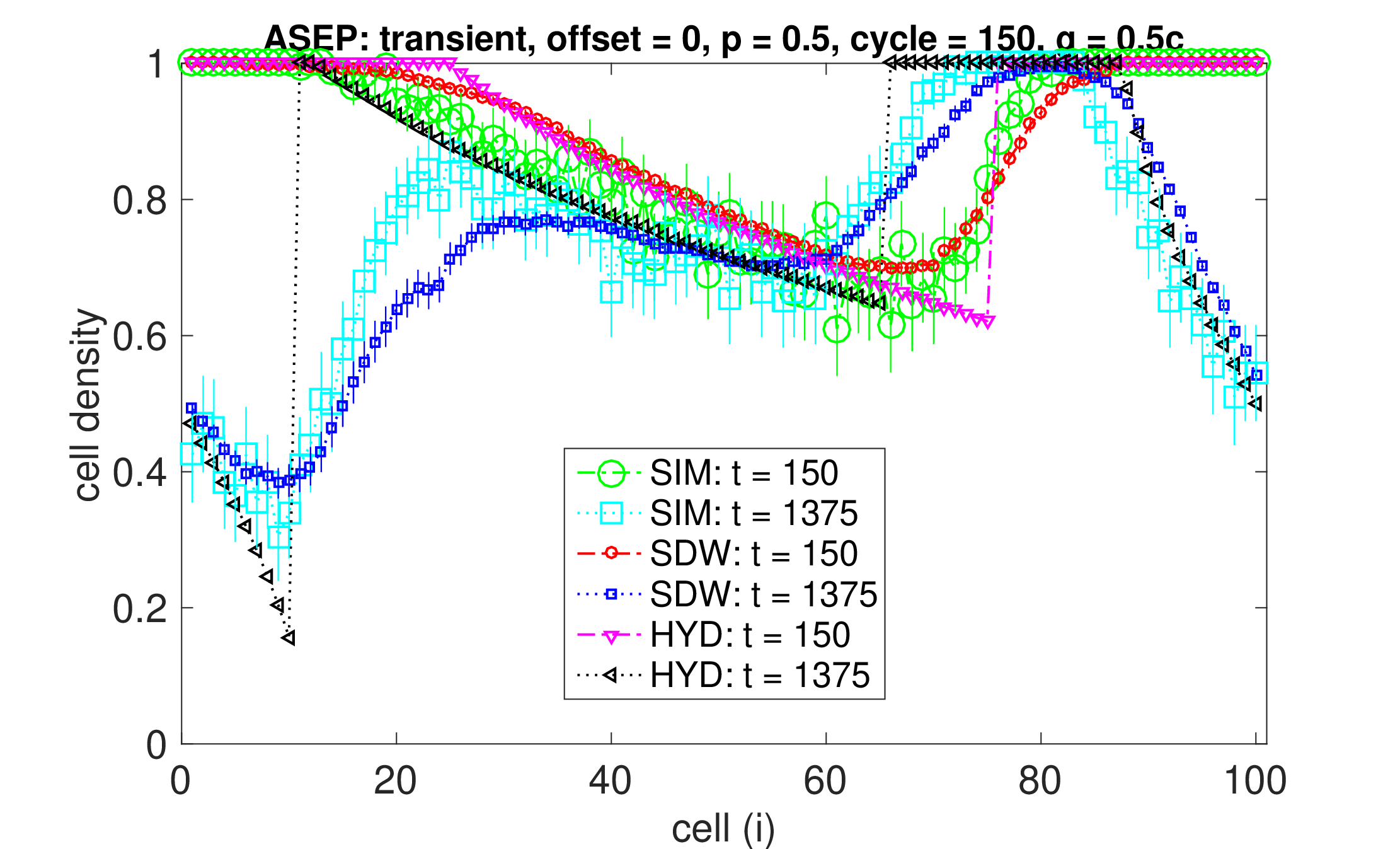}}%
\def\NaSchDensityPfiveEmptyTransient		{\includegraphics[scale=\onesize]{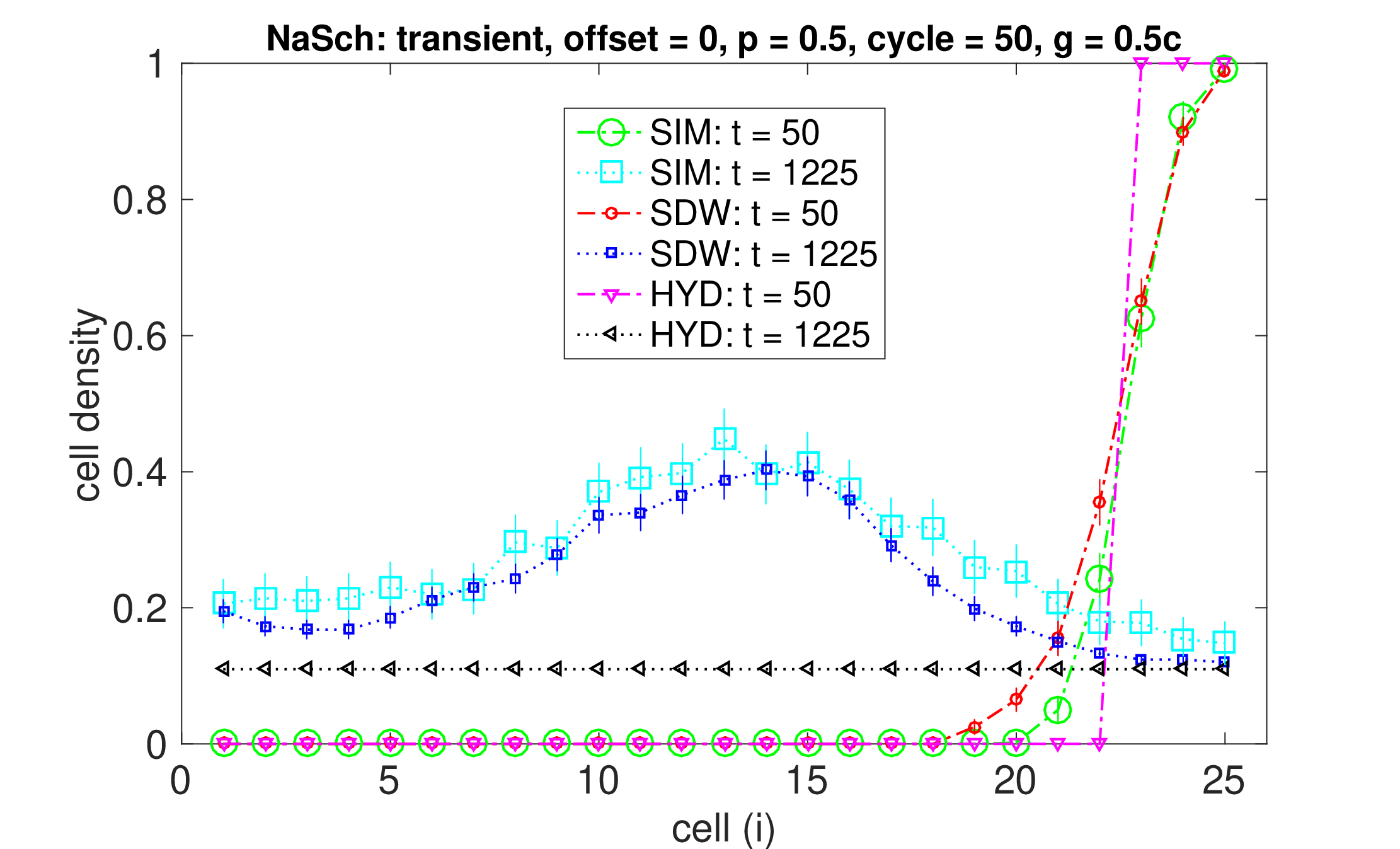}}%
\def\NaSchDensityPfiveEmptyTransientLongLarge	{\includegraphics[scale=\onesize]{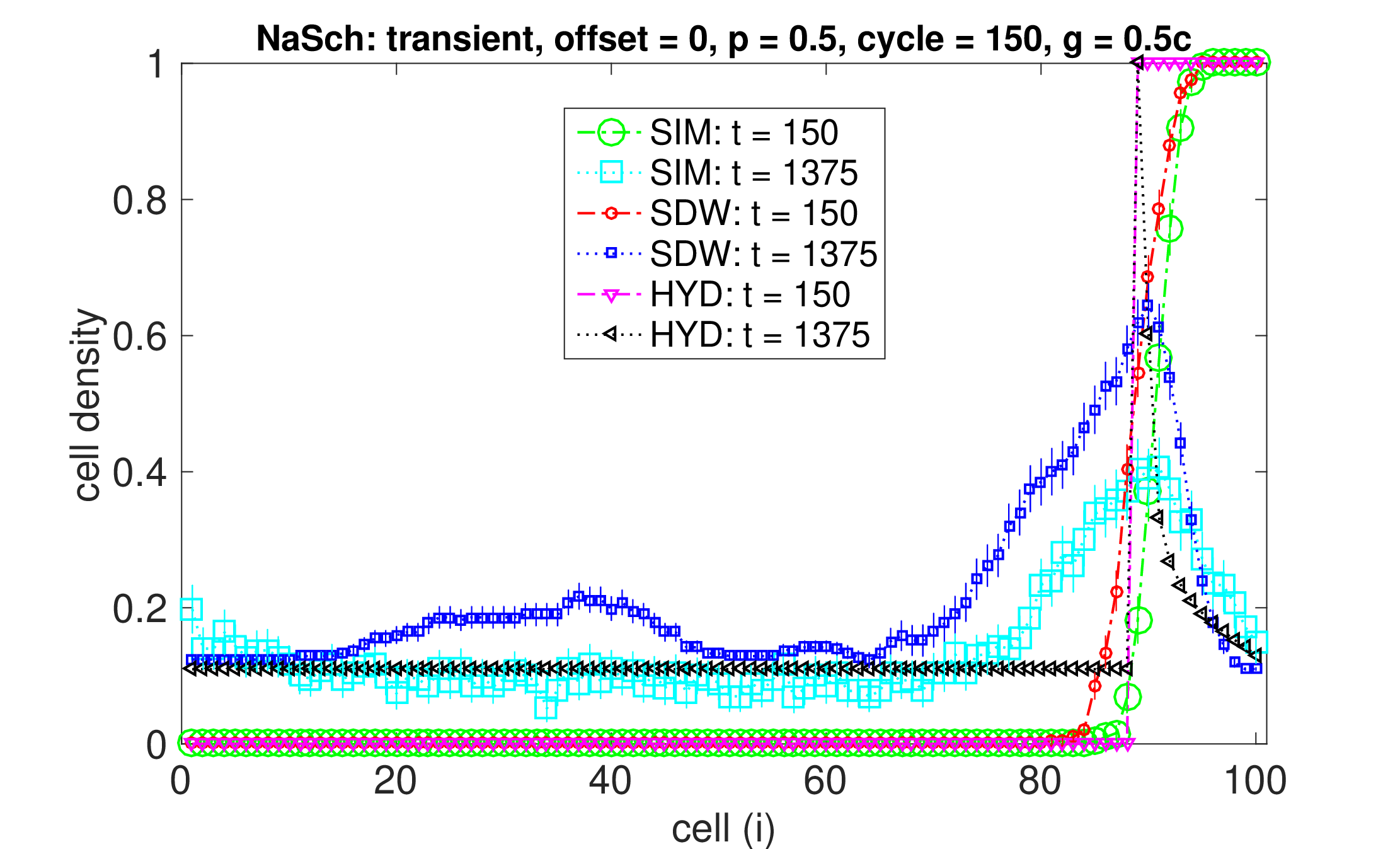}}%
\def\NaSchDensityPfiveRapidRed			{\includegraphics[scale=\onesize]{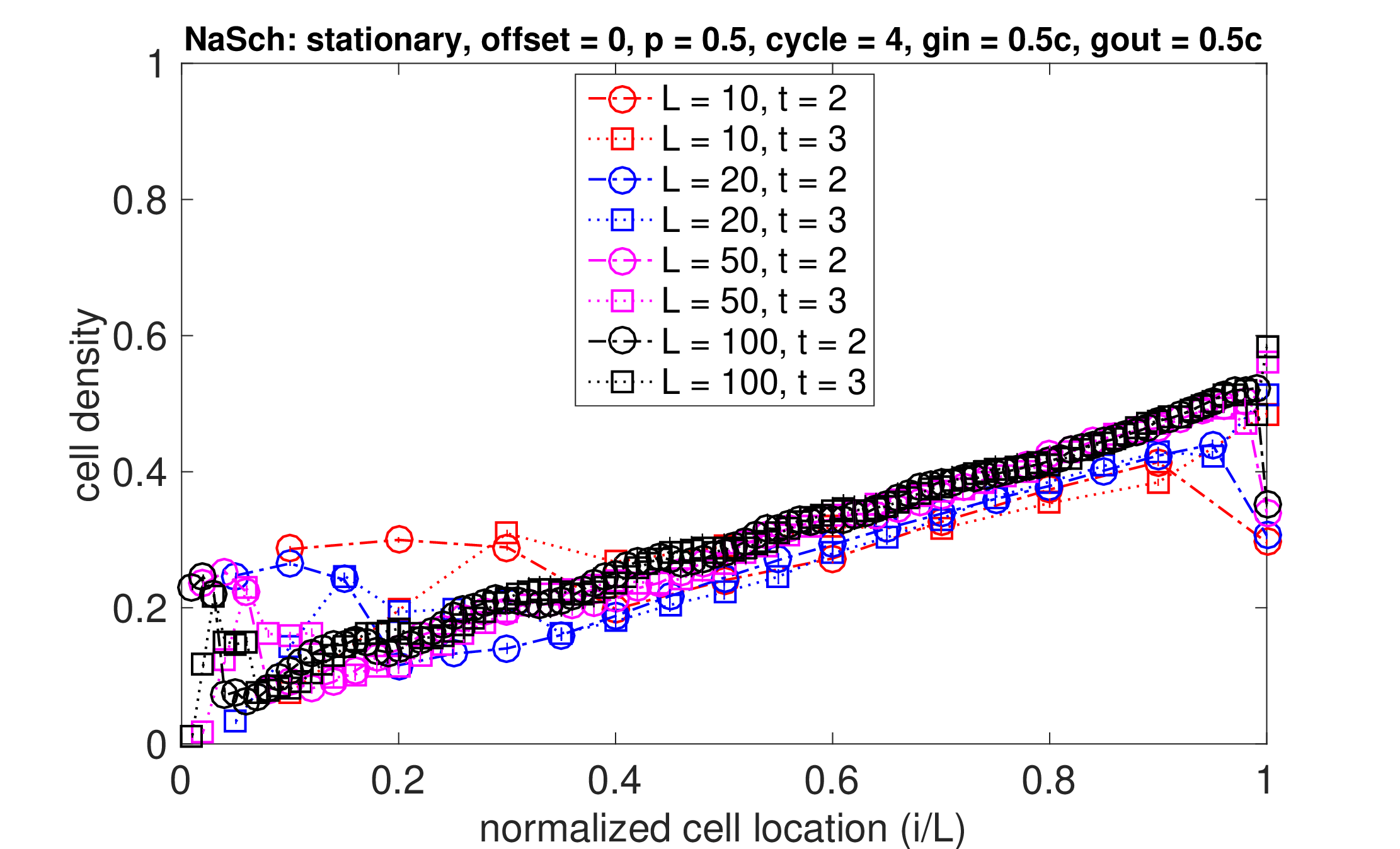}}%
\def\ASEPDensityPfiveRapidRed			{\includegraphics[scale=\onesize]{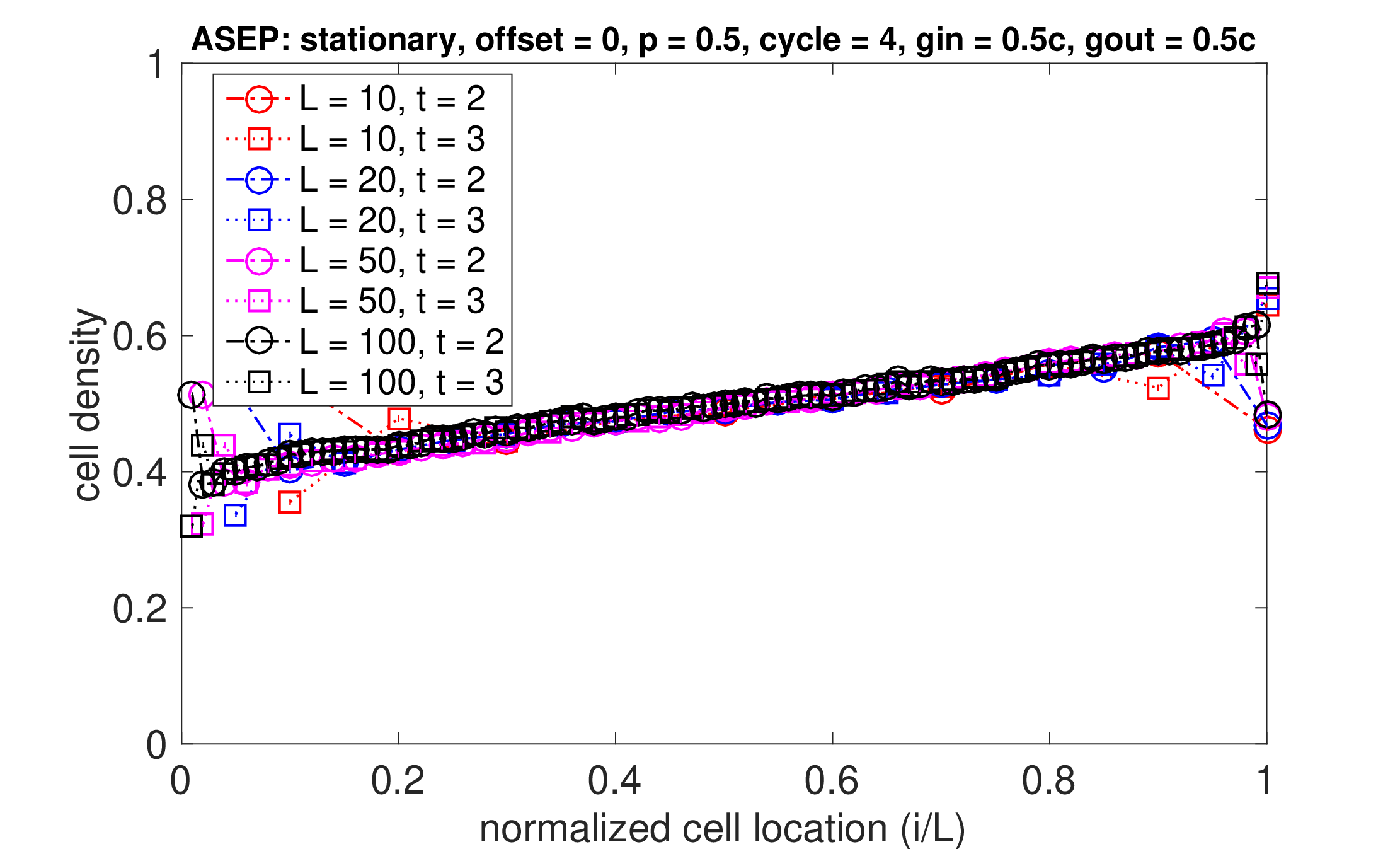}}%
\def\NaSchDensityPfiveRapidRedH			{\includegraphics[scale=\onesize]{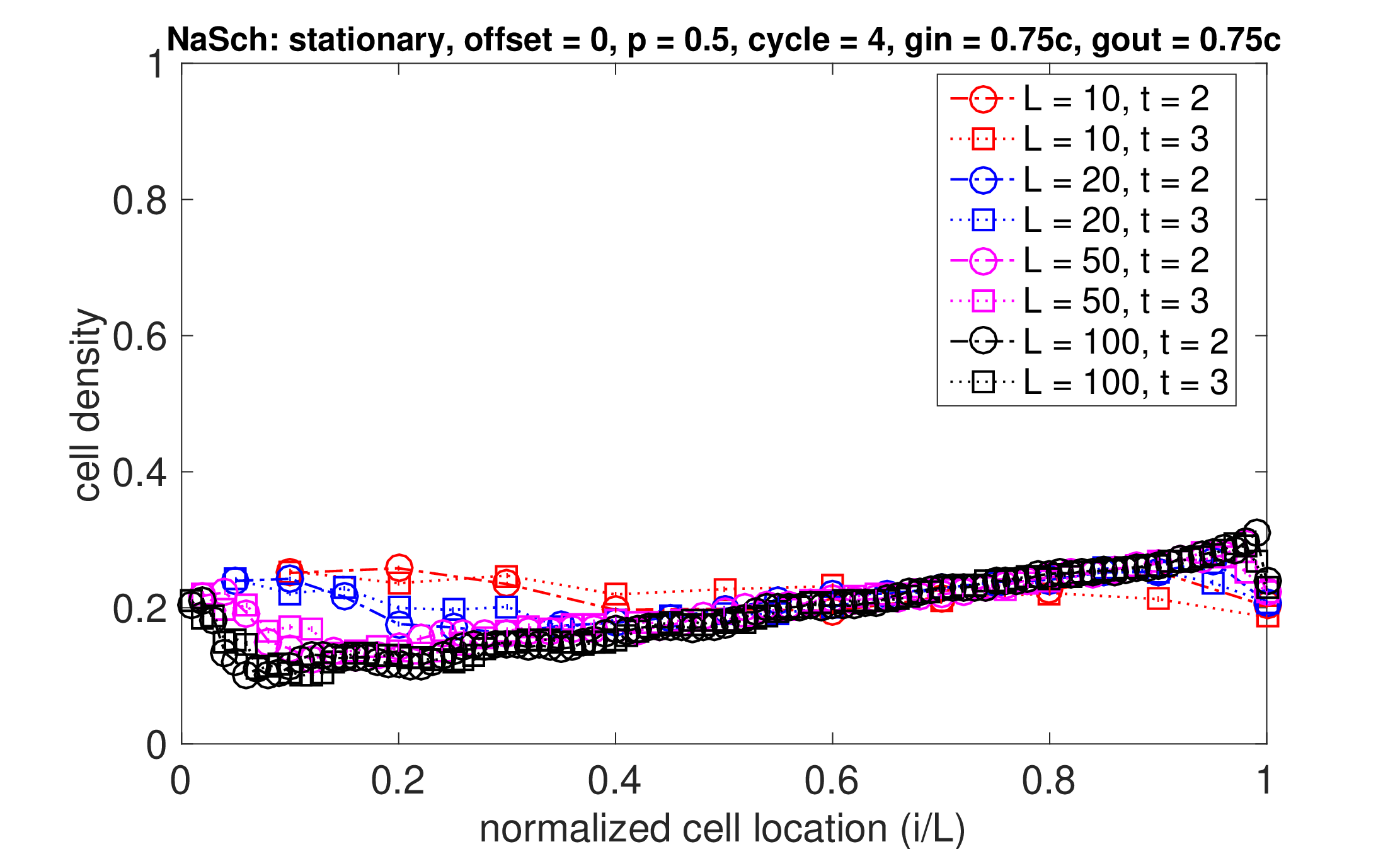}}%
\def\ASEPDensityPfiveRapidRedH			{\includegraphics[scale=\onesize]{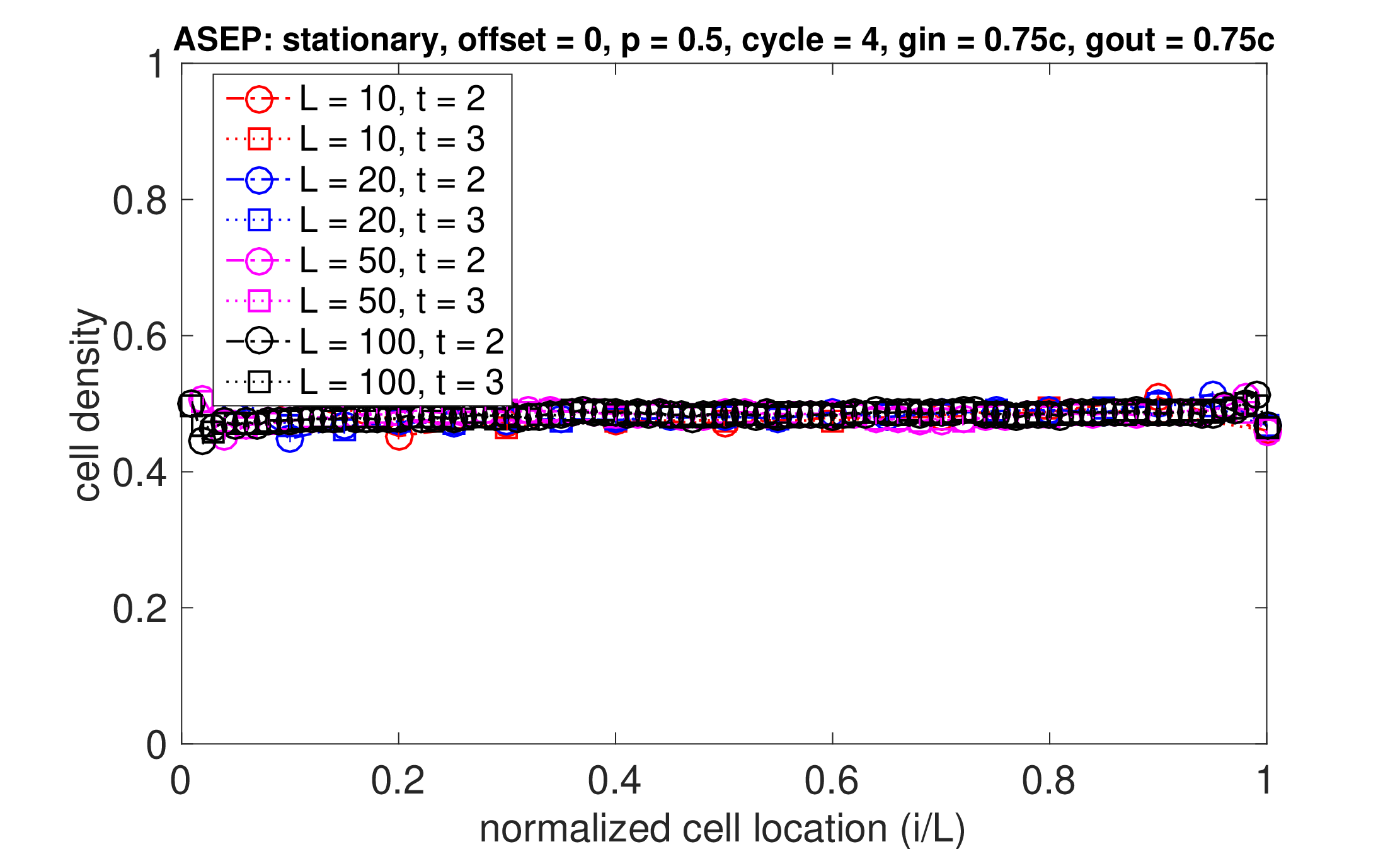}}%
\def\NaSchDensityPfiveRapidRedL			{\includegraphics[scale=\onesize]{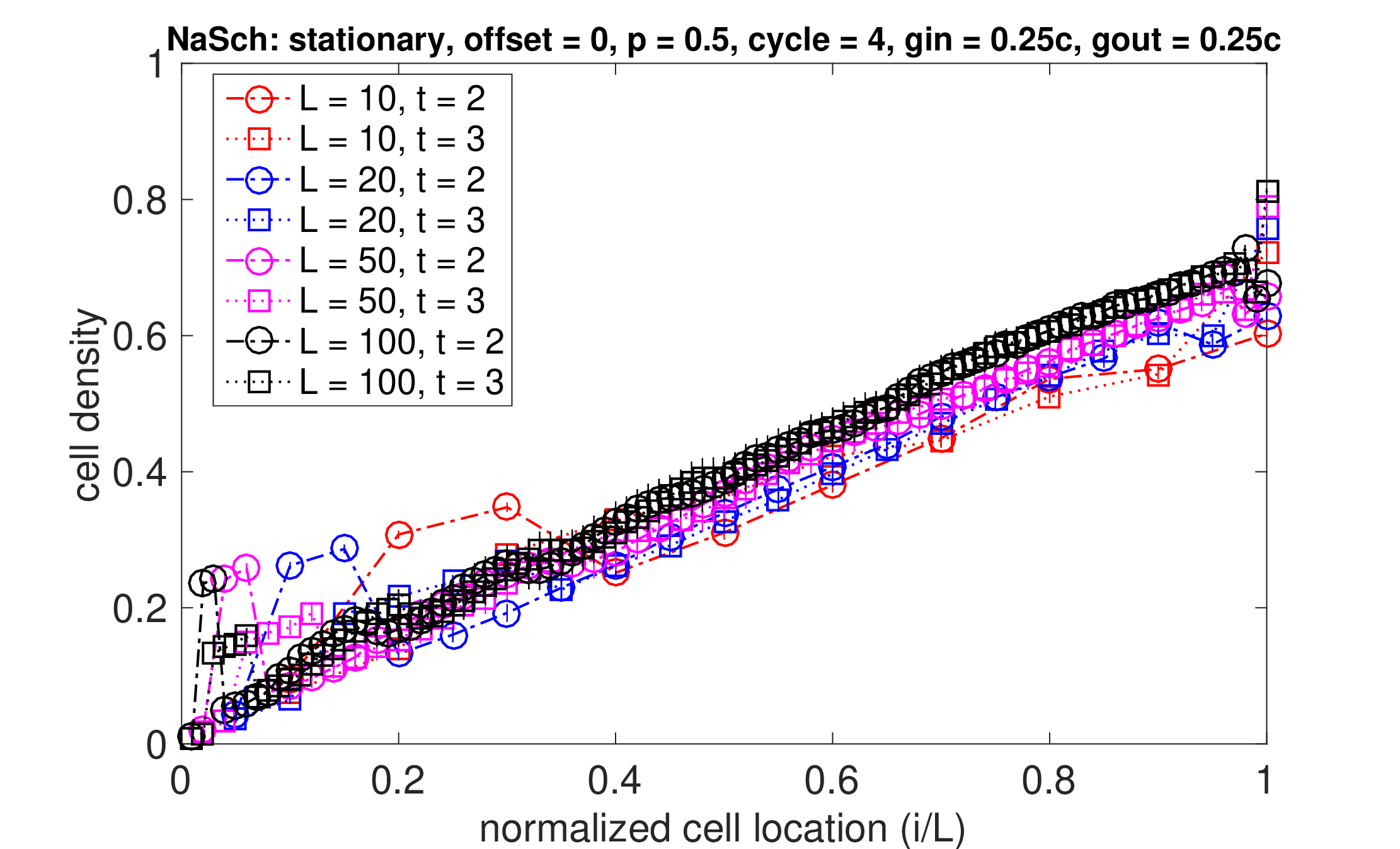}}%
\def\ASEPDensityPfiveRapidRedL			{\includegraphics[scale=\onesize]{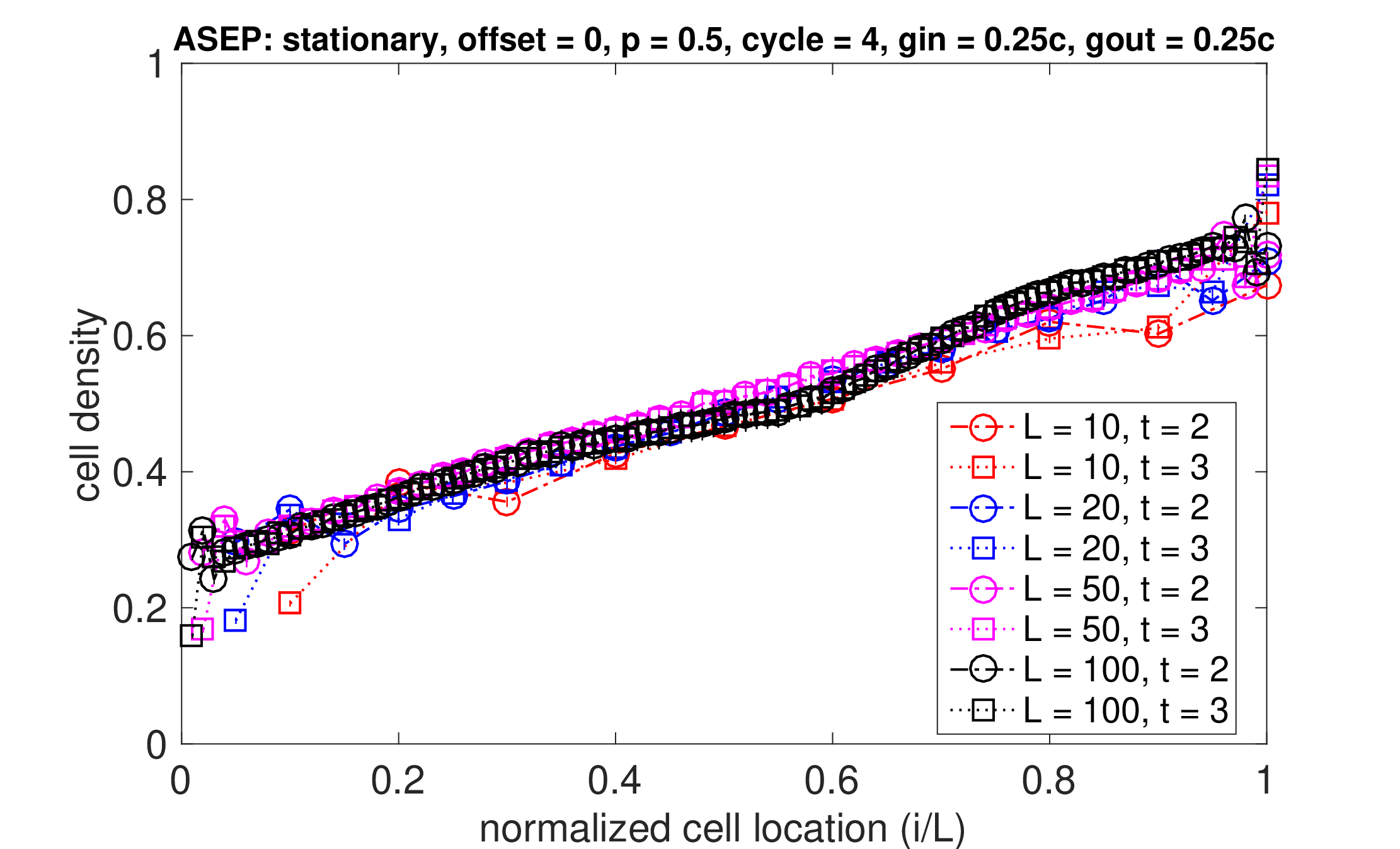}}%
\def\ASEPEffectiveInflow				{\includegraphics[scale=0.36]{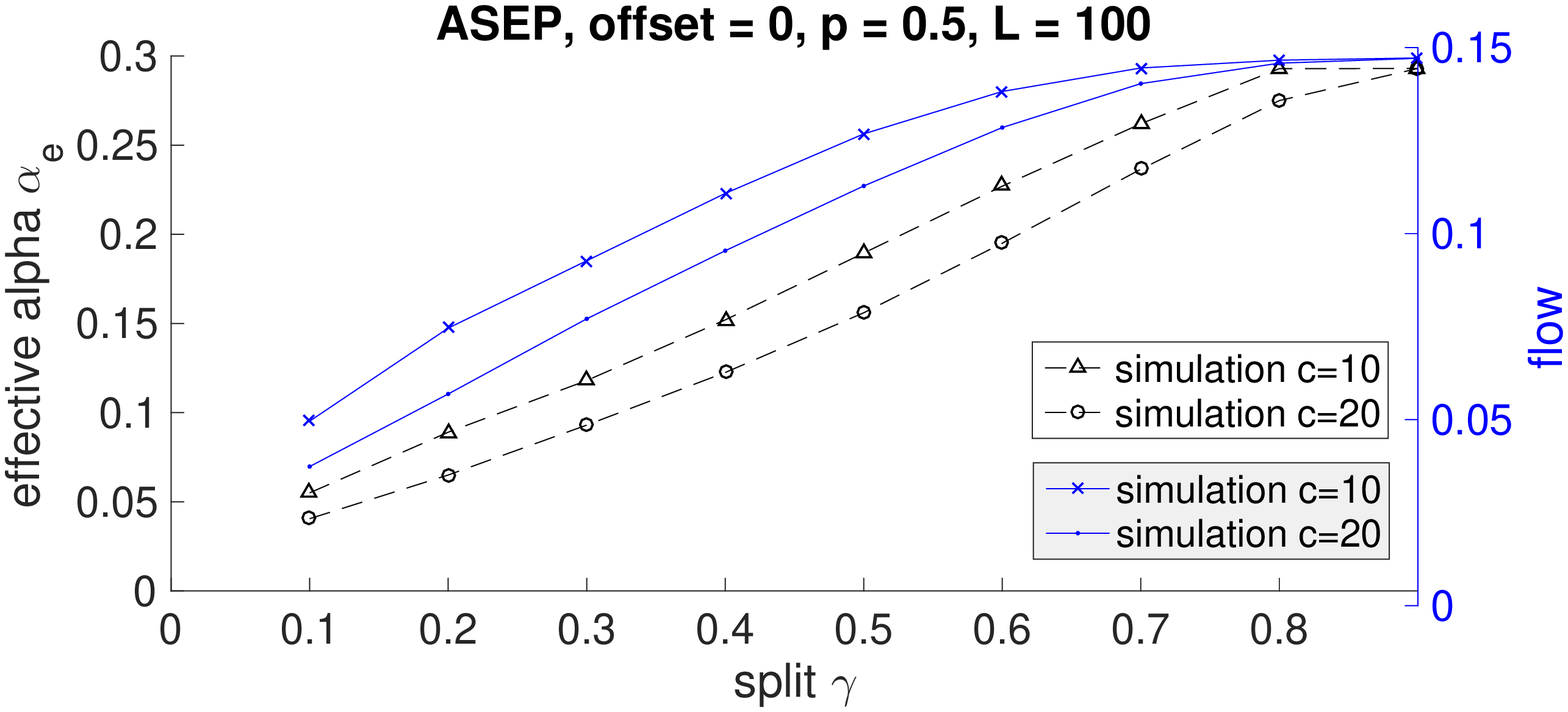}}%

\def\ASEPDensityPfiveRapidRedU			{\includegraphics[scale=\onesize]{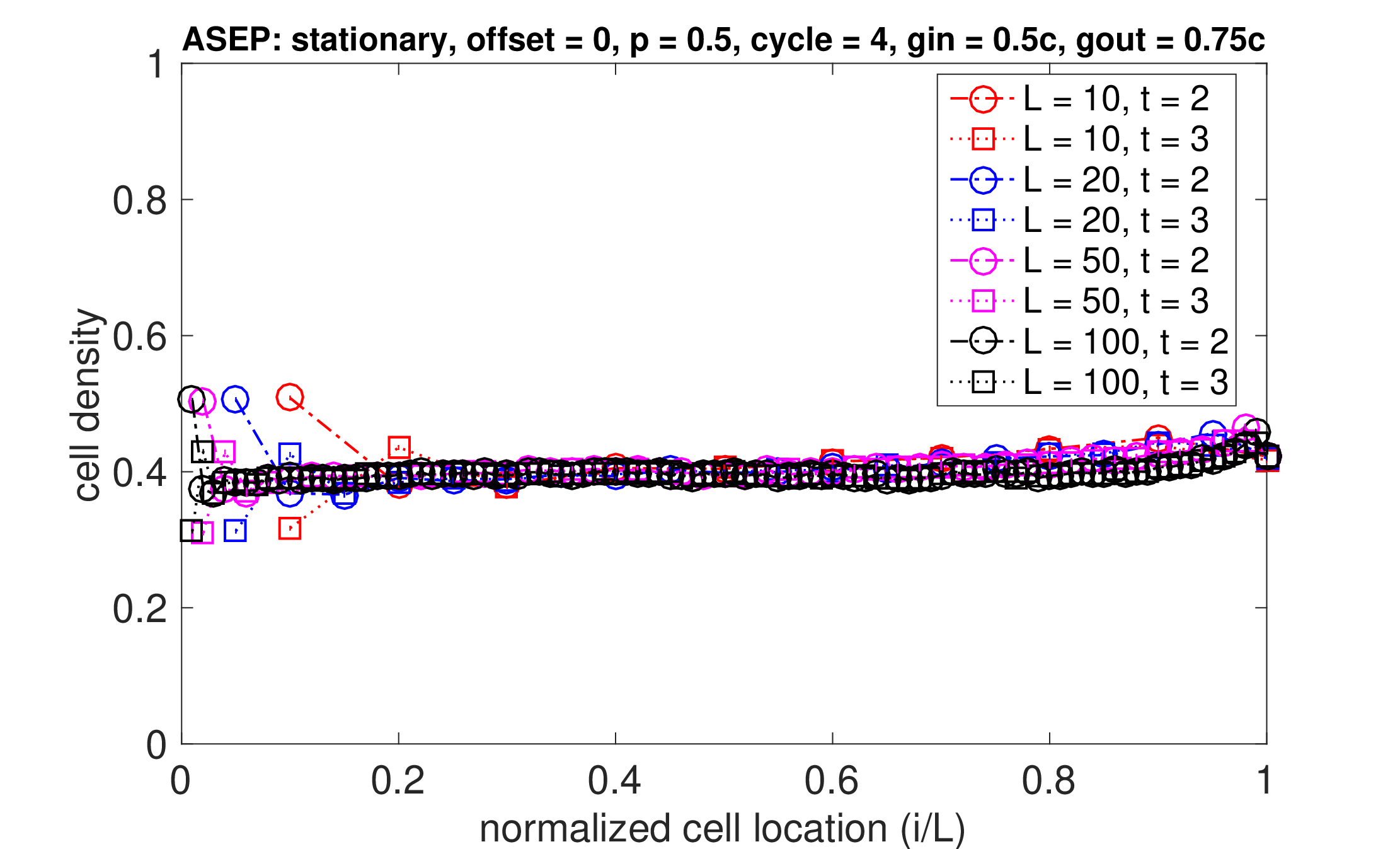}}
\def\NaSchDensityPfiveRapidRedU			{\includegraphics[scale=\onesize]{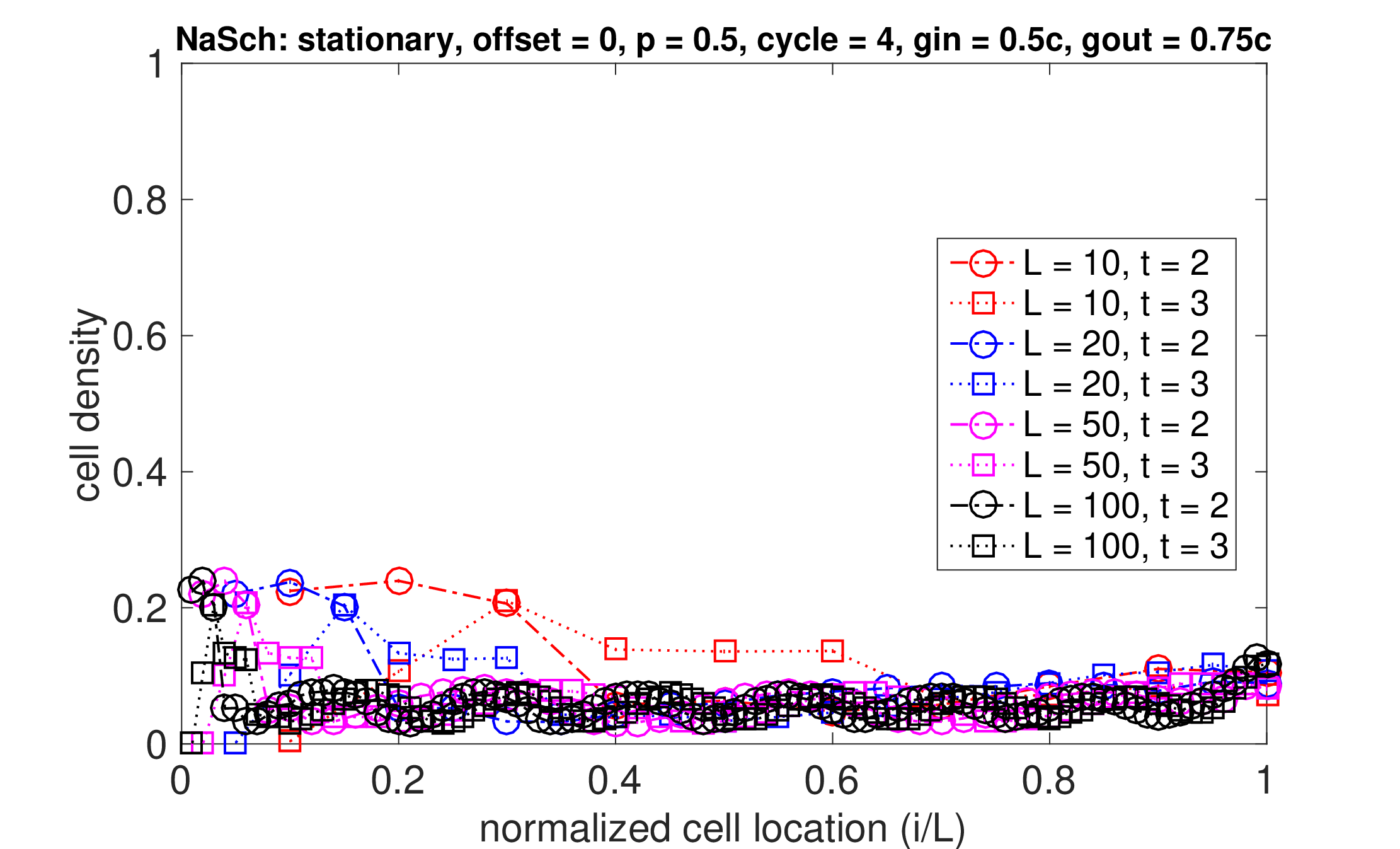}}

\def\ASEPDensityPfiveRescaleRedS			{\includegraphics[scale=\onesize]{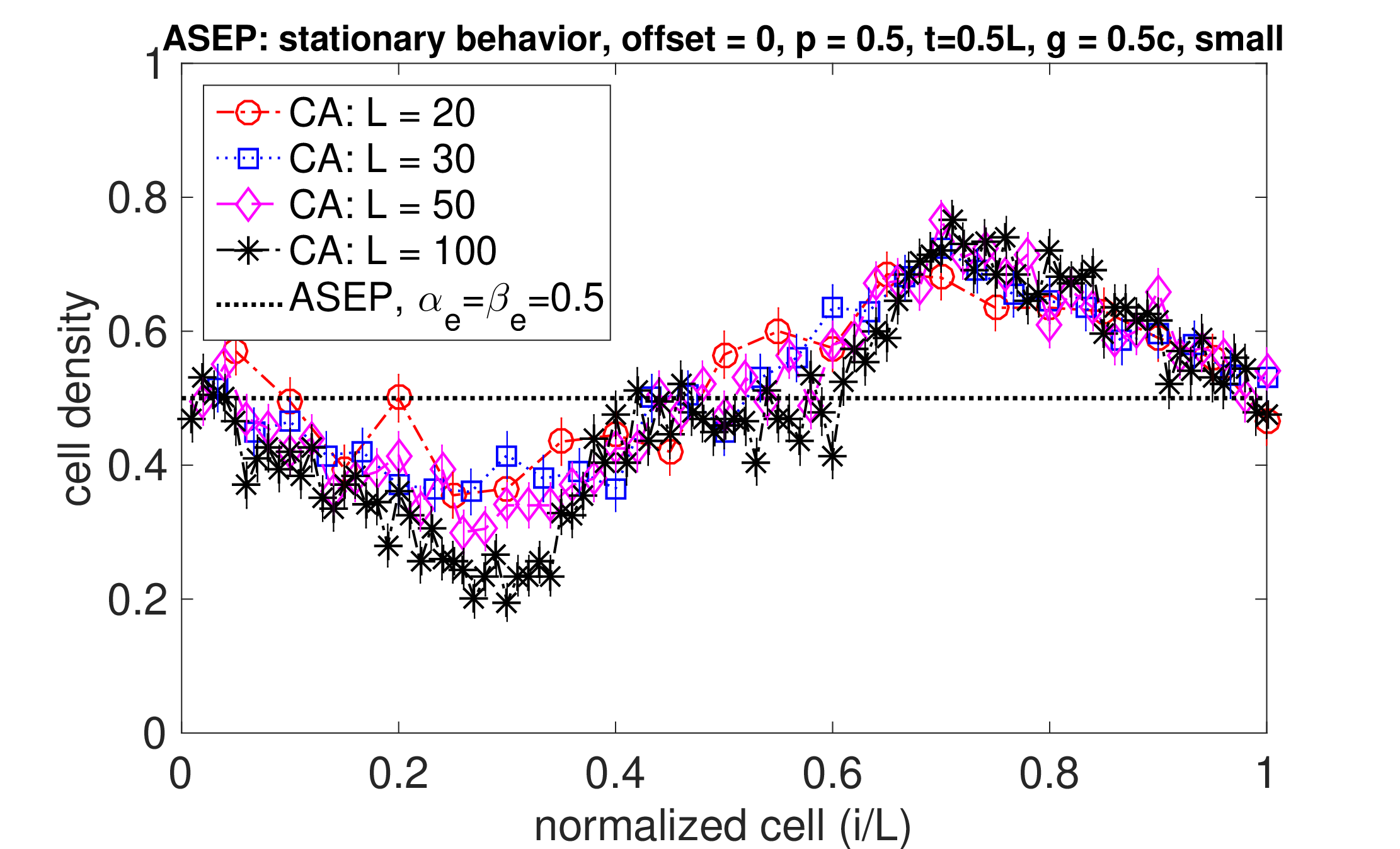}}
\def\ASEPDensityPfiveRescaleRedL 		{\includegraphics[scale=\onesize]{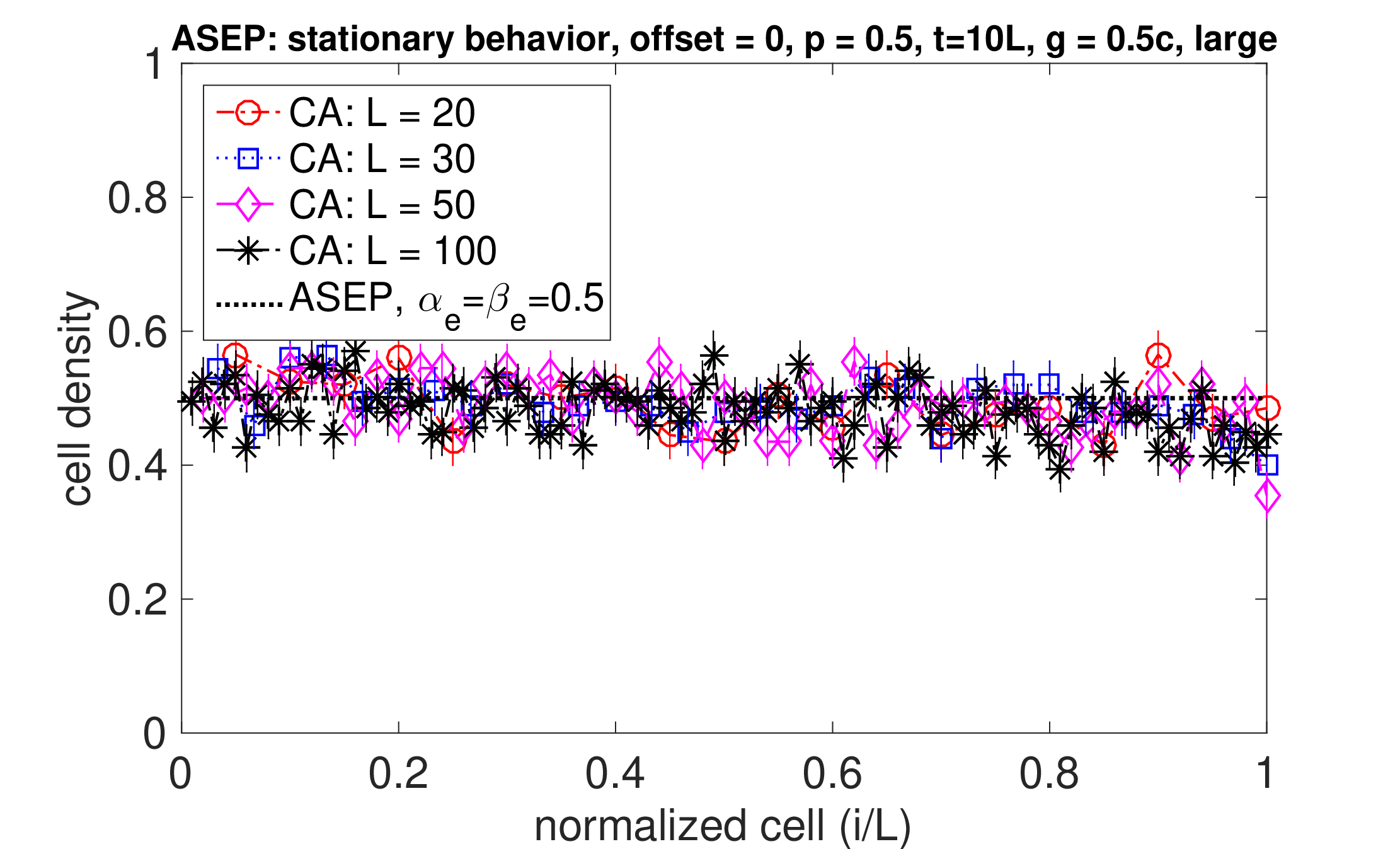}}

\def\ASEPDensityPfiveRescaleGreenS		{\includegraphics[scale=\onesize]{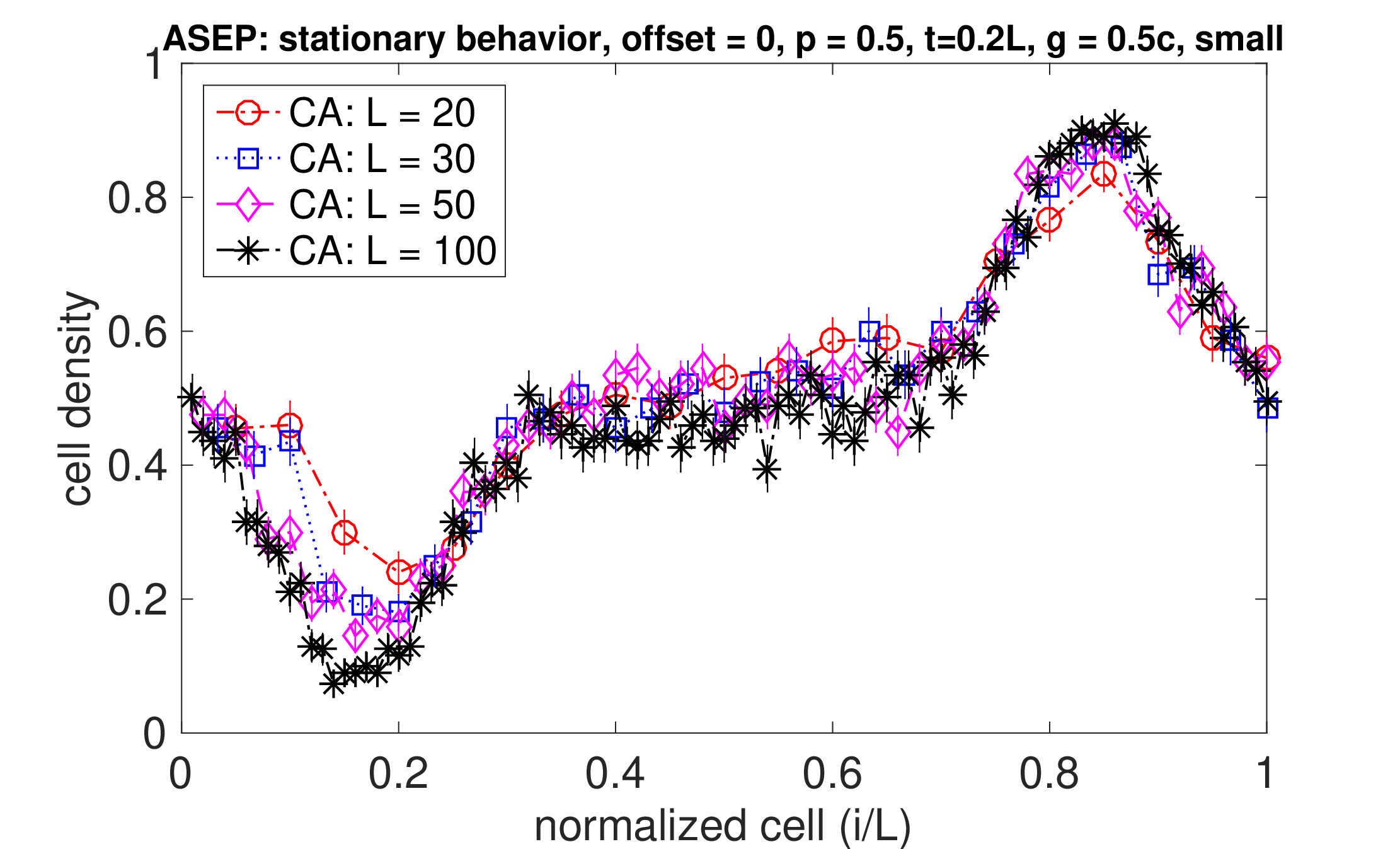}}
\def\ASEPDensityPfiveRescaleGreenL 		{\includegraphics[scale=\onesize]{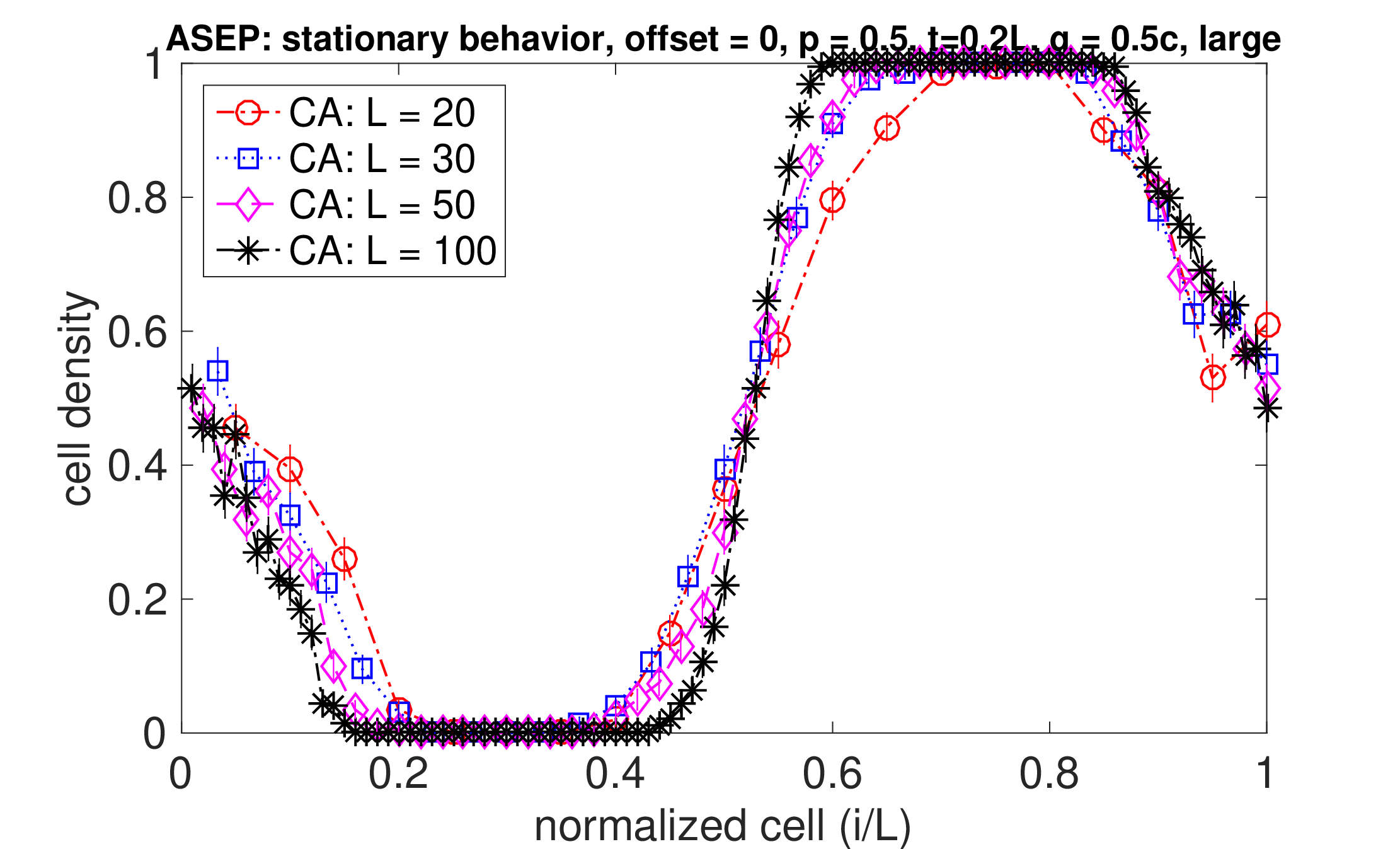}}

\def\ASEPDensityPfiveRescaleUnequalM		{\includegraphics[scale=\onesize]{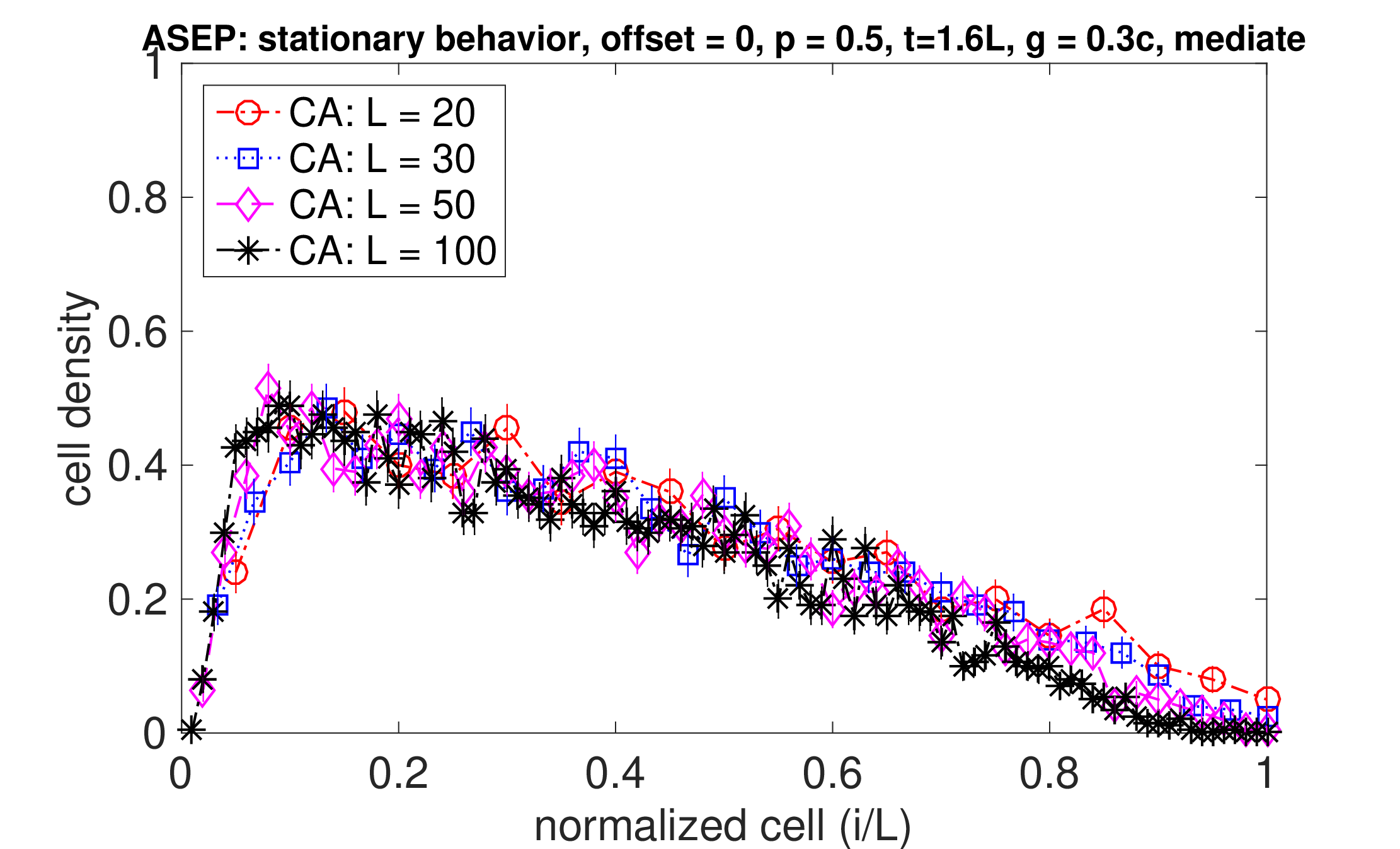}}
\def\ASEPDensityPfiveRescaleOffsetS		{\includegraphics[scale=\onesize]{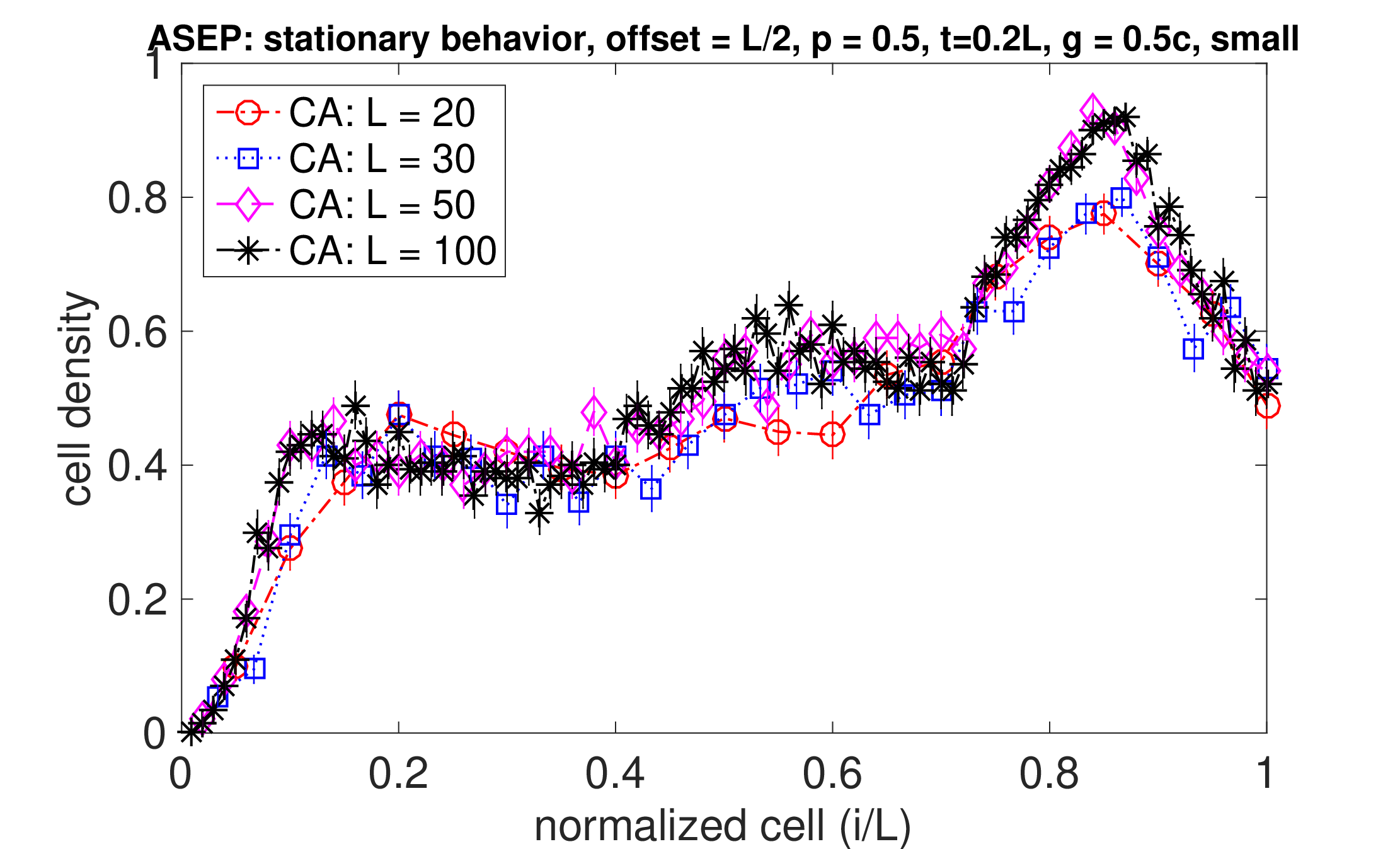}}
\def\NaSchDensityPfiveRescaleRedL		{\includegraphics[scale=\onesize]{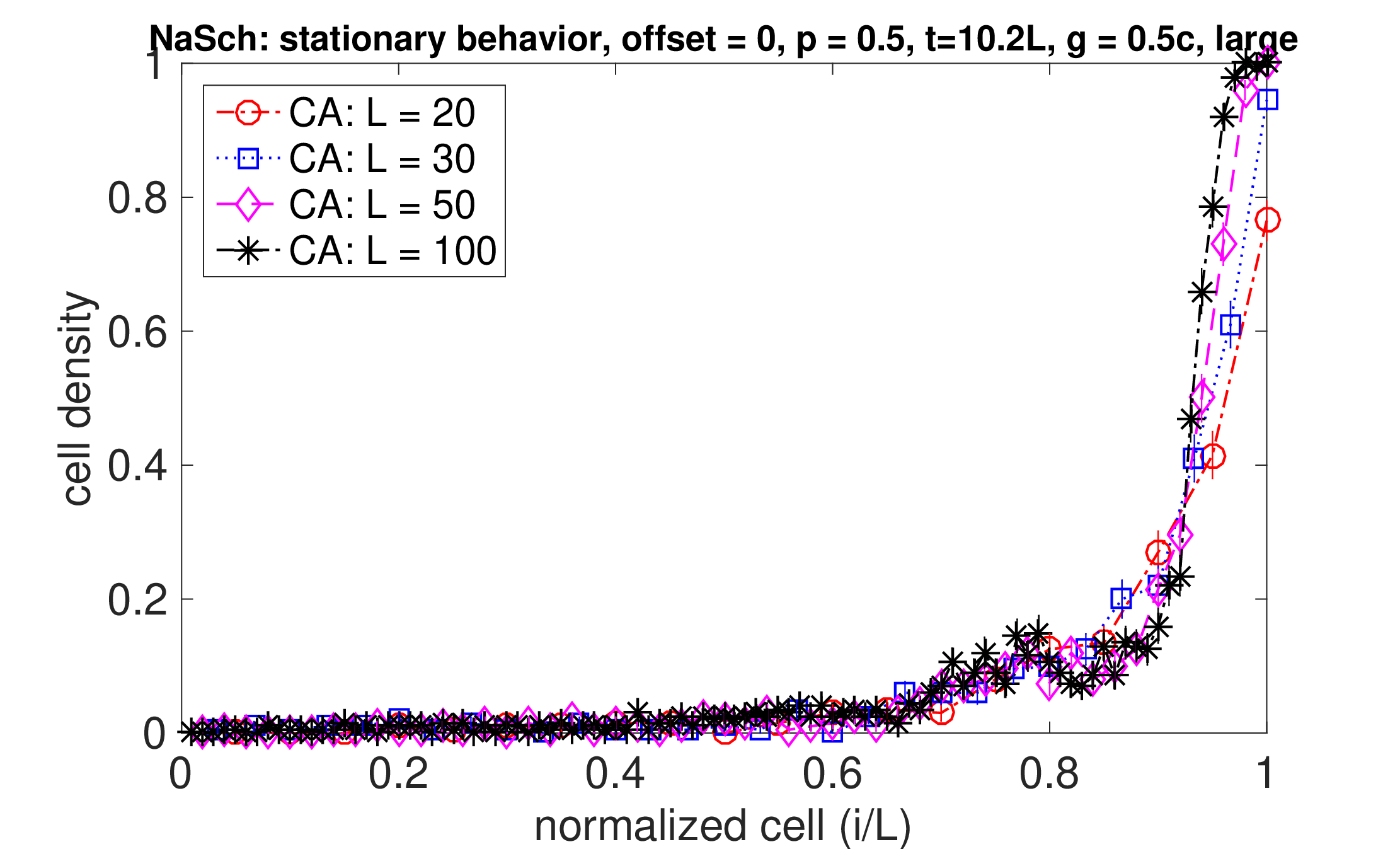}}
\def\NaSchDensityPfiveRescaleGreenL 		{\includegraphics[scale=\onesize]{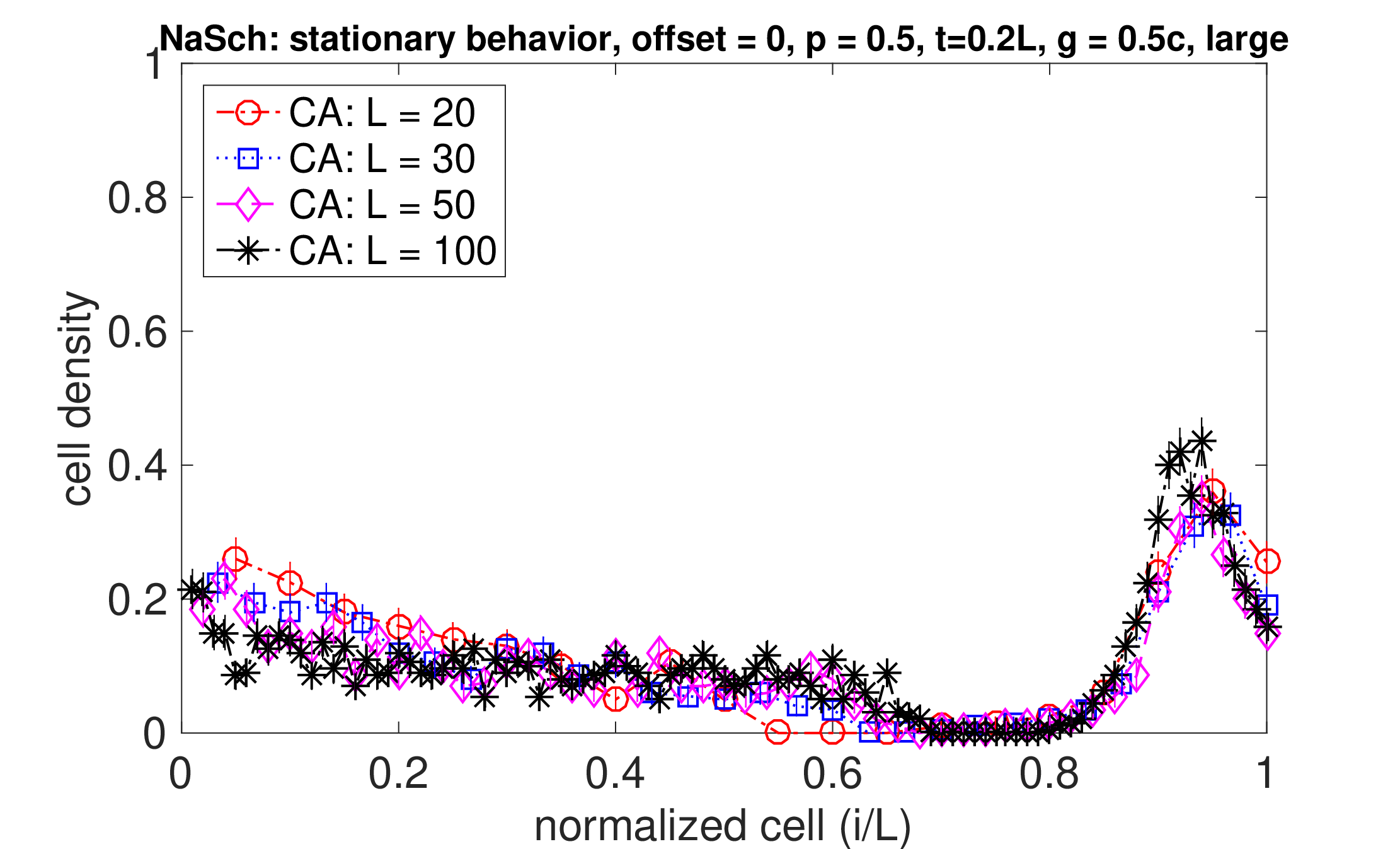}}

\begin{document}

\begin{frontmatter}
\title{Behaviour of traffic on a link with traffic light boundaries}

\author[unimelb,acems]{Lele Zhang\corref{cor1}}
\ead{lele.zhang@unimelb.edu.au}
\author[caley,unimelb]{Caley Finn}
\ead{caley.finn@lapth.cnrs.fr}
\author[monash,acems]{Timothy M. Garoni}
\ead{tim.garoni@monash.edu}
\author[unimelb,acems]{Jan de Gier}
\ead{jdgier@unimelb.edu.au}

\cortext[cor1]{Corresponding author}
\address[unimelb]{School of Mathematics and Statistics, The University of Melbourne, Victoria~3010, Australia}
\address[monash]{School of Mathematical Sciences, Monash University, Clayton, Victoria~3800, Australia}
\address[caley]{LAPTh, CNRS - Universit{\'e} Savoie Mont Blanc, 9 chemin de Bellevue,
BP 110, F-74941  Annecy-le-Vieux Cedex, France}
\address[acems]{ARC Centre of Excellence for Mathematical \& Statistical Frontiers (ACEMS)}

\begin{abstract}
This paper considers a single link with traffic light boundary conditions at both ends, and investigates the traffic evolution over time with various signal and system configurations. A hydrodynamic model and a modified stochastic domain wall theory are proposed to describe the local density variation. The Nagel-Schreckenberg model (NaSch), an agent based stochastic model, is used as a benchmark. The hydrodynamic model provides good approximations over short time scales. The domain wall model is found to reproduce the time evolution of local densities, in good agreement with the NaSch simulations for both short and long time scales. A systematic investigation of the impact of network parameters, including system sizes, cycle lengths, phase splits and signal offsets, on traffic flows suggests that the stationary flow is dominated by the boundary with the smaller split. Nevertheless, the signal offset plays an important role in determining the flow. Analytical expressions of the flow in relation to those parameters are obtained for the deterministic domain wall model and match the deterministic NaSch simulations. The analytic results agree qualitatively with the general stochastic models.
When the cycle is sufficiently short, the stationary state is governed by effective inflow and outflow rates, and the density profile is approximately linear and independent of time. 
\end{abstract}

\end{frontmatter}

%
\section{Introduction}

Cellular Automata (CA) are a popular approach to study freeway traffic. Particularly widely used is the famous one-dimensional Nagel-Schreckenberg (NaSch) model \cite{NagelSchreckenberg92}. This model may be considered on a ring or with reservoirs at the boundaries to model on- and off-ramps. A special case of the NaSch model is the asymmetric simple exclusion process (ASEP), in which car speeds are either 1 or 0.
 
A significant challenge in extending the model to an urban traffic \emph{network} is to incorporate intersections (nodes) with realistic traffic lights. It is well known that road network performance is strongly dependent on traffic signal parameters. The relation between them is complex and difficult to predict. Given that, we choose to study traffic behaviours on a link with traffic light boundaries, the most fundamental component of a road network.

In the sections below we will develop theory to describe a single link and implement traffic lights at the boundaries. Four important parameters determine the behaviour of traffic on the link; these are the cycle time, the split, the offset and the link length. The cycle time is the total time allocated to one green-red cycle of the traffic light, and the split determines how much green vs. red time is allocated within each cycle. The offset is the time difference between the cycle start times at the two boundaries.

Intuition suggests us that a short cycle time, or fast switching, will result in similar behaviour to that of a link without traffic lights but with effective in- and outflow rates proportional to the amount of green time at the boundaries. Large cycle times, compared to the relaxation time of the model, will lead to two pseudo-stationary regimes, one with a green light and one with a red light. The simulation results presented in this article confirm this intuition.

\subsubsection*{Cellular automata and density profiles} 

NaSch models with time-independent boundary conditions have been extensively studied, e.g., \cite{Barlovic2002, Jia2009, Schadschneider2011b}, and exact solutions have been obtained for the special case of ASEP corresponding to maximum speed $v_{\max}=1$. With time-independent boundary conditions, there are three general macro-states: low density, high density and maximum flow. Which state the model attains depends on the inflow and outflow rates at boundaries. 

Studies of systems with time-dependent boundaries have also been conducted using CA models, in particular NaSch, and mean field theories. The first such study appears to be \cite{Popkov2008}, who considered a semi-finite ASEP (i.e., assumed as $(-\infty,0]$) and time-dependent outflow rates and split time fixed at a half. There it was found that the density has a time-periodic stationary sawtooth structure. A similar problem was studied in \cite{Wood2009}. Both \cite{Popkov2008} and \cite{Wood2009} concluded that the most interesting behaviour, and most difficult to analyse, occurs when boundary parameters vary slowly, like realistic traffic lights with intermediate values of the cycle time. 
Woelki \cite{Woelki2013} considered ASEP with random sequential updates using a mean-field approximation. The author investigated density-dependent inflow rates, which models the scenarios of adaptive drivers and/or traffic control (such as ramp-metering). When the system density exceeds a threshold, the inflow rate reduces to a smaller value. As a result two additional phases may occur aside from the usual low/high density and maximum flow phases. When the inflow rate is in rapid alternation the controlled-density phase and co-existence phase come into existence. 

Deterministic NaSch systems with a traffic-light-controlled out-boundary were considered by \cite{Neumann2008, Neumann2009, Jia2011}. Compared with the sawtooth structure in \cite{Popkov2008}, a rectangular density structure was found in \cite{Neumann2009}, which was induced by the traffic light and was sustained due to the absence of stochasticity. In \cite{Neumann2008}, the authors focused on the discussion of travel delays caused by traffic lights. They found a dependence between the road length and the travel time. Jia and Ma \cite{Jia2011} studied a model similar to \cite{Neumann2008, Neumann2009}.
They derived a theoretical expression of the outflow if the red phase is longer than 1 time step. Otherwise, they found the system generated periodic orbits and the maximum flow could depend on the road length. Their study of the deterministic NaSch has a limited application, as the analysis was on the basis of assuming a saturated inflow, and is impossible to be generalised to the stochastic model.
Tobita and Nagatani \cite{Tobita2012} used another deterministic CA model to study the impact of the cycle time, split and offset on the traffic flow through a series of traffic lights in a deterministic ring system with respect to the dynamic transition and the fundamental diagram (FD). Their model is less realistic than NaSch, due to the assumption of infinite acceleration.

Ito and Nishinari \cite{Ito2014} studied the ASEP with parallel updates and a time-dependent outflow rate controlled by pedestrians crossing at an intersection. The pedestrian crossing is modelled as an $M/M/\infty$ queue with discrete time and thus the effective outflow rate becomes a function of the pedestrian arrival rate and exiting rate. They proposed two approximation methods for the study of traffic outflows and the phase diagram: an extended two-cluster approximation and isolated rarefaction wave approximation, which, respectively, produced results in good agreement with the simulations for sufficiently large and small pedestrian crossing speeds. When pedestrian signals are present, it was found that the average flow in the high density phase is approximately proportional to the effective green time over the cycle time. 
Other studies \cite{Huang2003, Toledo2004, Gartner2004} analysed traffic through a sequence of traffic lights.

\subsubsection*{Variational theory and macroscopic fundamental diagrams} 

Daganzo and Geroliminis \cite{Daganzo2008} studied the impact of traffic signal switching on road capacity using variational theory (VT) \cite{Daganzo2005}. In particular, they found the relation between the road capacity and green signal fraction (i.e., green split or ratio) for short and long roads. Their work and the following studies mainly focused on applying VT and the capacity formula to study aggregated route/network performance, see \cite{Gayah2014, Gayah2012, Hans2015}, in particular the estimation of the macroscopic fundamental diagram (MFD). 
Other studies investigated the MFD and critical density in relation to various network and signal parameters including: cycle length \cite{Geroliminis2012, Jin2013, Jin2015}, street length \cite{Geroliminis2012,  Daganzo2016, Laval2015, Leclercq2013, Leclercq2014}, signal phase split \cite{Geroliminis2012, Laval2015, Leclercq2013, Leclercq2014} and signal coordination \cite{Geroliminis2012, Daganzo2016, Girault2016}.

\subsection*{Stationarity and transient behaviour}

In this paper we provide a systematic investigation of the influence of four important parameters: cycle time, phase split, offset and road length, on traffic flows. Instead of explicitly developing traffic signals for optimising traffic flows (or travel times), e.g., \cite{Liu2011}, we aim to explore the general relationships between the traffic flow and traffic signal and road parameters. 

The application of the VT model mentioned above focused on the analysis of the network aggregated performance including the MFD in the stationary state. It is also important to understand more detailed traffic behaviour such as the density profile on a bulk link in the network. Therefore, we focus on the analysis of the time evolution of the traffic, in both stationary and transient states, on a single link with traffic light boundaries, which mimics a bulk link embedded in a traffic network.

The study of the local traffic density evolution between signalised intersections helps us capture spill-back queues and how they are affected by various signal and network parameters. It can be further used to develop appropriate strategies for spill-back prevention and traffic prediction. We consider three models: a simple hydrodynamic model, a stochastic domain wall model and a mesoscopic CA model to study the traffic evolution. Our models do not necessarily assume triangular FDs.\smallskip

We investigate the stationary flow as well as transient traffic behaviour. If the green and red lights are sufficiently long, so the boundary conditions remain invariant for a long time, then the system may settle in pseudo-green/red-light stationary states. Nevertheless, when the light changes, the system becomes transient once more. 

In this paper, we say that the system is at stationarity if the average flow over each cycle is (approximately) time-invariant. We define this flow to be the \textit{stationary flow}. There could be large fluctuations within a cycle but the average flow of one cycle is approximately the same as that of next cycle. There are two transient states of interest: one is before the flow reaches stationarity, and the other is the transient state (within a cycle) after a traffic light change and prior to the green/red-light stationarity (which may or may not occur) after the flow has obtained stationarity. 

\subsubsection*{Organisation of the paper} 

The paper is organised as follows. In Section~\ref{sec:model} we describe a hydrodynamic theory, a domain wall theory and the NaSch model. We incorporate traffic light boundaries into the hydrodynamic model, and improve the stochastic domain wall model, \cite{Zhang2014}, to allow walls to `decay'. In Section~\ref{sec:density} we investigate the traffic evolution before and after the traffic flow has reached a stationary state and compare the results produced by the three models. 
In Section~\ref{sec:stationary_flow} we study the influence on the stationary flow of the four key parameters: cycle length, split, offset and road length. We consider the situation when signals are switched very rapidly in Section~\ref{sec:rapid_switch}. Although the fast switching signal is unlikely to be implemented in most practical scenarios, the study is of interest as it bridges the results for the time-dependent and time-independent boundary conditions. In Section~\ref{sec:discussion} we summarise our findings.

\section{Models}\label{sec:model}

\subsection{System configurations}

\begin{figure}[!h]
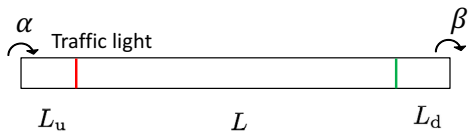

\centering
\system
\vspace{-0.2cm}
\caption{Illustration of the entire system. Parameters $\alpha$ and $\beta$ represent the traffic inflow and outflow rates respectively, and are both set to 1.}\label{fig:system}
\end{figure}

We consider a system of total size $L_{\rm u}+L+L_{\rm d}$. See Figure~\ref{fig:system}. For discrete models, the size is measured in terms of site (or cell) numbers. Two intersections with traffic lights $n_\ri$ and $n_\ro$ are placed at locations $L_{\rm u}$ and $L_{\rm u}+L$, respectively, which leaves a bulk link of length $L$ in-between. We refer to $n_\ri$ and $n_\ro$ as the upstream and downstream nodes. A vehicle is injected into the system with some inflow rate at the in-boundary (left-boundary). A vehicle exits from the out-boundary (right-boundary) with some outflow rate. This configuration corresponds to the simplest scenario of a link embedded in a traffic network. 

The inflow and outflow rates are both set to $1$. We do not consider stochasticity for left and right boundaries so that we can focus on the impact caused by the switching of signals at nodes $n_\ri$ and $n_\ro$. Given both inflow and outflow rates are $1$, the sizes of the upstream and downstream links become unimportant. We fix the lengths of the upstream and downstream links and vary the length of the bulk link. In this setting, the downstream link is a very high capacity road, for example a freeway, whereas the upstream link contains a strong traffic source. As we are only interested in the middle link, from now on, we refer to it as \textit{the} system. 
We consider varying cycle length $c$, green time $g$, split $\gamma=g/c$ and offset $d$ (the time gap between the start times of the green lights at the nodes). Subscripts $\ri$ and $\ro$ are used to distinguish the values at upstream and downstream nodes. We omit the subscripts if the parameter values for the two nodes are the same. 

In the following subsections, we shall describe three models: hydrodynamics, domain wall theory and a CA model. The CA model is discrete and so is the stochastic domain wall model. A particle (i.e., a vehicle) occupies one cell and can take $v_{\max}+1$ values for its speed $v =0,1,2,\dots,v_{\max}$. A \textit{bond} $i$ connects cells $i$ and $i+1$. Domain walls are localised on bonds.

\subsection{Hydrodynamic model}\label{ssec:hydro}

In this subsection, we describe the hydrodynamic model. The use of kinematic waves to study traffic flows dated back to 1950s \cite{Lighthill1955, Richards1956}. We briefly outline the hydrodynamic method, following \cite{BlytheE07, KrapivskyRBN10}, which can be used to compute time dependent density profiles $\rho(i,t)$, with $i$ the position on the lattice.  We treat the density as a continuous function $\rho(x,t)$, and assume it satisfies the continuity equation
\begin{equation}
    \frac{\partial \rho}{\partial t} + \frac{\partial J}{\partial x} = 0.
\end{equation}
Then, if we have a relation $J(\rho)$ between the local density $\rho$ and flow $J$, the continuity equation
becomes
\begin{equation}
    \frac{\partial \rho}{\partial t} + m(\rho) \frac{\partial \rho}{\partial x} = 0 \mbox{ with } m(\rho) = \frac{\partial J}{\partial \rho}.
    \label{eq:hydroCtJrho}
\end{equation}
The solutions $\rho(x,t)$ of \eqref{eq:hydroCtJrho} are constant along characteristics, $x(t)$,  with gradient
$m(\rho)$: with $\rho(t) = \rho(x(t),t)$,
\begin{equation*}
\begin{aligned}
    \frac{d \rho}{d t}
        & = \frac{\partial \rho}{\partial t} + \frac{d x}{d t} \frac{\partial \rho}{\partial x}
         = \frac{\partial \rho}{\partial t} + m(\rho) \frac{\partial \rho}{\partial x} = 0.
\end{aligned}
\end{equation*}
The idea is then, given the initial density profile, $\rho(x, 0)$, to find the density profile at later
times by tracing along the characteristics on which the density remains constant.

\begin{figure}[ht]
    \centering
    \subfigure[FD for discrete ASEP]{
        \includegraphics[width=0.54\textwidth]{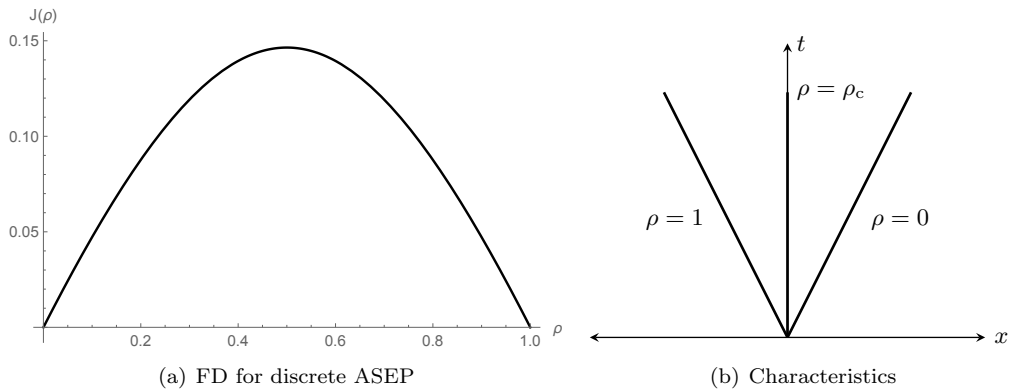}
    }
    \subfigure[Characteristics]{
    \begin{tikzpicture}[scale=1.3]
        \draw[>=stealth, line width=0.5pt, <->] (-2,0) -- (2,0); 
        \draw[>=stealth, line width=0.5pt, ->] (0,0) -- (0,3); 

        \draw[line width=1pt] (0,0) -- (-1.25,2.5);
        \draw[line width=1pt] (0,0) -- (0,2.5);
        \draw[line width=1pt] (0,0) -- (1.25,2.5);

        \node [left] at (-0.75, 1.2) {\small $\rho = 1$};
        \node [right] at (0, 2.5) {\small $\rho = \rho_{\rm c}$};
        \node [right] at (0.75, 1.2) {\small $\rho = 0$};

        \node [right] at (0, 3) {\small $t$};
        \node [right] at (2, 0) {\small $x$};
    \end{tikzpicture}
    }
    \caption{
        FD and example characteristics for the discrete time ASEP with parallel updates.
        \label{fig:dASEPfd}
    }
\end{figure}

To illustrate the method, we will take, as an example, the discrete time ASEP with parallel updates. Let $p$ be the deceleration probability. The flow-density relation of this model is \cite{SchadschneiderS93, SchreckenbergSNI95}
\begin{equation}
    J(\rho) = \frac{1 - \sqrt{1 - 4(1-p)\rho(1-\rho)}}{2},
    \label{eq:JrhoDASEP}
\end{equation}
giving
\begin{equation}
    m(\rho) = \frac{(1-p)(1 - 2\rho)}{\sqrt{1 - 4(1-p)\rho(1-\rho)}}.
    \label{eq:mrhoDASEP}
\end{equation}
This is plotted  in Figure~\ref{fig:dASEPfd} alongside typical characteristics.

We will consider other flow-density relations, but will always assume that
\begin{itemize}
    \item There is a unique maximum flow value $J_\mm$ occurring at some critical density $\rho_\mm$;
    \item The derivative of $J(\rho)$, $m(\rho)$, is continuous, and a strictly decreasing function of $\rho$;
    \item $J(0) = J(1) = 0$.
\end{itemize}

\hl{Hydrodynamics for traffic-light boundary}

{
We now give some examples, which will illustrate how the hydrodynamics method is applied to the system with  traffic lights we consider.

\rms{Rarefaction}

Consider first a downward step profile at $t = t_0$, with $\rho(x, t_0)$ stepping from $\rho_+$ to $\rho_-$ at
$x_0$, with $\rho_+ > \rho_-$.  This is illustrated in Figure~\ref{fig:hydroDownStep} alongside the
corresponding characteristics.
\begin{figure}[ht]
    \centering
    \subfigure[Profile with a downward step]{
    \begin{tikzpicture}[scale = 1.3]
        \draw[>=stealth, line width=0.5pt, ->] (0,0) -- (4,0); 
        \draw[>=stealth, line width=0.5pt, ->] (0,0) -- (0,2.2); 

        \draw[line width=0.5pt] (2,0) -- ++(0, -0.1);
        \node [below] at (2, -0.1) {\small $x_0$};

        \draw[line width=1pt] (0,1.5) -- (2, 1.5) -- (2, .25) -- (4, .25);

        \draw[line width=0.5pt] (0,0.25) -- ++(-0.1, 0);
        \node [left] at (-0.1, 0.25) {\small $\rho_-$};

        \draw[line width=0.5pt] (0,1.5) -- ++(-0.1, 0);
        \node [left] at (-0.1, 1.5) {\small $\rho_+$};

        \node [right] at (4, 0) {\small $x$};
        \node [right] at (0, 2.2) {\small $\rho(x,t_0)$};
    \end{tikzpicture}
    }
    \subfigure[Characteristics]{
    \begin{tikzpicture}[scale = 1.3]
        \draw[>=stealth, line width=0.5pt, ->] (0,0) -- (4,0); 
        \draw[>=stealth, line width=0.5pt, ->] (0,0) -- (0,3); 

        \draw[line width=0.5pt] (2,0) -- ++(0, -0.1);
        \node [below] at (2, -0.1) {\small $x_0$};

        \foreach \x in {0.5, 1, 1.5, 2}
            \draw[line width=1pt] (\x, 0) -- ++(-0.5, 2);

        \foreach \x in {2, 2.5, 3, 3.5}
            \draw[line width=1pt] (\x, 0) -- ++(0.75, 2);

        \draw[line width=1pt, dashed] (2, 0) -- ++(-0.3, 2);
        \draw[line width=1pt, dashed] (2, 0) -- ++(0, 2);
        \draw[line width=1pt, dashed] (2, 0) -- ++(0.4, 2);

        \node [right] at (4, 0) {\small $x$};
        \node [right] at (0, 3) {\small $t$};
    \end{tikzpicture}
    }
    \subfigure[Time evolution]{
        \includegraphics[width=0.55\textwidth]{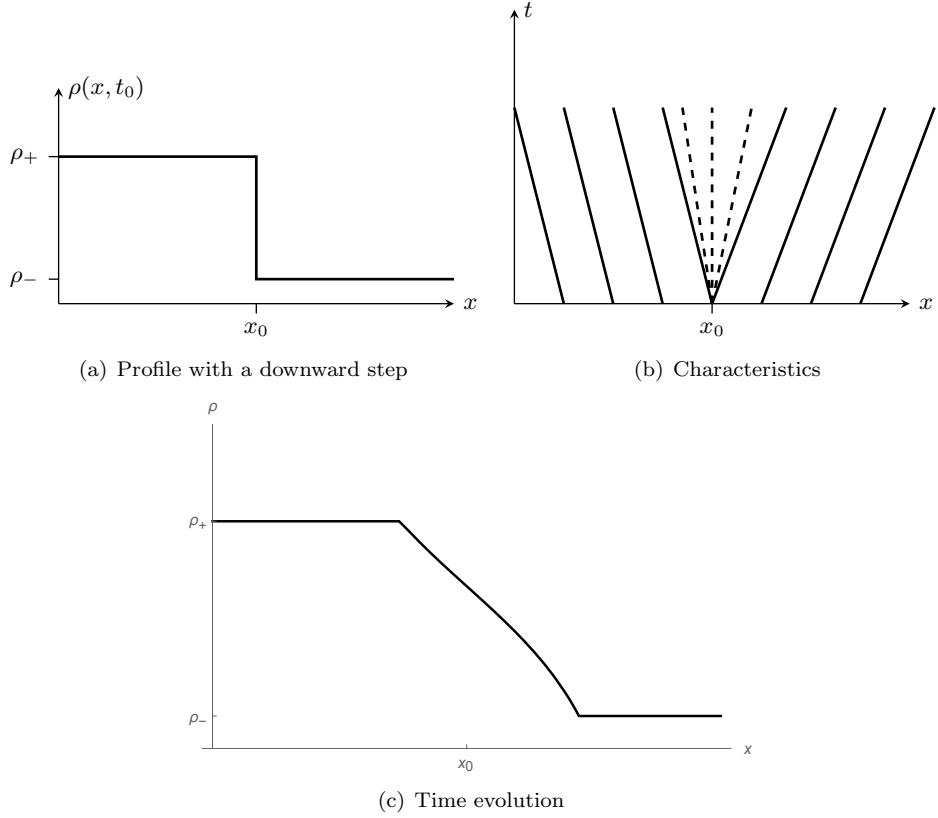}
        \label{fig:hydroDownTLater}
    }
    \caption{
        Rarefaction studied in the hydrodynamics setting
        \label{fig:hydroDownStep}
    }
\end{figure}

In this example, the $\rho_+$ characteristics angle left, while the $\rho_-$ characteristics angle right
(Figure \ref{fig:hydroDownStep}(b)), and the boundaries of the high and low density regions move left and right
accordingly, separated by a rarefaction region (Figure \ref{fig:hydroDownTLater}).

The gap between the high and low density characteristics is filled by a fan of straight line characteristics,
originating a $x = x_0$ (the dashed lines in Figure \ref{fig:hydroDownStep}(b)).
A point  $(x^*, t^*)$ in this fan region will have some density $\rho^*$, and this density will be constant
along the line
\begin{equation*}
    x^* = m(\rho^*)(t^* - t_0) + x_0.
\end{equation*}
Inverting, we find that the density must be
\begin{equation*}
    \rho^* = m^{-1}\left(\frac{x^* - x_0}{t^* - t_0}\right),
\end{equation*}
and this gives the density of any point $(x,t)$ in the rarefaction front. Thus for $t > t_0$,
\begin{equation*}
    \rho(x, t) = \begin{cases}
        \rho^+, & x < m(\rho_+)(t - t_0) + x_0;\smallskip
        \\
        \displaystyle m^{-1}\left(\frac{x - x_0}{t - t_0}\right), & m(\rho_+)(t - t_0) + x_0 < x < m(\rho_-)(t - t_0) + x_0;\smallskip
        \\
        \rho_-, & m(\rho^-)(t - t_0) + x_0 < x,
    \end{cases}
\end{equation*}
which is plotted in Figure~\ref{fig:hydroDownTLater}. For the NaSch model with $v_{\max} = 1$ and $m(\rho)$
given by \eqref{eq:mrhoDASEP}, we have
\begin{equation}
    m^{-1}(z) = \frac{1}{2}\left(1 - z\sqrt{\frac{p}{(1-p)(1-p-z^2)}}\right).
\end{equation}

\rms{Shocks}

If, on the other hand, we have an upward step representing a shock, the characteristics cross and a shock forms between the density regions (see Figure~\ref{fig:exshock}).
Either side of the shock the density propagates along the straight line characteristics from their respective
domains.  The shock forms a moving boundary between the two domains. Its position, $s(t)$, is determined by
conservation of mass and must satisfy
\begin{equation}
    \frac{d s}{d t} = \frac{J(\rho(s(t)^+)) - J(\rho(s(t)^-))}{\rho(s(t)^+) - \rho(s(t)^-)},
    \label{eq:hydroShock}
\end{equation}
where $s(t)^\pm$ indicates the limit of the shock position approached from the left or right domain.

\begin{figure}[!t]
    \centering
    \subfigure[Profile with upward step]{
        
        \begin{tikzpicture}[scale = 1.3]
        \draw[>=stealth, line width=0.5pt, ->] (0,0) -- (4,0); 
        \draw[>=stealth, line width=0.5pt, ->] (0,0) -- (0,3); 

        \draw[line width=0.5pt] (2,0) -- ++(0, -0.1);
        \node [below] at (2, -0.1) {\small $x_0$};

        \draw[line width=1pt] (2, 0) -- ++(-2., 1.5);
        \draw[line width=1pt] (2, 0) -- (2, 2.0);
        \draw[line width=1pt] (2, 2) -- (4, 2);
        \node [right] at (4, 0) {\small $x$};
        \node [right] at (0, 3) {\small $t$};
    \end{tikzpicture}
        \label{fig:exshock}
    }
    \subfigure[Characteristics and shock (thick line)]{
    \begin{tikzpicture}[scale = 1.3]
        \draw[>=stealth, line width=0.5pt, ->] (0,0) -- (4,0); 
        \draw[>=stealth, line width=0.5pt, ->] (0,0) -- (0,3); 

        \draw[line width=0.5pt] (2,0) -- ++(0, -0.1);
        \node [below] at (2, -0.1) {\small $x_0$};

        \draw[line width=1.5pt] (2, 0) -- ++(-0.75, 2.5);
        \draw[line width=0.5pt] (0, 0) -- ++(0.3, 2.5);
        \draw[line width=0.5pt] (0.5, 0) -- ++(0.6, 2.5);
        \draw[line width=0.5pt] (1, 0) -- ++(0.85, 2.5);
        \draw[line width=0.5pt] (1.25, 0) -- ++(1.1, 2.5);

        \draw[line width=1pt, opacity=0.5] (2.25, 0) -- ++(-1.25, 2.5);
        \draw[line width=1pt, opacity=0.5] (2.5, 0) -- ++(-1.25, 2.5);
        \draw[line width=1pt, opacity=0.5] (3, 0) -- ++(-1.25, 2.5);
        \draw[line width=1pt, opacity=0.5] (3.5, 0) -- ++(-1.25, 2.5);
        \node [right] at (4, 0) {\small $x$};
        \node [right] at (0, 3) {\small $t$};
    \end{tikzpicture}
    }\caption{Shocks studied in the hydrodynamics setting}
\end{figure}
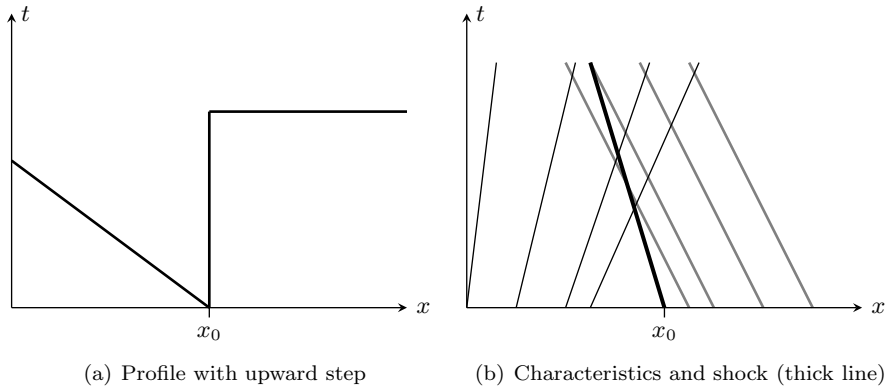

\rms{Boundary conditions}

Switching traffic lights are modelled as time-dependent boundary conditions
\begin{equation*}
    \rho(0, t) = \begin{cases}
        1, & \text{green light};
        \\
        0, & \text{red light};
    \end{cases}
    \qquad
    \rho(L, t) = \begin{cases}
        0, & \text{green light};
        \\
        1, & \text{red light}.
    \end{cases}
\end{equation*} 

The boundary condition at $x = L$ is imposed to either block all traffic, or allow free flow.  But the density
at site $L$ is obtained by taking the $x \to L$ limit from below.  When the traffic lights are placed
mid-lattice (as is the case here with upstream and downstream links) we break the lattice at each red light,
considering the up- and downstream sections as independent systems.  When the light switches to green, the
lattice sections are rejoined.

\rms{Solving the continuity equation}

To solve the continuity equation \eqref{eq:hydroCtJrho}, we start from an initial density profile
$\rho(x,t=0)$, and find the evolution by following the density along characteristics, inserting rarefaction
regions and solving \eqref{eq:hydroShock} for shock positions, where required.  The time-dependent boundaries
(the traffic lights) cause the characteristics originating from the boundaries to vary.

The chief difficulties in this process are solving the shock equation \eqref{eq:hydroShock} for general non-constant
density profiles, and keeping track of the crossing and spreading of characteristics, especially when
time-dependent boundaries are involved.  We overcome the first difficulty by solving the shock equation
numerically, although we then need to ensure we have sufficient numerical precision.  To keep track of the
characteristics, we have implemented code that automates the process.
}

\subsection{Domain wall model}\label{ssec:dwm}

The domain wall model has been shown to be capable of explaining the behaviour of NaSch in both the stationary \cite{Kolomeisky98} and transient regimes \cite{Zhang2014, SantenAppert02}. In this subsection, we explain how we extend it to take into consideration traffic-light boundaries. Domain walls separate two regions of different densities. We shall assume that the width of such walls is small compared to the macroscopic system size, so that to a good approximation we can consider the position of the domain wall to be localised on the bonds between adjacent cells.
The drift velocity of a wall $W_{\cal A|B}$ separating domains $\cal A$ and $\cal B$ with density and flow $(\rho_{\cal A}, J_{\cal A})$ and $(\rho_{\cal B}, J_{\cal B})$ is given by
\begin{equation}
V_{\cal A|B} = \frac{J_{\cal A}-J_{\cal B}}{\rho_{\cal A} - \rho_{\cal B}}.\label{equ:wall_velocity}
\end{equation}
Note that the domain wall drift velocity in \eqref{equ:wall_velocity} is the same as equation \eqref{eq:hydroShock}, when the densities in left and right domains in \eqref{eq:hydroShock} are constant.

The motion of a domain wall can be described as a biased simple random walk. 
See \cite{Zhang2014} for details. The density of cell $i$ at time $t$ becomes a well-defined stochastic process induced by the dynamics of the domain walls. We refer to this domain wall model as the \textit{stochastic domain wall} (SDW). A simple deterministic approximation of the stochastic domain wall model is to approximate the position of a domain wall 
at time $t$ as a linear function of the drift speed.
The densities of domains are assumed to be bulk quantities that are homogeneous between consecutive domain walls. For deterministic ASEP and NaSch, we investigate the \textit{deterministic domain wall} (DDW) model. The density is simply a linear function of the drift speed. 

\hl{SDW for traffic-light boundary}

We now list a few key properties of the SDW model tailored for the traffic-light boundary condition.

\rms{Walls generated by traffic-light changes}

Switching traffic lights at nodes generates domain walls. Three typical traffic regions (or domains) possibly exist in the system: i) A grid-locked region $\mc$, which is full of vehicles with no gaps. It is built from the stop line of node $n$ due to a red light; ii) An empty region $\me$ with no vehicles inside, which is formed at the downstream side of node $n$ when the red light is switched on; and iii) A maximum flow region $\mm$, which is created when the green light is switched on. The conditions for the formation of $\mm$ are a sufficient inflow from upstream and a non-congested regime in downstream, both realised due to the recently closed red lights. Readers may refer to \cite{Zhang2014} for an argument explaining the formation of the maximum flow domain.

At the upstream node $n_\ri$, when the light switches from green to red (red to green), a domain wall $W_{\me|{\cal A}}$ ($W_{\mm|{\cal A}}$) is placed at the node, provided that $\cal A\neq \me$ ($\cal A\neq \mm$). Domain $\cal A$ is the current state of the cell to the right of $n_\ri$. Likewise, a domain wall $W_{{\cal B}|\mc}$ ($W_{{\cal B}|\mm}$) is placed at the downstream node $n_\ro$, when the light changes from green to red (red to green), provided that $\cal B\neq \mc$ ($\cal B\neq \mm$). State $\cal B$ is the current state of the cell left to $n_\ro$.

\rms{Walls merging}

In order to model the effect of the traffic lights, we will need to allow for the possibility of multiple domain walls coexisting in the system simultaneously. We assume that the dynamics of each wall are independent of all the other walls, unless two or more walls choose to move to the same bond. In such a case the walls immediately merge and then form a single wall if two sides of the merged walls have different states. The number of walls present in the system at time $t$ is not conserved. 

\rms{Walls exiting the system}

The SDW model allows walls to exit the (bulk) system when they reach either of the nodes, as there are upstream and downstream links connected to the (bulk) system. For example, suppose that the system is divided by a wall $W_{\mm|\mc}$ into two domains $\mm$ and $\mc$, and the signals at $n_\ri$ and $n_\ro$, respectively, stay green and red. The wall will drift upstream, pass node $n_\ri$ and eventually move into the upstream link.

\rms{Unstable walls}

The stability of a domain wall is determined by the {\it collective velocity} (local velocity) $V_c ={d }J/{d} \rho$, \cite{Schadschneider2011}. An upward wall (i.e., shock) with  $\rho_{\rm left} < \rho_{\rm right}$ is stable, whereas a downward wall (i.e., rarefaction) with $\rho_{\rm left} > \rho_{\rm right}$ is unstable.  This is because, for the downward wall, $V_c(\rho_{\rm left}) <V_{\rm wall}< V_c(\rho_{\rm right})$, and the excess mass moves away from the wall. In our discussions, walls $W_{\mm|\me}$ and $W_{\mc|\mm}$ are unstable walls. They will smear out as they drift. 
Compared to ASEP, NaSch with $v_{\max}>1$ has a steeper FD. Larger maximum speeds lead to steeper FDs. As a result, walls $W_{\mm|\me}$ and $W_{\mc|\mm}$ are more stable in the $v_{\max}>1$ setting. 

In order to capture the characteristics of the unstable walls for ASEP, we extend our SDW model and allow unstable walls to split. The basic idea of the extended model is as follows. Suppose that a downward wall $W_{\cal A|B}$ drifts right. It hops right with probability $V_c(\rho_{\cal A})$ and stays still with probability $1-V_c(\rho_{\cal B})$. Otherwise, wall $W_{\cal A|B}$ splits into two walls $W_{\cal A|D}$ and $W_{\cal D|B}$ with $\rho_{\cal D} =(\rho_{\cal A}+\rho_{\cal B})/2$, and then wall $W_{\cal D|B}$ hops right. As a result, an unstable wall spreads out (into multiple walls). 
We remark that our SDW model for NaSch with $v_{\max}>1$ ignores the unstable downward domain walls, and so do not consider wall splitting. That means it only uses the empirical data for section $\mm$, and assumes a triangle-shaped FD. {The assumption is reasonable given the shape of the FD (see Figure~\ref{fig:nasch_fd}). In general, the proposed SDW does not rely on a triangle-shaped FD as it deals with both stable and unstable walls.}

\subsection{CA model -- NaSch}\label{ssec:ca}

A CA model such as NaSch, is discrete in space, time and state variables. In the NaSch model, each vehicle occupies one cell and has speed of $0,1,2,\dots,$ or $v_{\max}$ depending on local traffic conditions. In this paper, we apply the NaSch model with two signalised internal nodes to study the traffic behaviour on a link with traffic signals at both sides embedded in a network. We use the CA simulation results as a benchmark to compare with the domain wall and hydrodynamic approximations. When setting green splits at both nodes $\gamma_\ri=\gamma_\ro=1$, the system becomes a NaSch in the maximum flow regime. We let $J_{\max}$ denote the maximum flow obtained in such a case. We shall compare the stationary flows obtained with varying parameters with $J_{\max}$.

\rms{Parameters}

The NaSch model includes, each time step and for each vehicle, a random deceleration which is applied with probability $p$. The model becomes deterministic when $p=0$, and is stochastic otherwise. In this article, we focus on the stochastic case with $p=0.5$. We also consider the deterministic NaSch, whose simple analytic results help us qualitatively explain the observations for the stochastic case. Regarding the maximum speed $v_{\max}$ we consider two values: $v_{\max}=1$ and $4$. The FDs for $v_{\max}=1$ and $v_{\max}>1$ are significantly different. Unless specified otherwise, in the remainder of this paper, we refer to the NaSch with $v_{\max}=4$ as the NaSch case and that with $v_{\max}=1$ as the ASEP case.

\rms{FDs}

Both the SDW and hydrodynamic models rely on knowledge of the FD. For ASEP we use the analytical FD (see Figure~\ref{fig:dASEPfd}(a) and Eq.~(\ref{eq:JrhoDASEP})), whereas for NaSch we use approximations of the FD obtained numerically from simulations. See Figure~\ref{fig:nasch_fd} in \ref{app:fd}.

\section{Density Profiles}\label{sec:density}


As discussed in the introduction, the system with traffic-light boundaries has two types of transient states: before the traffic flow reaches stationarity (i.e., the flow remains invariant); and between a signal change and the reach of pseudo-stationarity, after the average flow remains invariant. In this section we study the transient behaviour in both states. For the deterministic case, the proposed DDW model can reproduce the transient solution of the CA simulation. For the stochastic case,  we analyse the density profile results provided by the hydrodynamic and the domain wall models and compare them with those obtained from the CA simulations. 



\subsection{Deterministic case}\label{ssec:deterministic_tr}

\begin{figure}[!t]
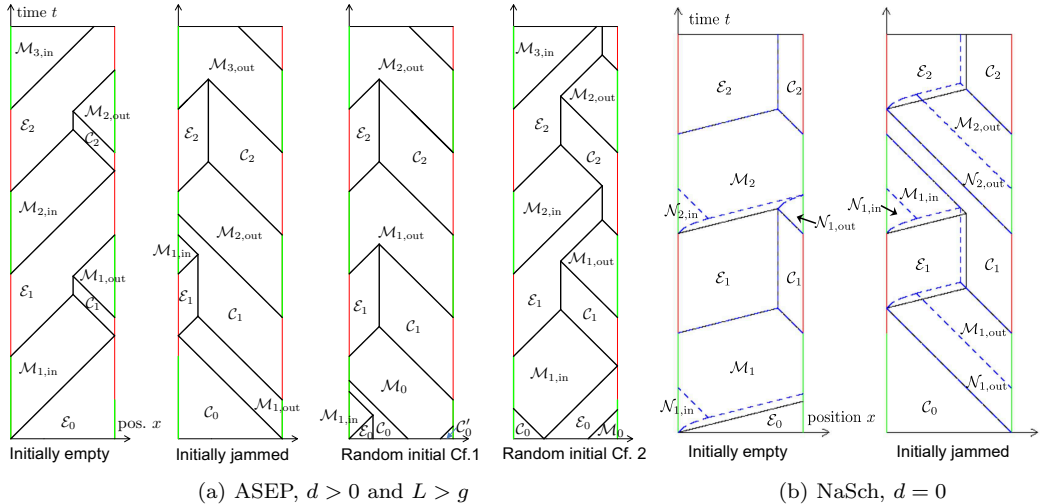

\centering
\begin{tabular}{cc}
\hspace*{-0.2cm}{\asepTrajectoryLarged}&\hspace{-0.4cm} {\naschTrajectorycmp}\\
{\footnotesize (a) ASEP, $d > 0$ and $L>g$}& {\footnotesize (b) NaSch, $d=0$}\\
\end{tabular}
\vspace{-0.4cm}
\caption{Domain wall trajectories for deterministic ASEP (a) and NaSch (b) with $p=0$ and $\gamma=0.5$ and different initial conditions. X-axis: position in the system; Y-axis: time. Green and red lines represent the traffic lights at nodes $n_\ri$ and $n_\ro$. (b) Black wall trajectories are obtained when ignoring the existence of the expansion domain, whereas blue dashed trajectories are obtained taking into consideration the expansion domain.
}\label{fig:deterministic_walls}
\end{figure}

For the deterministic case, the time evolutions of density profiles can be obtained from the domain wall trajectories produced by the DDW model. Examples are shown in Figure~\ref{fig:deterministic_walls}, where the wall trajectories produced by DDW for ASEP and NaSch with different initial conditions are plotted. In addition to domain walls initially existing in the system, five types of domain walls $W_{\mm|\mc}$, $W_{\mc|\mm}$, $W_{\mm|\me}$, $W_{\me|\mm}$ and $W_{\me|\mc}$ can be found in the system. Walls other than $W_{\me|\mc}$ are created by signal switching. For the ASEP system illustrated in Figure~\ref{fig:deterministic_walls}(a) with non-zero offsets and equal splits, we consider four initial configurations. In the two extreme cases, initially empty and full (jammed), stationarity is reached by the end of the first green light at $n_\ri$ and the end of the second green light at $n_\ri$ respectively. After that time, the state during the next cycle simply repeats the previous one. In the other two cases, random initial configurations 1 and 2, stationary is obtained at the end of the first green light at $n_\ri$. For the NaSch systems shown in Figure~\ref{fig:deterministic_walls}(b) with zero offsets and equal splits, stationarity is reached by the end of the first green light at $n_\ri$.

We would like to comment on two issues regarding the results for the deterministic variants. Firstly, we observe that, due to the fact that the system is completely deterministic, the stationary configuration depends on the initial condition. In Figure~\ref{fig:deterministic_walls}, in most cases, the stationary states are different for different initial conditions. 
We focus on the two extreme initial cases and use them as representatives. Nevertheless, we would like to comment on other initial cases. For an arbitrary initial configuration, the system can always be partitioned into sections $\me,\ \mm$ and $\mc$ with domain walls from the above-mentioned five types. Two examples of ASEP with randomly generated initial configurations are given in Figure~\ref{fig:deterministic_walls}(a). The random initial configuration 1 has four initial sections: $\me_0,\ \mc_0,\ \mm_0 $ and $ \mc'_0$. Domain walls separating those sections all disappear prior to the start of the first red light at $n_\ri$. Its stationary state is exactly the same as that obtained from the initial full condition. In fact, after wall $W_{\mc_0|\mm_0}$ arrives at $n_\ro$, the system starts repeating its stationarity pattern. This is in advance of the initial full case, in which the repeating pattern starts when wall $W_{\mc_1|\mm_{2,\text{out}}}$ reaches $n_\ro$, as the random configuration 1 is `closer' to the stationary state. Regardless of the initial conditions, the DDW can reproduce the CA simulations. Further discussion is given in \ref{app:reducible}.

Secondly, for NaSch, there is one sub-domain in addition to the three domains $\mm,\mc$ and $\me$. Suppose there exists a queue at the stop line during the red light. When the green light is switched on, vehicles start to accelerate, and form an \textit{expansion domain} $\mn$, which is a section of vehicles that are accelerating to $v_{\max} > 1$. Once they reach the maximum speed, a maximum flow domain $\mm$ forms and expands upstream and downstream. The expansion domain stays left to domain $\mm$ and drifts upstream with speed $1$ until the queue dissolves completely. This expansion domain takes time $t^\dagger=\sum_{i=2}^{v_{\max}} i$ to clear a node $n$, which is defined as the last second that a vehicle traverses a node $n$ with speed less than $v_{\max}$. The total number of vehicles that have traversed node $n$ by time $t^\dagger$ is $q^\dagger=\sum_{i=1}^{v_{\max}-1} i$. This means the average flow during that period is lower than the stationary maximum flow during a infinitely long green light, which is $v_{\max}/(v_{\max}+1)$. As a result, the stationary flow for small cycle times is lower than that for large cycle times, which we shall see in next section. Figure~\ref{fig:deterministic_walls}(b) compares the wall trajectories for NaSch both when considering and not considering the expansion domain $\mn$. The DDW produces exactly the same solution as the NaSch simulations if considering $\mn$. Otherwise, it gives good approximations in most scenarios except when the cycle time and/or the system size is too small. 

\subsection{Stochastic case: Before the flow reaches the stationary state}\label{ssec:before_stat}

We begin by studying the transient behaviour of the stochastic ASEP and NaSch at the early stage when the flow is not at stationarity. The splits at both nodes are set to $0.5$ and the deceleration rate is set to $0.5$. We choose two representative system sizes: $L= 25$ and $100$. We choose a short cycle length $c=50$ for the small size, and $c=150$ for $L=100$. These choices are reasonable as congestion is more likely to spread out on short roads, and choosing a short cycle length would prevent queues from spilling back to upstream intersections. {We choose the two extreme initial conditions: initially empty and full. For other initial conditions, intermediate results are expected. Figure~\ref{fig:density_p5_tr} compares the density profiles $\rho(i)$ vs position $i$ at two time steps $t$ during the early stage for different sizes, cycle times and initial conditions. The first time step $t$ is chosen to be the end of the first cycle, while the second one is chosen to be during a green light after the system has evolved for over 1200 iterations. 

\begin{figure}[!t]
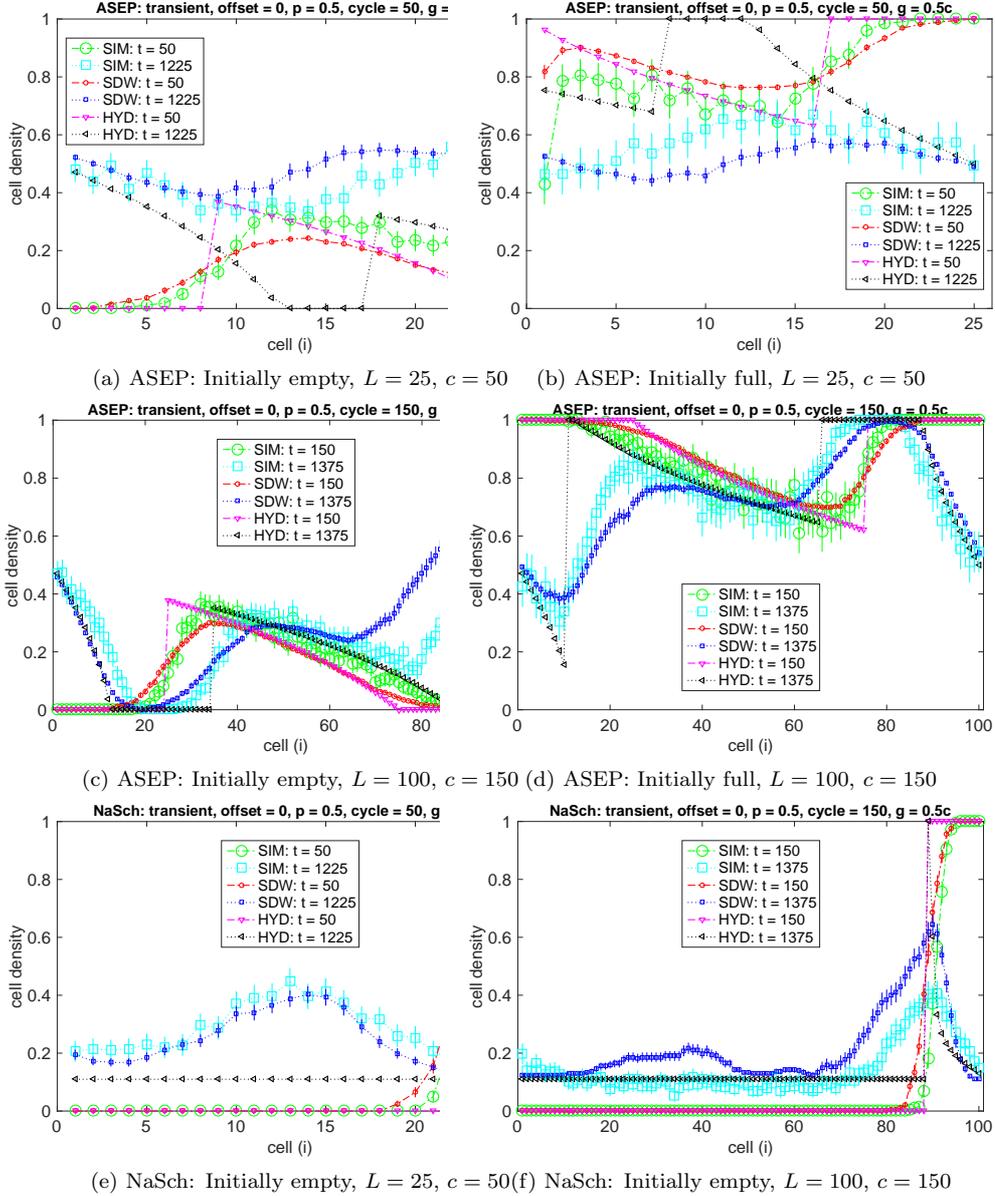

\centering
\begin{tabular}{cc}
\hspace*{-0.9cm}\ASEPDensityPfiveEmptyTransient&\hspace{-1.2cm} \ASEPDensityPfiveFullTransient\\
{\footnotesize (a) ASEP: Initially empty, $L=25$, $c=50$}&\hspace{-1.2cm}{\footnotesize (b) ASEP: Initially full, $L=25$, $c=50$}\smallskip\\
\hspace*{-0.9cm}\ASEPDensityPfiveEmptyTransientLongLarge&\hspace{-1.2cm}\ASEPDensityPfiveFullTransientLongLarge\\
{\footnotesize (c) ASEP: Initially empty, $L=100$, $c=150$}&\hspace{-1.2cm}{\footnotesize (d) ASEP: Initially full, $L=100$, $c=150$}\smallskip\\
\hspace*{-0.9cm}\NaSchDensityPfiveEmptyTransient&\hspace{-1.2cm}\NaSchDensityPfiveEmptyTransientLongLarge\\
{\footnotesize (e) NaSch: Initially empty, $L=25$, $c=50$}&\hspace{-1.2cm}{\footnotesize (f) NaSch: Initially empty, $L=100$, $c=150$}\smallskip\\
\end{tabular}
\vspace{-0.4cm}
\caption{Comparison of cell densities among CA simulations, domain wall and hydrodynamic predictions for stochastic ASEP and NaSch with $p=0.5$, $d=0$ and $\gamma=0.5$ at $t$ seconds after the system starts to evolve. CA: CA simulation; SDW: stochastic domain wall; HYD: hydrodynamics. Error bars corresponding to one standard deviation are shown.
}\label{fig:density_p5_tr}
\end{figure}

The results for ASEP are given in Figures~\ref{fig:density_p5_tr}(a)-(d). The hydrodynamics provide good approximations in terms of predicting the shape and position of rarefaction. The main deviations from the CA curves occur at the locations of shocks (vertical dotted lines in Figures~\ref{fig:density_p5_tr}(a)-(d)). Unlike SDW, the hydrodynamic model considers that shocks move deterministically. Thus a step profile is maintained at the (time-varying) position of the shock. This is also the reason behind the discrepancies between the CA and hydrodynamic curves at latter times except the effect is much more pronounced. In Figure~\ref{fig:density_p5_tr}(a), in the cell range $i\in[13, 17]$, the hydrodynamic model predicts that the cell densities at $t=1225$ are (close to) zero, suggesting the rarefaction from $n_{\ri}$ is catching up with the shock from $n_\ro$. On the other hand, both the SDW and CA results show that the cell densities are about 0.3, suggesting that the two domain walls are likely to have met and merged by then. A similar scenario occurs in the cell range $i\in [8,14]$ in Figure~\ref{fig:density_p5_tr}(b). 

Next we consider the domain wall results. The SDW predictions are in good agreement with the simulation results for both the representative configurations. The agreement is better at the very early stage $t=50$ or $150$. At latter time, for $L=100$, moderate discrepancies between SDW and CA are observed in the cell ranges $i\in[65,85]$ and $[20,30]$ under two different initial conditions respectively. These regions correspond to the ``density dispersal'' of shocks (i.e., upward domain walls). In the SDW model, shocks do random walks, and unlike rarefaction (i.e., downward unstable walls), they do not split. The density dispersal is a result of the random walk and the merge with other (unstable) walls. The discrepancies suggest that SDW overestimates the dispersing speed. Those discrepancies are less pronounced or absent for $L=25$, likely because the system is too small for the ``overestimated scattering" to accumulate. By contrast, the deterministic essence in the hydrodynamic model underestimates the density dispersal near shocks.

The results plotted in Figure~\ref{fig:density_p5_tr}(e)-(f) are for the NaSch system with an empty initial condition. The performance of both the SDW and hydrodynamic models improves. For the small system, the SDW and CA curves almost overlap each other. For the large system, the SDW prediction at $t=150$ is very close to the CA simulation. At the latter time step, Figure~\ref{fig:density_p5_tr}(f) shows that the densities in two cell regions $i\in[20,45]$ and $[70,90]$ are overestimated. This suggests that, contrary to the result for ASEP, the SDW model underestimates the density dispersal near shocks. One possible explanation for the density dispersal overestimation for ASEP and underestimation for NaSch is the mechanism by which SDW treats unstable walls. Recall that our SDW model  considers unstable walls splitting for ASEP and neglects their instability for NaSch. Nevertheless, the discrepancy is not significant, since the effect of the unstable walls is not strong due to the shape of the FD of NaSch.

With respect to the hydrodynamic predictions, most deviations from the CA results occur near shocks at $t=50$, for example the cell range $i\in[80,89]$ at $t=1375$ in Figure~\ref{fig:density_p5_tr}(f), which are essentially the same as the observations made for the results of ASEP. Another significant difference is at time $t=1225$ in Figure~\ref{fig:density_p5_tr}(e). It seems the hydrodynamic model overestimates the propagation of the maximum flow region and hence produces a flat density profile, whereas the SDW and CA results both suggest that the red-light created high density has not dissipated completely yet. We therefore examined the time evolution of the density profile from $t=1$ to $1225$, and found that this discrepancy is accumulated along time due to the deterministic nature of hydrodynamics.

The hydrodynamic model is simple and can provide a reasonable prediction of the transient profile on short time scales. However, due to lack of stochasticity, it does not perform well for long time scales. Comparing its performance for ASEP for different system sizes at a relatively late stage (i.e., $t=1225$ or $1375$), the hydrodynamic model performs better for the larger system. One reason is that the smaller system has a shorter relaxation time, that is, the time that the systems takes to obtain stationarity. At the chosen time step the smaller system is closer to the stationary state than the larger system. As a result, the performance of hydrodynamics is worse. 

\subsection{Stochastic case: After the flow reaches the stationary state}\label{ssec:after_stat}

In this subsection we explore the transient behaviour within a cycle when the flow has obtained stationarity. To ensure the flow is stationary, we simulate the models for a long time (up to 900,000 iterations) prior to measurements. We then study the cell density profile at different time steps of a cycle. The selected time steps correspond to the start and middle of green and red lights. The investigation focuses on the initial condition of an empty system, which suffices due to the ergodicity of stochastic NaSch. 

\begin{figure}[!h]
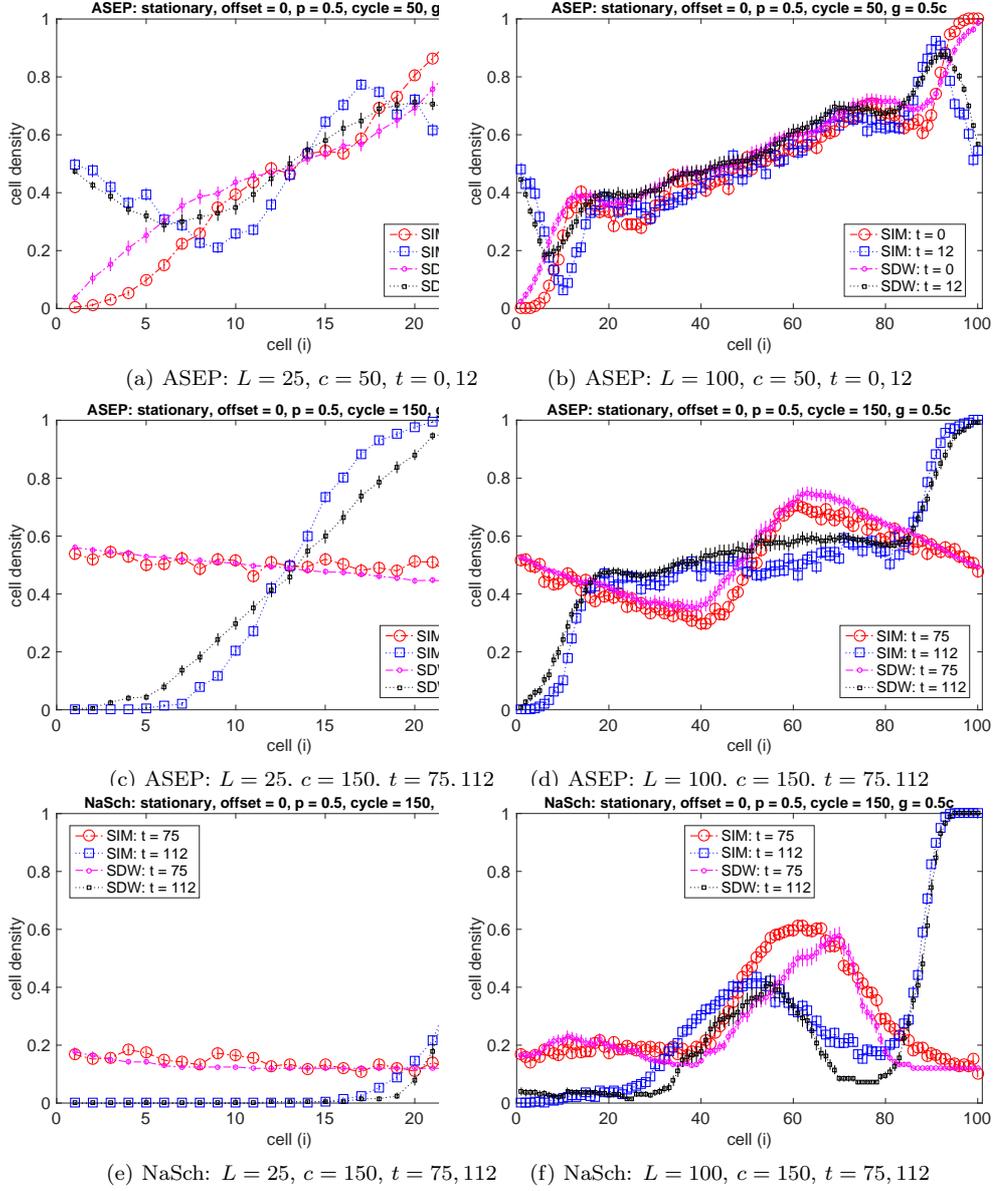

\centering
\begin{tabular}{cc}
\hspace*{-0.95cm}{\ASEPDensityPfiveCmpGreen}&\hspace{-1.15cm}{\ASEPDensityPfiveCmpLongGreen}\\
{\footnotesize (a) ASEP: $L=25$, $c=50$, $t=0,12$}&\hspace{-1.15cm}{\footnotesize (b) ASEP: $L=100$, $c=50$, $t=0,12$}\smallskip\\
\hspace*{-0.95cm}{\ASEPDensityPfiveCmpLargeRed}&\hspace{-1.15cm}{\ASEPDensityPfiveCmpLongLargeRed}\\
{\footnotesize (c) ASEP: $L=25$, $c=150$, $t=75,112$}&\hspace{-1.15cm}{\footnotesize (d) ASEP: $L=100$, $c=150$, $t=75,112$}\\
\hspace*{-0.95cm}{\NaSchDensityPfiveCmpLargeRed}&\hspace{-1.15cm}{\NaSchDensityPfiveCmpLongLargeRed}\\
{\footnotesize (e) NaSch: $L=25$, $c=150$, $t=75,112$}&\hspace{-1.15cm}{\footnotesize (f) NaSch: $L=100$, $c=150$, $t=75,112$}\\
\end{tabular}
\vspace{-0.2cm}
\caption{Comparison of cell densities between CA simulations and stochastic domain wall predictions for stochastic ASEP and NaSch, $p=0.5$, $d=0$ and $\gamma=0.5$ at $t$ seconds after a cycle starts (when the system flow is at stationarity). The system is initially empty. CA: CA simulation; SDW: extended stochastic domain wall. 
}\label{fig:density_p5_st}
\end{figure}

\rms{Comparison between CA and SDW}

In this section we compare the results of the SDW theory with the CA simulations. Figure~\ref{fig:density_p5_st} depicts the density profiles for ASEP and NaSch respectively, at different stages of a cycle when the system is in the stationary state. The CA and SDW results are in relatively good agreement. The agreement for NaSch is slightly worse compared to ASEP. This is likely due to the following reasons. The first is the neglect of the expansion domain. Since it takes a constant time for the expansion domain to traverse a node, the larger the cycle length is, the less impact (the neglect of) the expansion domain has. The fact that the agreement improves with increasing cycle length provides evidence in support of this hypothesis. The second reason is the approximation of the FD for NaSch and the neglect of unstable walls. The neglect of unstable walls implies that we assume a triangle-shaped FD and the same collective velocity within each individual regime (low or high density). In fact, the empirical FD suggests that the maximum flow region is approximately in the density range $\rho\in[0.12, 0.135]$ ($\rho_\mm=0.12$ is used for the SDW model.). The collective velocity in the maximum flow region is small and very different from that when density is close to $0$ or $1$. If the system is expected to derive a maximum flow section, it is likely that the density stays at values higher than $\rho_\mm$. As a result, the valleys of the SDW predictions are usually lower than the CA simulation results.

\rms{Finite-size scaling}

We next investigate the density profiles for a variety of system sizes. We fix the splits at both nodes. We plot the density against the normalised cell location at selected time steps after the start of a cycle. For better illustration, only the results obtained by the CA simulations are shown. 

\begin{figure}[!t]
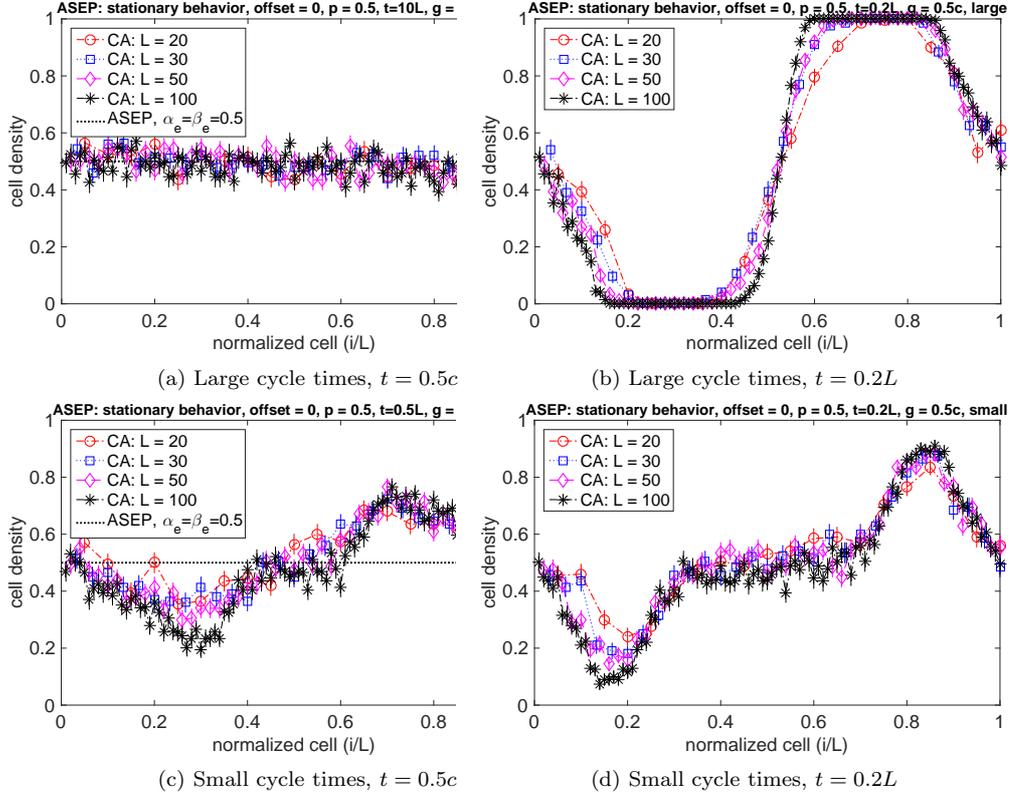

\centering
\begin{tabular}{cc}
\hspace*{-0.95cm}{\ASEPDensityPfiveRescaleRedL}&\hspace{-1.15cm}{\ASEPDensityPfiveRescaleGreenL}\\
{\footnotesize (a) Large cycle times, $t=0.5c$}&\hspace{-1.15cm}{\footnotesize (b) Large cycle times, $t=0.2L$}\smallskip\\
\hspace*{-0.95cm}{\ASEPDensityPfiveRescaleRedS}&\hspace{-1.15cm}{\ASEPDensityPfiveRescaleGreenS}\\
{\footnotesize (c) Small cycle times, $t=0.5c$}&\hspace{-1.15cm}{\footnotesize (d) Small cycle times, $t={0.2L}$}\smallskip\\
\end{tabular}
\vspace{-0.2cm}
\caption{Cell densities against normalised location ($i/L$) for stochastic ASEP with $p=0.5$ at $t$ seconds after a cycle starts (when the system flow is at stationarity). The time steps are rescaled as $t=t'L/10$. For $L=20,30,50,100$, cycle lengths are chosen to be (a)-(b): large cycles, (c)-(d): small cycles. }\label{fig:asep_density_p5_rescale}
\end{figure}

We first look into the stochastic ASEP with equal splits $\gamma=0.5$ and zero offsets $d=0$; see Figures~\ref{fig:asep_density_p5_rescale}. For different system sizes, $L=20,30,50,100$, the time steps are chosen to be proportional to $L$. We consider two ranges of cycle times: small cycles $(c=L)$ and large cycles ($c=20L$). The choices of cycle times correspond to two scenarios: $i$) All the systems are in the transient regime at the light change; and $ii$) All the systems have reached the pseudo-stationary state prior to the end of green light. We choose $c$ proportional to $L$.

With the choice of the long cycles, given the largest system size $L=100$ and $\gamma=0.5$, the green time is greater than $L^{\frac32}$, and thus the pseudo-green-light stationary state is obtained. The relaxation time is expected to be $\propto L^{\frac32}$ \cite{Derrida1993, Pierobon2005}. Figure~\ref{fig:asep_density_p5_rescale}(a) demonstrates this. It shows the density profiles at $t=0.5c$, that is, prior to the start of the red light. The profiles are flat with cell densities approximately $0.5$, which correspond to the ASEP stationary profile in maximum-flow regime. 
In that case, we observe good agreements in the density profiles with different $L$s at different time steps. For example, Figure~\ref{fig:asep_density_p5_rescale}(b) gives the profile results at $0.2L$ seconds after the start of the green light.
When the cycles are relatively short, the systems do not reach pseudo stationarity, as suggested by Figure~\ref{fig:asep_density_p5_rescale}(c). We again see good data collapse; see Figures~\ref{fig:asep_density_p5_rescale}(c) and (d). Later we shall see that this result is analogous to when signals switch much more rapidly in Section~\ref{sec:rapid_switch}.

\begin{figure}[!t]
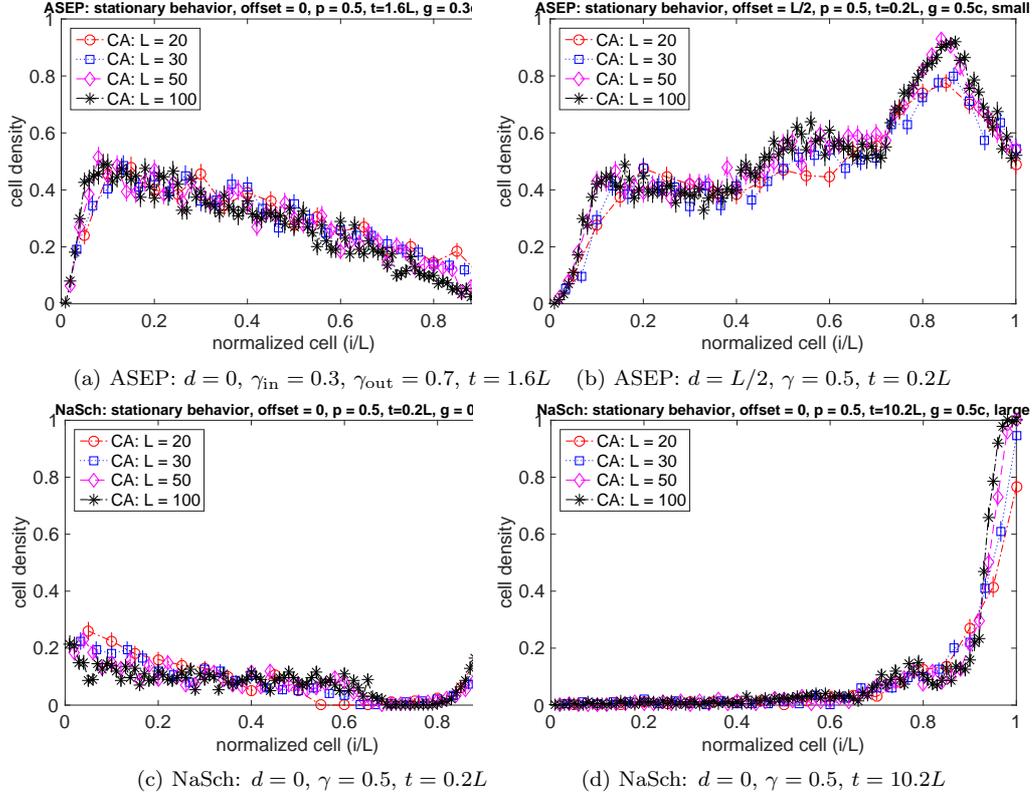

\centering
\begin{tabular}{cc}
\hspace*{-0.95cm}{\ASEPDensityPfiveRescaleUnequalM}&\hspace{-1.15cm}{\ASEPDensityPfiveRescaleOffsetS}\\
{\footnotesize (a) ASEP: $d=0$, $\gamma_\ri=0.3$, $\gamma_\ro=0.7$, $t={1.6L}$}&\hspace{-1.15cm}{\footnotesize (b) ASEP: $d=L/2$, $\gamma=0.5$, $t={0.2L}$}\smallskip\\
\hspace*{-0.95cm}{\NaSchDensityPfiveRescaleGreenL}&\hspace{-1.15cm}{\NaSchDensityPfiveRescaleRedL}\\
{\footnotesize (c) NaSch: $d=0$, $\gamma=0.5$, $t=0.2L$}&\hspace{-1.15cm}{\footnotesize (d) NaSch: $d=0$, $\gamma=0.5$, $t={10.2L}$}\smallskip\\
\end{tabular}
\vspace{-0.2cm}
\caption{Cell densities against normalised location ($i/L$) for stochastic ASEP and NaSch with $p=0.5$ at $t$ seconds after a cycle starts (when the system flow is at stationarity). The time steps are rescaled as $t=t'L/10$. For $L=20,30,50,100$, cycle lengths are chosen to be (a): $c=5L$, (b): small cycles and (c)-(d): large cycles.}\label{fig:density_p5_rescale}
\end{figure}

For unequal splits, Figure~\ref{fig:density_p5_rescale}(a) displays the density profiles for stochastic ASEP with $\gamma_\ri=0.3$, $\gamma_\ro=0.7$, $c=5L$ and $d=0$. The time is $t=1.6L$, that is, when the light at $n_\ri$ just switches to red and the light at $n_\ro$ is still green. With the choices of cycle lengths and signal splits, the green light at $n_{\ri}$ is too short for any system to reach pseudo stationarity. 
For the non-zero offset case, the offset is also set to be proportional to the system size in order to see the profile resemblance. An example is given in Figure~\ref{fig:density_p5_rescale}(b). It displays the density profiles for stochastic ASEP with non-zero offsets $d=L/2$, small cycles and $\gamma=0.5$. The time is $t=0.2L$, that is, when the red light at $n_\ri$ is on for $0.2L$ seconds and the green light at $n_\ro$ is on for $0.2L$ seconds. For both the unequal split and the non-zero offset cases, the density profiles for different $L$s are very similar.

Finally, we investigate the simulation results for NaSch. Figures~\ref{fig:density_p5_rescale}(c) and (d) depict the density profiles at $t=0.2L$ seconds after the green and red lights start, respectively. The other parameters are the same as in Figures~\ref{fig:asep_density_p5_rescale}(a) and (b). The NaSch density profiles show good agreements among various $L$s. Moderate discrepancy is found close to $n_\ro$ due to the finite-size effect. The time for NaSch to reach the pseudo-green-light stationary state is shorter compared to ASEP. Therefore, with the choice of the large cycle lengths, all the systems are at the pseudo-stationarity at late of green light. For the small cycle lengths we also find good data collapse. The results observed for ASEP are qualitatively the same for NaSch.

\section{Stationary Flows}\label{sec:stationary_flow}

The hydrodynamic and the domain wall models are based on prior knowledge of the FD. Thus, one can easily use the FD to obtain the transient and the stationary flows from the density profiles produced by the two models, as discussed in the previous section. In this section we consider the stationary flow, the flow aggregated over a cycle after the system has reached stationarity, since it is important and can be considered as the link capacity. We focus on discussing the impact of the four parameters, $c$, $d$, $\gamma$ and $L$ on the flow.


\subsection{Equal splits and zero offsets}\label{ssec:flow_equal}

Figure~\ref{fig:asep_nasch_flow} displays the CA simulation results of the stationary flows for deterministic and stochastic ASEP and NaSch vs cycle times. For the deterministic case, the results obtained from the DDW model are the same as the CA results. As for the stochastic case, the SDW model predictions are in most cases in good agreement with the CA models (within two standard errors). The discrepancies between the SDW and CA results are mainly derived from the approximation of the FD, and are slightly larger for small system sizes such as $L=10$. Therefore, for a clear illustration of the impact of the parameters on traffic flows, Figure~\ref{fig:asep_nasch_flow} (and Figure~\ref{fig:asep_nasch_flow_2}) depicts the CA results only. We start with the simple case of offset $d=0$ and splits $\gamma_\ri=\gamma_\ro =\gamma= 0.5$, for which the results are given Figures~\ref{fig:asep_nasch_flow}(a)-(d).

\begin{figure}[!t]
\centering
\begin{tabular}{x{5.50cm}x{5.5cm}}
\hspace*{-1.6cm}{\ASEPflowcycleC}&\hspace*{-0.3cm}{\ASEPStflowcycle}\\
{\footnotesize (a) Deterministic ASEP}&{\footnotesize (b) Stochastic ASEP}\\
\hspace*{-1.6cm}{\NaSchflowcycle}&\hspace*{-0.3cm}{\NaSchStflowcycle}\\
{\footnotesize (c) Deterministic NaSch}&{\footnotesize (d) Stochastic NaSch}\\
\hspace*{-1.6cm}{\NaSchflowcycleLB}&\hspace*{-0.3cm}{\NaSchStflowcycleLA}\\
{\footnotesize (e) Deterministic NaSch: $L=25$}&{\footnotesize (f) Stochastic NaSch: $L=10$}\\
\end{tabular}
\vspace{-0.2cm}
\caption{Stationary flow against cycle length for ASEP and NaSch with $\gamma=0.5$ and $p=0, 0.5$. (a)-(d): $d=0$, $L=10,25,50,100$ and (e)-(f): $d=0,20,40,60$. Only even cycle times are considered.}\label{fig:asep_nasch_flow}
\end{figure}

Let us first investigate the results for the deterministic case. We recall that $J_{\max}$ is the maximum flow when setting $\gamma=1$, and $J_{\max}\approx 0.15$  for ASEP and $J_{\max}\approx 0.32$ for NaSch with $v_{\max}=4$. For ASEP, shown in Figure~\ref{fig:asep_nasch_flow}(a), the stationary flow equals $\gamma J_{\max}$ and is independent of the system size $L$ and cycle length $c$ {\footnote{This is true provided that only even green times are considered as the deterministic system is rather sensitive to the parity of parameters $L$, $c$ and $g$. In order to make the arguments simpler, for deterministic NaSch, we only consider even $L$, $c$ and $g$.}}.
This result is intuitively reasonable and agrees with the conclusion made in \cite{Daganzo2008} that the capacity (i.e., the flow) is proportional to the split. The flow during the green time is $J_{\max}$. Thus the average flow over a cycle is $gJ_{\max}/c=\gamma J_{\max}$. For NaSch, due to the existence of the expansion domain, the flow at the start of a green light is lower than $J_{\max}$, which we have discussed in Subsection~\ref{ssec:deterministic_tr}. So the flow is always less than $\gamma J_{\max}$, and approaches $\gamma J_{\max}$ as the cycle time $c$ grows; see Figure~\ref{fig:asep_nasch_flow}(c). A sawtooth structure is observed in the flow, which is essentially the same as the sawtooth (or rectangular) structures found in \cite{Jia2009, Neumann2008, Neumann2009, Jia2011}, due to the deterministic hopping mechanism. Once the green time exceeds $t^\dagger$, each tooth has a constant width of $v_{\max}+1$. The analytical expression of the flow for this NaSch case is given in \ref{app:expansion}.

For the stochastic case, shown in Figures~\ref{fig:asep_nasch_flow}(b) and (d), when the cycle time is small, the average stationary flow $J$ exceeds $\gamma J_{\max}$.  The difference is more pronounced for ASEP. To understand this, let us again start with ASEP.
Suppose we have an initially empty system. The inflow rate is $1$. The system is semi-infinite (i.e., assumed as $[0,+\infty)$), which is assumed simply to eliminate the impact of the outflow rate. We compute the distribution of the accumulated flow $Q(t)$ by time $t$, measured at the in-boundary, for the first few iterations, and obtain the transient flow $J_t$, that is, the average flow over the first $t$ iterations.

\begin{align*}
\mathbb{P}(Q(1)=1) &= 0.5  &&& \Rightarrow J_1&=0.5\\
\mathbb{P}(Q(2)=1) &= 0.75&&& \Rightarrow J_2&=0.375\\
\mathbb{P}(Q(3)=1) &= 0.8125,&\mathbb{P}(Q(3)=2) &= 0.0625& \Rightarrow J_3&=0.3125\\
\mathbb{P}(Q(4)=1) &= 0.7344,&\mathbb{P}(Q(4)=2) &= 0.2031& \Rightarrow J_4&=0.2852\\
\vdots
\end{align*}

The transient flows at the very beginning are much larger than the maximum stationary flow $J_{\max}\approx 0.15$. A very similar scenario occurs when a green light is switched on. The previous red light clears (the end of) the downstream link, and produces a queue on the upstream link, which leads to a large input flow rate for (the start of) the green time. The transient flow during (the start of) the green time is larger than the maximum flow $J_{\max}$. As a result, the average flow exceeds $\gamma J_{\max}$. Although this argument concerns the flow measured at the node, not in the bulk, the stationary flows are the same. Such \textit{start-up} effects are strong when the cycle time and the green time are short. On the other hand, this effect also occurs for higher speed limits $v_{\max} > 1$ but is less pronounced. We observe from Figure~\ref{fig:asep_nasch_flow}(d) that the stationary flows for very small cycle times, less than 20 iterations, are larger than those for large cycle times, although the difference is not significant due to the relatively large statistical fluctuations in the average flow. The reason is because the high flow due to the start-up effect is off-set by the low flow of the expansion domain. However, if the acceleration rate is larger, for example, the infinite acceleration rate model discussed in \cite{Tobita2012}, the start-up effect will be more noticeable and further improve the intersection capacity when relatively short cycle time is implemented.


\subsection{Unequal splits and non-zero offsets}\label{ssec:flow_unequal}

Next we consider non-zero offsets and unequal splits. Figure~\ref{fig:asep_nasch_flow_2} plots the average flows against offsets for deterministic and stochastic ASEP and NaSch. The results for the deterministic cases are included for comparison and explanation purposes, as they are analogous to the stochastic variants. 
We first analyse the equal split results with non-zero offsets given in Figure~\ref{fig:asep_nasch_flow_2}(a), (c) and (e). For zero-offsets, the stationary flow stays approximately at $\gamma J_{\max}$ independent of the cycle length. For $\gamma > 0$, however, if the system is relatively small, namely, $L<\gamma cJ_{\max}$, the flow varies with offsets and cycle lengths. Figures~\ref{fig:asep_nasch_flow_2}(a), (c) and (e) suggest that:

\begin{itemize}
\item The flow maximum is obtained by a range of offsets instead of only one optimal offset. This agrees with the result found in \cite{Geroliminis2012}.
\item The flow starts to decrease once the offset exceeds the \textit{leading offset}, which is defined to be $d_l = L/v_{\rm free}$, where $v_{\rm free}$ is the free flow speed. 
The leading offset is the time it takes for the traffic platoon to travel from nodes $n_\ri$ to $n_\ro$.
\item The flow reaches the minimum $L\rho_{\rm jam}/c = L/c$ at offset $d=g-d_d$. The \textit{dissipation offset} $d_d = L/v_{\rm hole}$ is the time it takes for the hole (i.e., empty cell) to travel from nodes $n_\ro$ to $n_\ri$ through a queue. We remark that, for the deterministic case with the triangular FD, the leading and dissipation offsets are equivalent to the forward and backward offsets, respectively, used in \cite{Daganzo2016}.
\item Further increasing the offset from offset $d = c-g-d_l$ raises the flow linearly. By offset $d = c-d_d$ the flow returns to the maximum $\gamma J_{\max}$. 
\end{itemize}

\begin{figure}[!t]
\centering
\begin{tabular}{x{5.50cm}x{5.5cm}}
\hspace*{-1.6cm}{\ASEPflowoffset}&\hspace*{0.05cm}{\ASEPflowratio}\\
{\footnotesize (a) Deterministic ASEP: $c=140$, $\gamma=0.5$}&{\footnotesize (b) Deterministic ASEP: $c=140$, $L=10$}\\
\hspace*{-1.6cm}{\NaSchflowoffset}&\hspace*{-0.25cm}{\NaSchflowratio}\\
{\footnotesize (c) Deterministic NaSch: $c=140$, $\gamma=0.5$}&{\footnotesize (d) Deterministic NaSch: $c=140$, $L=25$}\\
\hspace*{-1.6cm}{\NaSchStflowoffset}&\hspace*{-0.25cm}{\NaSchStflowratioLA}\\
{\footnotesize (e)  Stochastic NaSch: $c=180$, $\gamma=0.5$}&{\footnotesize (f) Stochastic NaSch: $c=180$ and $L=10$}\\
\end{tabular}
\vspace{-0.2cm}
\caption{Comparison of average flow against offset in relation to cycle lengths, system sizes, splits for ASEP and NaSch.   Left: fixed cycle lengths, equal splits $\gamma=0.5$ and various system sizes $L=10,25,50,100$; Right: fixed system sizes, fixed cycle lengths and various splits $\gamma_\ri=0.25,0.5,0.75$, $\gamma_\ro=0.5$.}\label{fig:asep_nasch_flow_2}
\end{figure}

To obtain a stationary flow of $\gamma J_{\max}$, the flow through at least one of the nodes during the green time needs to be $J_{\max}$. In other words, spill-backs cannot reach the upstream node $n_\ri$ during its green time, and low-flow regions, including the empty region, cannot be at the downstream node $n_\ro$ during its green time, since both scenarios cause wasted green times. In terms of the traffic condition prior to the green light, there are two extreme cases: The bulk link is completely i) empty (i.e., no residual queue at $n_{\rm out}$) and ii)  jammed. For case i), in order to avoid wasted green times, node $n_\ro$ should start its green light $d_l = L/v_{\rm f}$ iterations after node $n_\ri$ does, that is when the vehicle platoon arrives at $n_\ro$. For case ii), node $n_\ro$ should start its green light $d_d=L/v_{\rm hole}$ iterations ahead of node $n_\ri$. This is the time when the first hole travels from node $n_\ro$ to $n_\ri$, and also the time for the queue to dissipate. 

On the other hand, the minimum stationary flow corresponds to $L$ vehicles traversing through the node every $c$ seconds. That is to say, the average flow at stationarity is never below $L/c$. For the case of no residual queue, if node $n_\ro$ switches on the red light when the platoon arrives, then the link can store up to $L$ vehicles. In that case the offset is $d_l+c-g$. For the fully jammed case, if node $n_\ri$ switches on the red light when the first hole arrives, then only $L$ vehicles can traverse node $n_\ro$ every cycle. The corresponding offset is $g-d_d$.

In terms of maximising flows, the optimum is obtained in the offset range $[0, d_l]\cup[c-d_d, c)$. The optimal offset has a wider range if the system size is larger. However, in terms of minimising travel times or delays on the bulk link, the optimal offset is $d_l$, for which no vehicles experience delays induced by red signals at node $n_\ro$. That corresponds to the optimal \textit{synchronization delay} for the low density region in \cite{Huang2003}. Zero offset (also known as simultaneous offset) always optimises the system flow with arbitrary  $L$ and $c$. This can be seen from the analytical expressions of the system flow described in \ref{app:deterministic}. The black wall trajectories in Figure~\ref{fig:deterministic_walls} produced by DDW may help understand this result. For $d=0$, no matter how one varies $L$ and $c$ and what the initial condition is, wasted green times never occur at either of the nodes. So the stationary flow will not be lower than $\gamma J_{\max}$. 
To some extent, our finding agrees with the result in \cite{Girault2016} that for grid networks good signal coordination provides little advantage over simultaneous offsets. 

Figures~\ref{fig:asep_nasch_flow}(e) and (f) plot the flow against cycle length for small NaSch systems with $\gamma=0.5$. We choose small system sizes so that we can see the impact of the offset and cycle length. The capacity of the system is constrained by the split $\gamma$. If the cycle length $c$ is relatively small, then the flow can be close to its $\gamma J_{\max}$. Once the cycle length exceeds a threshold, which depends on $L$ and $d$, the flow decreases rapidly with $c$. Regardless of $L$ one can tune either cycle length or offset to obtain the maximum flow. The results provide a preliminary guidance in choosing cycle lengths and offsets for a given system size. The analytical expression for the flow in relation to $c,\ d,\ \gamma$ and $L$ for the deterministic case is given in Eq.~\eqref{equ:asep_deterministic_flow}. One can extend it qualitatively to explain the stochastic case.

\medskip

Although we have focused on the discussion of equal splits $\gamma=0.5$, the cases with $\gamma\neq 0.5$ are qualitatively the same. Now let us fix the split $\gamma_\ro$ at node $n_\ro$ at $0.5$ and vary the split $\gamma_\ri$ at node $n_\ri$. Figures~\ref{fig:asep_nasch_flow_2}(b), (d) and (f) compare the simulation results of the average flow against offset for $\gamma_\ri=0.25,0.5$ and $0.75$. The general shapes are similar to the curves with equal splits in Figures~\ref{fig:asep_nasch_flow_2}(a), (c) and (e). But the values of the plateaus and turning points change. The maximum flow is determined by the node with the smaller capacity, that is, $\min\{{\gamma_\ri},{\gamma_\ro}\}J_{\max}$. The minimum flow becomes the larger value of $ (\gamma_\ri-\gamma_\ro)J_{\max}$ and $L/c$. So if the split $\gamma_\ri$ is significantly larger than $\gamma_\ro$, the minimum flow can be higher than that with the equal split. The four turning points also change and depend on both $\gamma_\ri$ and $\gamma_\ro$. The analytical expressions for the corresponding offset values are given in Eqs.~\eqref{equ:point_A}-\eqref{equ:point_D}. Again the stochastic results are qualitatively the same as the deterministic ones.


\section{Rapid Switching}\label{sec:rapid_switch}

So far, we have focused on regular (relatively large) cycle lengths and green times in the realistic ranges $20\leq c\leq 200$ and $5\leq g \leq 150$. One may wonder how the system would behave if the signal is switched very rapidly, for example every 2 seconds. One practical implementation of rapidly switched signals is in ramp-metering. This kind of signal switching frequency is not practical for arterial networks, for reasons of safety.  Nevertheless, it is of interest to see what the impact of rapid switching has on the traffic behaviour. We only consider the stochastic case in this section.


\subsection{Equal splits}

Comparing the results for different cycle times in Figure~\ref{fig:density_p5_st}, we find that, for equal splits $\gamma=0.5$, the bulk density tends to be more linear when the cycle gets shortened. When we switch the lights fast enough, e.g., $c=4$ and $g=2$ as chosen for Figure~\ref{fig:density_p5_rapid}(a) and (b), the density profile becomes linear, except for cells close to the nodes. The systems with varying sizes have the same profile. One can determine the density at the middle site using a domain wall argument. Every switch from red light to green light creates a downward wall, $W_{\mm|\me}$ at $n_{\ri}$ and $W_{\mc|\mm}$ at $n_{\ro}$. 
Walls $W_{\mm|\me}$ and $W_{\mc|\mm}$ meet at the middle of the system since their drift velocities are approximately the same. The average density at the middle can be then computed by $\rho_m=((\rho_\mm+\rho_\me)/2+(\rho_\mm+\rho_\mc)/2)/2 = \rho_\mm/2+0.25$, which is $0.5$ for ASEP and about $0.32$ for NaSch.

\begin{figure}[!t]
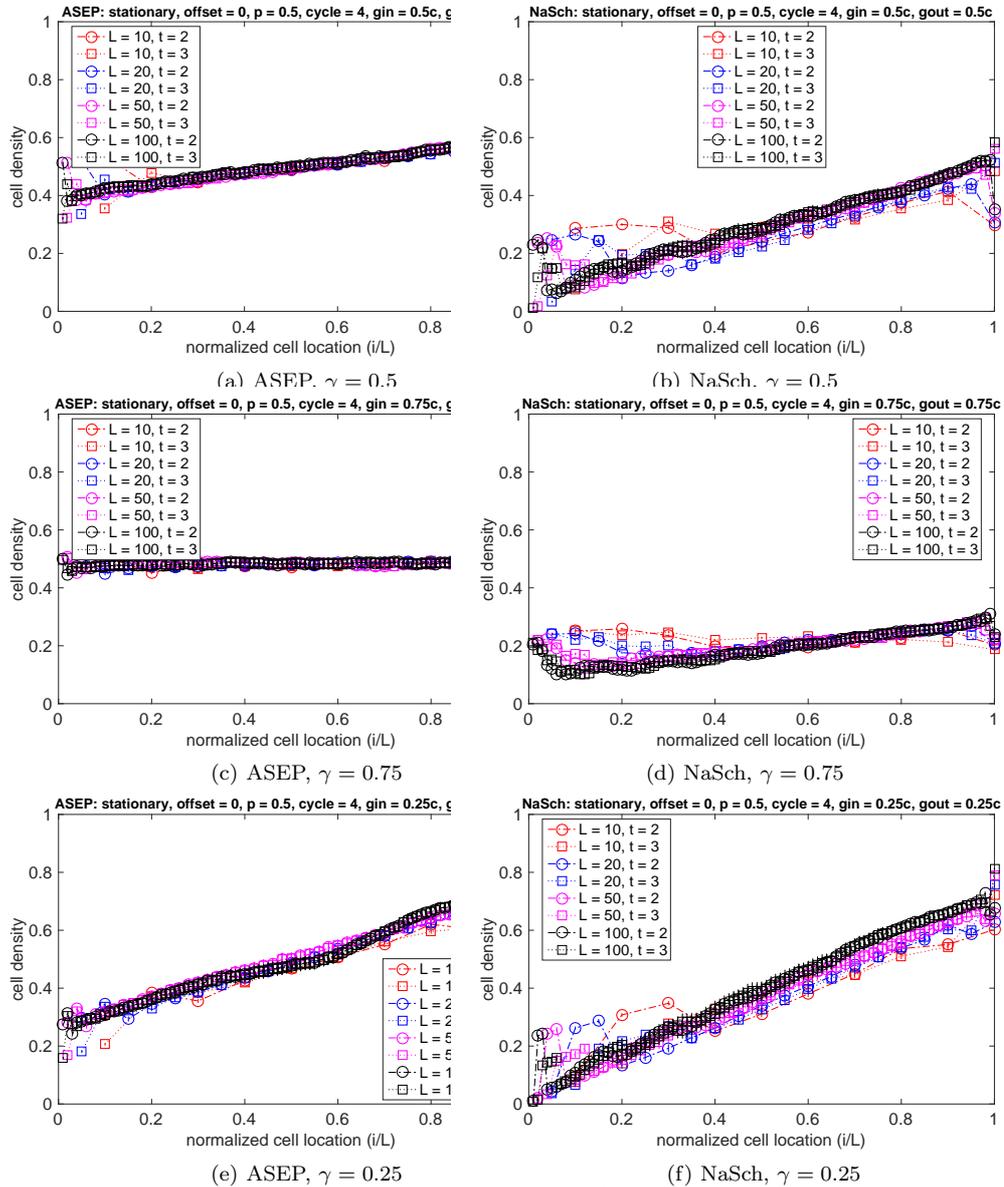

\centering
\begin{tabular}{cc}
\hspace*{-0.95cm}{\ASEPDensityPfiveRapidRed}&\hspace*{-1.1cm}{\NaSchDensityPfiveRapidRed}\\
{\footnotesize (a) ASEP, $\gamma=0.5$}&\hspace*{-1.0cm}{\footnotesize (b) NaSch, $\gamma=0.5$}\\
\hspace*{-0.95cm}{\ASEPDensityPfiveRapidRedH}&\hspace*{-1.1cm}{\NaSchDensityPfiveRapidRedH}\\
{\footnotesize (c) ASEP, $\gamma=0.75$}&\hspace*{-1.0cm}{\footnotesize (d) NaSch, $\gamma=0.75$}\smallskip\\
\hspace*{-0.95cm}{\ASEPDensityPfiveRapidRedL}&\hspace*{-1.1cm}{\NaSchDensityPfiveRapidRedL}\\
{\footnotesize (e) ASEP, $\gamma=0.25$}&\hspace*{-.50cm}{\footnotesize (f) NaSch, $\gamma=0.25$}\\
\end{tabular}
\vspace{-0.4cm}
\caption{Cell densities against normalised cell location $i/L$ for stochastic ASEP and NaSch with various system sizes $L=10, 20, 50, 100$, $p=0.5$, $d=0$, $c=4$ and $\gamma=0.5, 0.75, 0.25$ at $t=2,3$ seconds after a cycle starts when the system at stationarity.
}\label{fig:density_p5_rapid}
\end{figure}

The linear profile coincides with the linear profile for ASEP and NaSch with constant boundaries when inflow and outflow rates are equal and less than the critical values for obtaining maximum flow \cite{Schadschneider2011}. When traffic lights change fast, the nodes `transform' the boundary conditions to some effective inflow rate $\alpha_{\rm e}$ and outflow rate $\beta_{\rm e}$. Since the splits $\gamma$ are equal, the effective inflow and outflow rates are the same. With the growth of $\gamma$, the effective rates increase, and the slope of the profile drops as shown in Figures~\ref{fig:density_p5_rapid}(c) and (d).  Once they are above the critical values, the profile becomes flat, and the system is in maximum flow. See Figure~\ref{fig:density_p5_rapid}(c). On the other hand, if $\gamma$ decreases, the slope of the profile increases. See Figures~\ref{fig:density_p5_rapid}(e) and (f). 

\begin{figure}[!h]
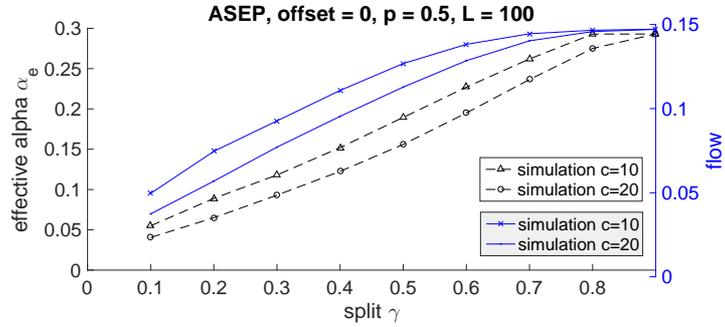

\centering
{\ASEPEffectiveInflow}
\vspace{-0.4cm}
\caption{Average flow $J$ and effective inflow $\alpha_{\rm e}$ against split $\gamma$ for stochastic ASEP, $p=0.5$, $L=100$, $d=0$ and $c=10,20$. The effective rate $\alpha_{\rm e}$ and the flow are bounded by the critical rate $\alpha_{\rm c}$ and $J_{\max}$ respectively.}\label{fig:asep_effective_a}
\end{figure}

\smallskip
For ASEP, we know the analytical expression of the flow in terms of the inflow and outflow rates (see Eq.~\eqref{eq:JrhoDASEP}). Thus, we can reverse-engineer the effective rate $\alpha_{\rm e}$ ($=\beta_{\rm e}$) for varying $\gamma$ according to the average system flow. Figure~\ref{fig:asep_effective_a} plots the average flow and the effective inflow rate against the split. In order to obtain more data points, relatively long cycles $c=10,20$ and a large system size $L=100$ are considered. Due to the start-up effect discussed in Section~\ref{sec:stationary_flow}, a smaller $c$ leads to a higher average flow, and thus a larger $\alpha_{\rm e}$. The results show some linear relation between $\alpha_{\rm e}$ and $\gamma$ for intermediate values of $\gamma$ for both cycle lengths. Of course, when the split $\gamma$ is sufficiently large, $\alpha_{\rm e}$ will simply plateau at the critical value $\alpha_{\rm c}=1-\sqrt{1-p}$. For small $\gamma$, the start-up effect raises the flow and $\alpha_{\rm e}$ to higher values. 


\subsection{Unequal splits}

When the splits at two nodes are different, the system may reach either low or high density depending on which node has the smaller split, provided that the effective inflow and outflow rates do not exceed the critical values. For example, Figure~\ref{fig:density_p5_rapid_unequal}(a) shows that the stochastic ASEP, $\gamma_\ri=0.5$ and $\gamma_\ro=0.75$ produces an approximately flat density profile in low density. Figure~\ref{fig:density_p5_rapid_unequal}(b) displays the result for the stochastic NaSch under the same setting. The resulting density is very low, less than $0.1$, and oscillation waves are present in the density profile. Such oscillation waves are essentially domain walls, and so are affected by the cycle length, the smaller split and the system size. The node with the smaller split (and hence the lower capacity) determines the system capacity and the amplitude of the oscillation. The cycle length determines the width between two waves. The system size determines how many oscillation waves are in the system.

\begin{figure}[!t]
\centering
\begin{tabular}{cc}
\hspace*{-0.95cm}{\ASEPDensityPfiveRapidRedU}&\hspace*{-1.1cm}{\NaSchDensityPfiveRapidRedU}\\
{\footnotesize (a) ASEP}&\hspace{-1.0cm}{\footnotesize (b) NaSch}\\
\end{tabular}
\vspace{-0.4cm}
\caption{Cell densities against normalised cell location $i/L$ for stochastic ASEP and NaSch with various system sizes $L=10, 20, 50, 100$, $d=0$, $c=4$ and $\gamma_\ri=0.5,\gamma_\ro= 0.75$ at $t=2,3$ seconds after a cycle starts when the system is at stationarity.}\label{fig:density_p5_rapid_unequal}
\end{figure}

Light switching creates domain walls. The more rapid the switching is, the closer those walls are. They smear out if they are unstable while they drift, which reduces the density contrast between two domains and smooths the local densities. The smoothing process is faster when walls are closer to each other. As a result, the rapid switching leads to a linear density profile. We note that the smoothing does not require all walls to be unstable. However, if all domain walls are stable, they preserve their own profiles while they travel. An extreme case is when $p=0$. One can plot the domain wall trajectories using the DDW model as we did in Figure~\ref{fig:deterministic_walls} and find the oscillation in the density profile. 
We recall that domain walls are stable if they are upward walls or the collective velocity is the same as the drift velocity. The FD of NaSch at very low density is basically linear, and so walls with both domains from that density region are stable. 
These stable walls result in the oscillation waves in the density profile. This explains why we observe the oscillation at low densities for NaSch in Figures~\ref{fig:density_p5_rapid}(b), (d), (f) and \ref{fig:density_p5_rapid_unequal}(b). It also explains why such oscillation is absent for ASEP. With the increase (or decrease) of $p$, the linear section on the FD of NaSch reduces (or grows), and as a result the oscillation becomes weak (or strong) in terms of magnitude. Finally, domain walls perform random walks for $0<p<1$. Thus, the oscillation amplitude always reduces, though possibly very slowly, e.g., the density profiles for $L=100$ from upstream to downstream in Figure~\ref{fig:density_p5_rapid_unequal}(b). If letting $L\rightarrow \infty$, the density profile at the downstream side would eventually become flat.

\section{Discussion}\label{sec:discussion}

We have studied the traffic behaviour, in both transient and stationary states, on a single link with traffic lights at both ends, for a variety of parameters. In particular we have studied the impact of various cycle lengths, split times and offsets. We have extended the domain wall and hydrodynamic theories to predict the traffic behaviour for time-dependent boundary conditions. {We have considered both the deterministic and stochastic models.}\medskip

Our main findings including the following:
\begin{itemize}
	\item The maximum stationary flow is mainly determined by the smaller split, and the minimum is determined by the system size, cycle length and the difference in splits. If the cycle length is relatively large, greater than the system size divided by the maximum flow $J_{\max}$ (with the absence of traffic lights), the stationary flow depends on the offset. Specifically, it varies linearly with the offset when it is not at the extrema. The relation of the stationary flow with those system parameters for the deterministic ASEP is derived analytically, which is (qualitatively) extended to describe deterministic and stochastic NaSch. For the stochastic models, the clearance of downstream traffic during red lights helps improve the flow at the start of green light. Therefore, the stationary flows for smaller cycle times can be larger than the product of the split and $J_{\max}$.
	\item For the deterministic case, the deterministic domain wall model can produce exactly the same results for both the stationary flow and the transient density profile as the deterministic ASEP and NaSch model. As for the stochastic case, the hydrodynamic model can approximate the density evolution for short-time scales, but not for long-time scales due to the lack of stochasticity. The stochastic domain wall model provides quantitative descriptions of the density evolution in both transient and stationary states. 
	\item For the stationary stochastic ASEP and NaSch, the cell density profiles scale with the system size, provided that the systems are in the same traffic regime at the traffic signal change.
	\item If the signal switching is sufficiently rapid, the traffic system behaves like the system with some effective time-invariant boundaries. For equal splits, the system is on coexistence line. For moderate split values, there exists a linear relation between the split and the effective in- and outflow rates.

\end{itemize}


Our study fills the gap between one-dimensional traffic systems and two-dimensional traffic networks. The results of the impact of the cycle length, split and offset relative to the system size on the traffic flow and the derivation of formulas for estimating the effective capacity provide useful insight in choosing key parameters for traffic signal control. This work can be generalised to study urban traffic network flows in relation to signal settings and the application of traffic management such as perimeter control \cite{Keyvan-Ekbatani2013,Kutadinata2016}. The analysis of the local density transitional behaviour for various signal settings allows us to estimate queues on links (in the circumstance of large inflow and outflow), which would be instructive for signal control design to prevent spill-backs. In addition to the comprehensive study of how traffic signals affect traffic flow and local density, this paper provides an evaluation of three models with various complexities. The macroscopic domain wall model is validated by the CA model in terms of its ability in predicting traffic behaviours with signals in both long and short terms. The hydrodynamic model is simple and efficient, and can provide good short-term forecasts of traffic evolutions.

\section*{Acknowledgments}
This work was supported under the Australia Research Council (ARC) Centre of Excellence for Mathematical and Statistical Frontiers (ACEMS). This research was undertaken with the assistance of resources provided at the NCI National Computational Merit Allocation Scheme supported by the Australian Government. 

\appendix
\section{Fundamental diagrams for NaSch}\label{app:fd}

Figure~\ref{fig:nasch_fd} shows the simulated FD of NaSch with $v_{\max}=4$. The SDW model considered in this paper for NaSch with $v_{\max}=4$ ignores the instability downward domain walls, and does not consider wall splitting. Therefore, we only use the empirical data of critical density and maximum flow $(J_{\mm},\rho_{\mm}) = (0.32,0.12)$.  That is equivalent to assuming that the FD has a perfect triangular shape. For the hydrodynamic model, we fit the simulated FD to
\begin{equation}
    J(\rho) = \begin{cases}
        \displaystyle J_{a_{\text{LD}}}\left(\frac{\rho - \rho_\mm}{\rho_\mm}\right)
        & \rho \le \rho_{\mm},\smallskip
        \\
        \displaystyle J_{a_{\text{HD}}}\left(\frac{\rho - \rho_\mm}{1 - \rho_\mm}\right)
        & \rho > \rho_{\mm},
    \end{cases}
\end{equation}
where
\begin{equation}
    J_a(\rho) = \frac{J_{\mm}}{1 - \sqrt{1 - a}}\left(
                    1 - \sqrt{1 - a(1 - \rho)(1 + \rho)}
                \right),
\end{equation}
with $a_{\text{LD}}$ and $a_{\text{HD}}$ the fit parameters.  This form meets the requirements of the hydrodynamics calculations, as outlined at the end of Subsection \ref{ssec:hydro}.  We take $(J_{\mm},\rho_{\mm})$ as before and from the fit obtain $(a_{\text{LD}}, a_{\text{HD}}) = (0.88, 0.99)$.

\begin{figure}[!h]
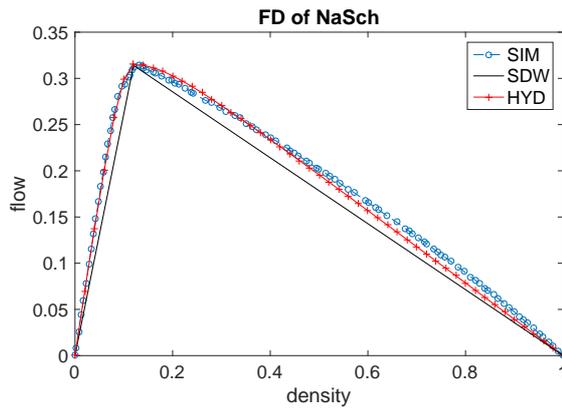

\centering
{\naschFD}
\vspace{-0.4cm}
\caption{Fundamental diagram for stochastic NaSch with $v_{\max}=4$ and $p=0.5$. Blue circles: the simulated FD; Black line: the FD used by SDW; Red crosses: the FD used by hydrodynamics.}\label{fig:nasch_fd}
\end{figure}

\section{Deterministic variants}\label{app:deterministic}

In this section we present some supplementary results produced by the CA simulations and the DDW model for deterministic ASEP and NaSch. These results are analogous to and thus help understand the findings for the stochastic variants.

\subsection{Stationary flows}\label{app:flow}

The stationary flow variations in relation to $c$, $d$, $L$, $\gamma_\ri$ and $\gamma_\ro$ have been discussed in Section~\ref{sec:stationary_flow}.  Figure~\ref{fig:asep_trajectory_L_d} plots an example of the wall trajectories produced by the DDW model, which corresponds to the result of the flow against offset shown in Figure~\ref{fig:asep_nasch_flow_2}(a) for short cycle times. The wall trajectories provide a direct illustration of the impact of offsets. Following the discussion of the relation between the stationary flow and the system parameters in Subsection~\ref{ssec:flow_equal}, we can derive the analytical expression of the stationary flow for equal splits $\gamma$ as below.


\begin{figure}[!t]
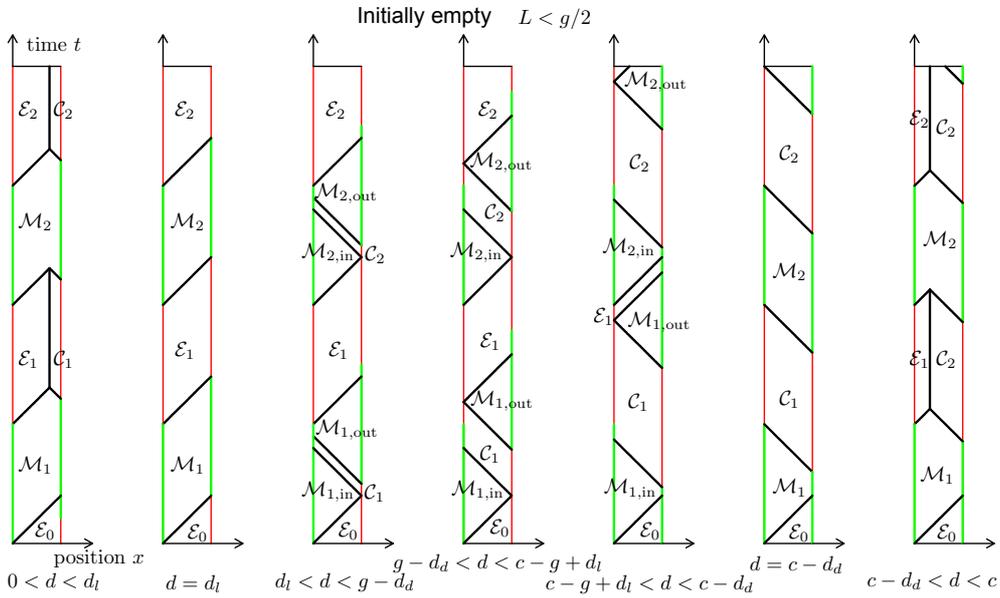

\centering
{\asepTrajectorySmallAlld}\\
\vspace{-0.4cm}
\caption{Domain wall trajectories for deterministic ASEP with $\gamma=0.5$, $L<g/2$ and various offsets. X-axis: position in the system; Y-axis: time. Green and red lines represent the traffic lights at nodes $n_\ri$ and $n_\ro$. From left to right, the offsets are: $0<d<d_l$,  $d = d_l$, $d_l < d < g-d_d$, $g-d_d\leq d \leq c-g+d_l$, $c-g+d_l<d<c-d_d$, $d=c-d_d$ and  $c-d_d<d<c$, which match the offset ranges shown in Figure~\ref{fig:asep_nasch_flow_2}(a).}\label{fig:asep_trajectory_L_d}
\end{figure}

\begin{equation}
J=\begin{cases} \gamma J_{\max}, &
\mbox{if }\displaystyle c \leq \frac{L}{\gamma J_{\max}}\mbox{ or } d \in [0, d_l]\cup [c-d_d,c);\smallskip\\
\displaystyle\frac{J_{\max}({\gamma c-d_d-d}+L)}{c},&
d\in(d_l, \gamma c-L);\smallskip\\
\displaystyle\frac{L}{c},&
d\in[\gamma c-d_d, c(1-\gamma)-d_l];\smallskip\\
\displaystyle\frac{J_{\max}(d-d_l-(1-\gamma)c+L)}{c},&
d\in (c(1-\gamma)+d_l,c-d_d).\\
\end{cases}
\label{equ:asep_deterministic_flow}
\end{equation}

For the case $\gamma_\ri\neq \gamma_\ro$, the result for ASEP of a small system $L=10$ is given in Figure~\ref{fig:asep_nasch_flow_2}(b). Using the DDW theory, we find the offset values for the four points, labelled as $A,B,C$ and $D$ in Figure~\ref{fig:asep_nasch_flow_2}(b), where the flow starts to increase, decrease and stay constant. Let $(x)^+ = x$ if $x> 0$ and $0$ otherwise. Points A and B are at
\begin{eqnarray}
d_A&=&c(\gamma_\ri-\gamma_\ro)^++d_l,\label{equ:point_A}\\
d_B&=&c-c(\gamma_\ro-\gamma_\ri)^+-d_d.
\end{eqnarray}
The two corners of the minimum flow plateau, points C and D are at:
\begin{eqnarray}
d_C\!\!&=\!&\left\{\begin{array}{*2{>{\displaystyle}l}}
\gamma_\ri c-\frac{L}{J_{\max}}+d_l,&\mbox{ if } \frac{L}{cJ_{\max}} > \gamma_\ri-\gamma_\ro,\medskip\\
\gamma_\ro c+d_l,&\mbox{ otherwise;}\\
\end{array}\right.\label{equ:point_C}\\
d_D\!\!&=\!&\left\{\begin{array}{*2{>{\displaystyle}l}}
c(1-\gamma_\ro)+\frac{L}{J_{\max}}-d_d,&\mbox{ if } \frac{L}{cJ_{\max}} >\gamma_\ri-\gamma_\ro,\medskip\\

c(1+\gamma_\ri-2\gamma_\ro)-d_d,&\mbox{ otherwise.}\\
\end{array}\right.\label{equ:point_D}
\end{eqnarray}
The above expressions match the results for the case with $\gamma=0.5$ illustrated in Figure~\ref{fig:asep_nasch_flow_2}(a), noticing that $L/J_{\max}=d_l+d_d$.

\subsection{Expansion domain for deterministic NaSch with $v_{\max}>1$}\label{app:expansion}

The expansion domain $\mn$ exists for NaSch with $v_{\max}>1$. Due to its existence, the stationary flow of the deterministic NaSch is lower than $\gamma J_{\max}$ for small cycle lengths (see Figure~\ref{fig:asep_nasch_flow}). For $d=0$ and $\gamma=0.5$, we can write the average flow as a function of the green time $g$. Recall that $t^\dagger=\sum_{i=2}^{v_{\max}} i$ is the time for $\mn$ to clear a node $n$ and is also its full length, and $q^\dagger$ is the total number of vehicles that have traversed the node by time $t^\dagger$. 
If $g < t^\dagger$, we let $q=\max\{j: \sum_{i=1}^{j} (i+1) - 1< g\}$,  and then have
\begin{equation}
J=\frac{1}{c}\left(\sum_{i=1}^q i + g-\sum_{i=1}^q (i+1) - 2\right).\label{equ:nasch_flow_d0_a}
\end{equation}
Otherwise, for $L\% (v_{\max}+1) = 0$ ($\%$ is the modulo operation.), 
\begin{equation}
J=\frac{q^\dagger}{c}+\frac{1}{c}\left[\left\lfloor \frac{g-t^\dagger}{v_{\max}+1}\right\rfloor+ g-t^\dagger\%(v_{\max}+1)-1\right].\label{equ:nasch_flow_d0_b}
\end{equation}
For other $L$ values, the flow may slightly deviate from Eq.~(\ref{equ:nasch_flow_d0_b}) as the flow result is sensitive to the parity of the parameters. 

\medskip
For the definitions of the leading and dissipation offsets, we have used free flow speed $v_{\rm f}$ and hole speed $v_{\rm hole}$. For deterministic NaSch, regardless of the value $v_{\max}$, $v_{\rm hole} = 1$. If $v_{\max}=1$, $v_{\rm f}=1$. Otherwise, given some travel time $t$,
\begin{equation*}
 v_{\rm f} = \left\{\begin{array}{*2{>{\displaystyle}l}}
 \frac{1}{t}{\sum_{i=1}^t i}&\mbox{ if } t< v_{\max},\medskip\\
 \frac{1}{t}\!\!\left[\sum_{i=1}^{v_{\max}-1} i+v_{\max}(t-v_{\max}+1)\right]&\mbox{ otherwise.}\\
 \end{array}\right.
 \end{equation*}


\subsection{Initial-condition-dependent stationary states} \label{app:reducible}

Mentioned Subsection~\ref{ssec:deterministic_tr}, when two splits are equal, the stationary density profile for deterministic NaSch may also depend on the initial condition. Examples are shown in Figures~\ref{fig:deterministic_walls} and \ref{fig:asep_trajectory_L_d}. For the systems illustrated in Figure~\ref{fig:deterministic_walls}, in most cases, different initial conditions result in different wall trajectories and stationary states. As for the system given in Figure~\ref{fig:asep_trajectory_L_d}, although the initial conditions for varying offsets are all empty state, one can choose arbitrary other initial states and will find that the stationary configurations are the same for respective offset values.
For the special case of ASEP and equal splits, we find that the stationary state is unique if $L\leq g-d$, $g-(c-d)$ or $g/2$.  


\clearpage
\newpage
\section*{References}
\bibliographystyle{elsarticle-num}
\bibliography{td_boundary}

\end{document}